\documentclass[journal]{IEEEtran}
\usepackage{amsmath,cite}
\usepackage{amsfonts}
\usepackage{amsthm}
\usepackage{amssymb}
\usepackage{extarrows}
\usepackage{multirow}
\usepackage{graphicx}
\usepackage{fancyhdr}
\usepackage[justification=centering]{caption} 
\usepackage{CJK}
\usepackage{array,graphicx}
\usepackage{color}
\usepackage{subfigure}
\usepackage[linesnumbered,ruled,vlined]{algorithm2e}
\usepackage{booktabs}
\usepackage{diagbox}
\usepackage{longtable}
\usepackage{rotating}
\usepackage{multirow}
\usepackage{epstopdf}

\usepackage[hang,flushmargin]{footmisc}
\usepackage{lipsum}
\makeatletter
\newcommand{\algorithmfootnote}[2][\footnotesize]{%
	\let\old@algocf@finish\@algocf@finish
	\def\@algocf@finish{\old@algocf@finish
		\leavevmode\rlap{\begin{minipage}{\linewidth}
				#1#2
		\end{minipage}}%
	}%
}

\allowdisplaybreaks

\definecolor{MyColor}{RGB}{182,222,192}

\newtheorem{definition}{Definition}
\newtheorem{theorem}{Theorem}

\makeatletter
\newcommand{\thickhline}{%
    \noalign {\ifnum 0=`}\fi \hrule height 1pt
    \futurelet \reserved@a \@xhline
}
\newcolumntype{I}{!{\vrule width 1pt}}
\makeatother

\begin{document}
\title{Privacy-preserving Medical Treatment System through Nondeterministic Finite Automata}
\author{
Yang Yang,~\IEEEmembership{Member,~IEEE,}
Robert H. Deng,~\IEEEmembership{Fellow,~IEEE,}
Ximeng Liu, Yongdong Wu,\\
Jian Weng,
Xianghan Zheng,
Chunming Rong
\thanks{Y. Yang is with 
	School of Information Systems, Singapore Management University, Singapore; 
	College of Mathematics and Computer Science, Fuzhou University, Fujian, China; 
	State Key Laboratory of Integrated Services Networks (Xidian University);
	Guangdong Provincial Key Laboratory of Data Security and Privacy Protection,
	Guangzhou, China; Fujian Provincial Key Laboratory of Information Processing and Intelligent Control (Minjiang University), Fuzhou, China (e-mail: yang.yang.research@gmail.com). R. Deng is with the School of Information Systems, Singapore Management University, Singapore. Y. Wu and J. Weng are with Department of Computer Science, Jinan University,
Guangdong, China. X. Liu and X. Zheng are with College of Mathematics and Computer Science, Fuzhou University, Fujian, China. C. Rong is with Department of Electronic Engineering and Computer Science, University of Stavanger, Norway; and MingByte Technology, Qingdao, China.}
}

\maketitle

\begin{abstract}
In this paper, we propose a privacy-preserving medical treatment system using nondeterministic finite automata (NFA), hereafter referred to as P-Med, designed for the remote medical environment. P-Med makes use of the nondeterministic transition characteristic of NFA to flexibly represent the medical model, which includes illness states, treatment methods and state transitions caused by exerting different treatment methods. A medical model is encrypted and outsourced to the cloud to deliver telemedicine services. Using P-Med, patient-centric diagnosis and treatment can be made on-the-fly while protecting the confidentiality of a patient's illness states and treatment recommendation results.
Moreover, a new privacy-preserving NFA evaluation method is given in P-Med to get a confidential match result for the evaluation of an encrypted NFA and an encrypted data set, which avoids the cumbersome inner state transition determination. We demonstrate that P-Med realizes treatment procedure recommendation without privacy leakage to unauthorized parties. We conduct extensive experiments and analyses to evaluate efficiency.
\end{abstract}


\IEEEpeerreviewmaketitle

\section{Introduction}
\IEEEPARstart{T}{he} aging of population and prevalence of chronic illnesses have exacerbated many social problems.  Remote diagnosis and treatment systems, which make use of information technology to provide accessible, cost-effective, and high-quality clinical healthcare services remotely, can be deployed
to alleviate some of the problems. Such a system makes it possible for continued treatment in a home environment and increases patient adherence to medical recommendations \cite{Young18}.  The medical Internet of Things (mIoT) plays a critical role in distant medical diagnosis and treatment by deploying
wireless wearable (or implantable) sensors on a patient to collect the vital signs and physiological data \cite{Verma17,YangYTIFS15,Kumar18}.
The monitored physiological parameters are sent to hospital for medical diagnosis, which supplies rich longitudinal health records than the brief illness description. Using the detailed monitoring data, physicians can make a much better prognosis for the patient and recommend treatment, early intervention and drug adjustment that are effective for disease recovery.
The key factor for the accuracy of remote medical diagnosis and treatment is the physician's expertise and professional experience.
A medical model is designed in accordance with objective and measurable observation to provide clinically useful information about the course of the illness over time
and direct specific treatments for the condition, which plays a significant role in regulating the treatment process
and providing premium rate healthcare services.

Finite automata (FA) \cite{Sipser13} is one of the mainstream technologies that can be
used to represent medical models.
Compared with the flow diagram or block diagram based model, a FA-based medical model has the advantage of regularized representation, flexibility in illness state evaluation and good expansibility \cite{Gambheer16,Alkhaldi18}. FA can be categorized into two types: deterministic finite automata (DFA) and nondeterministic finite automata (NFA).
The term ``deterministic" in DFA means that it can only transit to one state at a time (i.e. for some given input); ``nondeterministic" in NFA means it can transit to multiple states at once. Hence, DFA can be regarded as a special case of NFA; NFA is powerful to represent the nondeterministic state transitions and allows empty string input ($\epsilon$-move), which is more practical. NFA is a competent modeling tool and applicable to various fields in practice, such as regular language processing, program lexer, and medical modeling. 
NFA-based medical models have been used in healthcare monitoring \cite{Gambheer16,Alkhaldi18,YangY18FGCSFusion}, diagnosis and treatment of diseases \cite{Ruiz17}, virus genome detection \cite{Sasakawa14}, etc. 

Due to the high availability, accessibility and powerful computation capability of cloud, NFA-based medical models can be outsourced to a cloud platform to make diagnosis decisions and recommend the treatment methods on-the-fly
according to patient's physiological data that are monitored by mIoT.
Such as an approach could tremendously improve patients' healthcare, reduce cost, and enhance the accuracy of diagnosis due to its nature of quantitative analysis.
Despite the trememdous  benefits that can be brought by the remote diagnosis and treatment technology,
healthcare providers and patients are hesitant to adopt it
without adequate security and privacy protections \cite{YangY18FGCS}. Since a high quality NFA-based medical model is often regarded as the intellectual property and core competitiveness of a medical institution, one of the main challenges is to protect the privacy of the model and strictly prohibit it from disclosure
during online medical services. On the other hand, it is required in many jurisdictions to protect the
confidentiality of patients' health states and prevent them from unauthorized access. Moreover, treatment methods for patients are highly sensitive and must be kept confidential by the cloud platform or any other third party.

In this paper, we propose a \textbf{p}rivacy-preserving \textbf{med}ical treatment system using NFA-based medical model, hereafter referred to as \textbf{P-Med}. In a medical model, the illness states are expressed as the NFA states; an illness state transition caused by a therapeutic intervention is expressed as an NFA state transition; the diversity of therapeutic responses is expressed by exploring the nondeterministic
characteristic of NFA. To protect the privacy of the medical model, the NFA-based model is encrypted and before it is outsourced to a cloud server for remote medical service.
In order to perform privacy-preserving diagnosis and treatment, a patient uploads his or her recent (e. g. several days of) illness states in encrypted form to the cloud server which performs computations over encrypted data. Specifically, the proposed system achieves the following. 

$\bullet$ \textbf{\emph{Privacy-preserving NFA evaluation}}. Privacy-preserving NFA evaluation is essential to realize outsourced regular expression matching and pattern matching while without privacy leakage to the cloud server which performs the computation: given an encrypted NFA model and an encrypted set of symbols, the encrypted matching result is obtained through secure computation over encrypted data. Since NFA may contain many states, transitions, loops and self-loops, the challenge is to determine the inner state transitions caused by the input symbols in a privacy-preserving way. Previous schemes \cite{Troncoso07,Blanton10,Mohassel12,Sasakawa14} inevitably involve the interaction between a server (DFA/NFA holder) and a client (string holder) to determine the inner state transition. To overcome this disadvantage, we propose, for the first time, a secure NFA evaluation method to eliminate the client-server interaction such that the inner state transition can be computed by the server alone in a privacy-preserving way. P-Med converts the state transition problem into string matching. 
It firstly traverses the NFA model to find the regular language. Then, an elegant protocol is designed for privacy-preserving string matching and weight calculation (for weighted NFA). Lastly, a delicate ranking protocol is proposed to get the best-matching patterns. In addition, to deal with the approximate search, we design a privacy-preserving error-tolerant NFA evaluation method (in Section. \ref{SubSec:PGene}) and show its application in gene searching, which avoids the client-server paradigm of interaction and number of tolerated errors can be pre-defined. We believe that the proposed secure NFA evaluation methods are not only applicable to healthcare, but also provide a useful tool to other DFA/NFA based applications.

$\bullet$ \textbf{\emph{Secure automatic medical diagnosis on-the-fly}}. 
P-Med leverages privacy-preserving protocols to calculate on patient's encrypted illness states and healthcare provider's encrypted NFA-based medical model.
Illness state match protocol helps to find the match state in the NFA-based medical model according to patient's multi-dimension quantitative physiological data monitored by mIoT. Based on patient's successive illness states, 
the condition of disease is diagnosed in the cloud through secure outsourced computation.

$\bullet$ \textbf{\emph{Secure treatment procedure recommendation}}.
In an NFA-based medical model, we assign a weight to each state transition according to the therapeutic effect of the corresponding treatment method,
which is also encrypted before outsourcing.
Based on the diagnosis result, P-Med traverses all the possible treatment procedures and
calculates a weighted recommendation for each of them. Our secure best treatment procedure
selection protocol chooses the top-$k$ recommended procedures, which
preserves data privacy throughout the selection process.

$\bullet$ \textbf{\emph{Transparent operation for patient}}. P-Med does not require a patient
to perform any complex pre-processing before requesting remote medical service.  The interaction between a patient and a cloud server is simply a single round - the patient only needs to submit his or her  illness states in encrypted form and wait to get the best treatment procedure recommendation also in encrypted form.

\section{Preliminaries}
\label{Sec:Preliminary}

This section outlines the definition of FA, NFA and weighted NFA. Also, we introduce basic cryptosystem and privacy-preserving protocols used as building blocks of P-Med. 
Table \ref{Tab:Notations} lists the important notations in this paper.

\renewcommand\arraystretch{1.2}
\begin{table}[thbp]\centering \caption{Notations}
	\label{Tab:Notations}
	\begin{tabular}{ll}
		\hline Notation & Description \\
		\hline
		$\mathcal{L}(X)$/$[\![X]\!]_{pk}$ & bit length of $X$/ciphertext of $X$ (encrypted by $pk$)\\
		$\mathbb{M}/[\![\mathbb{M}]\!]_{pk}$ & plaintext/encrypted weighted NFA-based medical model\\
		$q/\phi$ & illness state in $\mathbb{M}$/illness state of a patient\\
		$q_0/\mathcal{F}$ & initial state/accept state set in $\mathbb{M}$\\
		$y/\mathcal{Y}$ & treatment method/treatment method set\\
		$w/W$  & transition weight in $\mathbb{M}$/weight of treatment procedure\\
		$\mathcal{TP}/\mathbb{TP}$ & treatment procedure/treatment procedure set\\
		$\mathcal{WTP}/\mathcal{ETP}$ & weighted/expanded treatment procedure\\
		$\mathbb{WTP}/\mathbb{ETP}$ & weighted/expanded treatment procedure set\\
		$array_{i,j}$ & the element in the $i^{th}$ row and $j^{th}$ column in $array$\\
		\hline
	\end{tabular}
\end{table}

\subsection{Finite Automata (FA)}
\label{SubSec:FA}

\begin{figure}[htp]
	\begin{center}
		\includegraphics[width=3.5in]{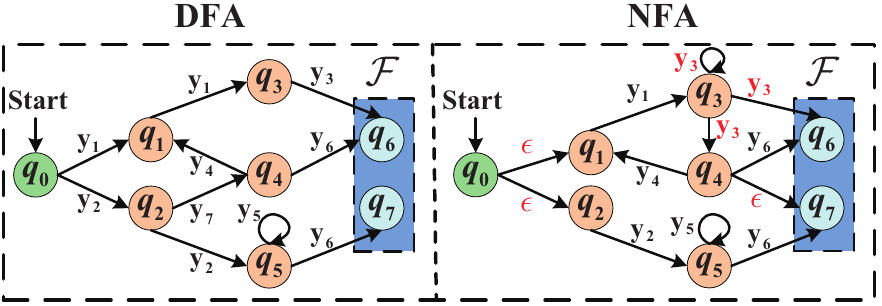}
		\caption{Examples of Finite Automata}
		\label{Fig:DFA-NFA}
	\end{center}
\end{figure}

In computation theory \cite{Sipser13}, DFA and NFA are two most important finite automata (Fig. \ref{Fig:DFA-NFA}).
In DFA, the next state is deterministic when a source state and an input symbol are given;
inputting a symbol is necessary for each state transition.
In NFA, the next state is nondeterministic given a source state and input symbol, and
several choices may exist for the next state at any point.
NFA also allows an empty string $\epsilon$ as a possible input,
and the transitions without consuming an input symbol are called $\epsilon$-transitions.
Formally, an NFA $\mathbb{M}_0$ \cite{Sipser13}
is a 5-tuple $(\mathcal{Q},\Sigma,q_0,\mathcal{F},\delta)$:
$\mathcal{Q}=(q_0,\cdots,q_{n_1})$ is a finite set of states;
$\Sigma=(y_{\sigma_1},\cdots,y_{\sigma_{n_2}})$ is a finite set of symbols;
$q_0\in \mathcal{Q}$ is the initial state;
$\mathcal{F}=(q_{\varrho_1},\cdots,q_{\varrho_{n_3}})\subseteq \mathcal{Q}$ is a set of accept states;
$\delta$ is the transition function, $\delta: \mathcal{Q}\times\Sigma_{\epsilon}\rightarrow \mathcal{P}(\mathcal{Q})$,
$\mathcal{P}(\mathcal{Q})$ is the power set\footnote{$\mathcal{P}(Q)$ is a collection of all subsets of $\mathcal{Q}$, and called the power set.} of $\mathcal{Q}$, $\Sigma_{\epsilon}=\Sigma\cup\{\epsilon\}$.

Let $\mathcal{Y}=(y_1, \cdots, y_l)$ be a set of symbols and $\mathcal{Y}\subseteq\Sigma$.
The NFA $\mathbb{M}_0$ accepts $\mathcal{Y}$ if there exists a sequence of states
$(r_0,r_1,\cdots,r_{n_0})$ in $\mathcal{Q}$ and satisfies:
1) $r_0=q_0$,
2) $\delta(r_i,y_{i+1})=r_{i+1}$ for $i=0$ to $n_0-1$,
3) $r_{n_0}\in \mathcal{F}$.
It is denoted as $ACCEPT(\mathbb{M}_0,\mathcal{Y})$ if the automata $\mathbb{M}_0$ accepts $\mathcal{Y}$.
Otherwise, it is denoted as $REJECT(\mathbb{M}_0,\mathcal{Y})$. An NFA $\mathbb{M}_0$ recognizes a language
$L$ if $\mathbb{M}_0$ accepts all $\mathcal{Y}\in L$ and rejects all $\mathcal{Y}\notin L$,
which is called \textbf{regular language}. An example of a DFA and an example of NFA are shown in Fig. \ref{Fig:DFA-NFA}.

\begin{figure}[htp]
	\begin{center}
		\includegraphics[width=3.5in]{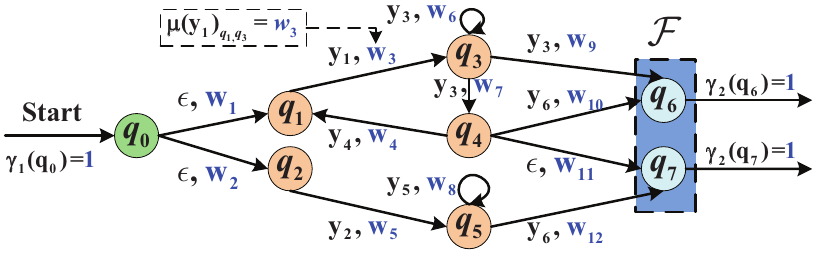}
		\caption{Example of Weighted NFA}
		\label{Fig:weightNFA}
	\end{center}
\end{figure}

Weighted NFA \cite{Weighted09} enables the transitions to carry weights, which models the cost involved when executing a transition (such as the amount of resources or time needed), or the probability or reliability of its successful execution. A weighted NFA $\mathbb{M}$ (over semiring\footnote{Semiring $S$ is an algebraic structure $(S,+,\cdot,0,1)$ satisfying the usual axioms for a (not necessarily commutative) ring, but without the requirement that addition be invertible. Important examples of semiring include: the natural numbers $(Z_N,+,\cdot,0,1)$ with the usual addition and multiplication; the Boolean semiring $\mathbb{B}=(\{0,1\},\lor,\land,0,1)$ \cite{Weighted09}.} $S$) is defined as $(\mathcal{Q},\Sigma,q_0,\mathcal{F},\delta,\mu, \gamma_1,\gamma_2)$: $\mathcal{Q},\Sigma,q_0,\mathcal{F},\delta$ are defined the same as NFA; $\mu: \Sigma\rightarrow S^{\mathcal{Q}\times\mathcal{Q}}$ is the transition weight function; $\gamma_1,\gamma_2: \mathcal{Q}\rightarrow S$ are weight functions for entering and leaving a state, respectively. Here, $\mu(y)$ is a $(|\mathcal{Q}|\times|\mathcal{Q}|)$-matrix whose $(q_i,q_j)$-entry $\mu(y)_{q_i,q_j}\in S$ indicates the weight $w$ of transition $q_i\stackrel{y}{\longrightarrow}q_j$, and the weighted transition can be represented as $q_i\stackrel{y,w}{\longrightarrow}q_j$.
For simplicity, we set $\gamma_1(q_0)=1$, $\gamma_2(q_{\varrho_i})=1$ ($\forall q_{\varrho_i}\in\mathcal{F}$) to omit the functions $\gamma_1,\gamma_2$ in this paper, and briefly denote the weighted NFA as $\mathbb{M}=(\mathcal{Q},\Sigma,q_0,\mathcal{F},\delta,\mu)$. Figure \ref{Fig:weightNFA} shows an example of weighted NFA, where the transition weights are labeled in blue.

\subsection{Threshold Paillier Cryptosystem}
\label{SubSec:Basic}

P-Med utilizes Paillier cryptosystem with threshold decryption (PCTD) \cite{Paillier99,Bresson03}
as the basic crypto primitive. PCTD includes the following algorithms: KeyGen, encryption (\texttt{Enc}),
decryption with weak secret key (\texttt{WDec}), decryption with master secret key (\texttt{SDec}),
master secret key splitting (\texttt{SKeyS}), partial decryption Step-1 (\texttt{PD1}), partial decryption Step-2 (\texttt{PD2})
and ciphertext refresh (\texttt{CR})
(the concrete algorithms are shown in supplemental materials A-1). The ciphertext of $X\in Z_N$ (encrypted by $pk$) is denoted as $[\![X]\!]_{pk}$. PCTD is additive homomorphic
($[\![m_1]\!]_{pk}\cdot [\![m_2]\!]_{pk} = [\![m_1+m_2]\!]_{pk}$) and
scalar-multiplicative homomorphic ($([\![m]\!]_{pk})^r=[\![r\cdot m]\!]_{pk}$, $\forall r\in Z_N$).

\subsection{Privacy-preserving Protocols}
P-Med makes use of the following protocols as the primitive privacy-preserving tools,
which are introduced in our previous work
\cite{YangY_TIFS17, LiuX_TIFS16,YangY_INS18} and supplemental material A.
Let $pk_A$ and $pk_B$ be the public keys of users $A$ and $B$.
$pk_\sigma$ is an authorization public key from user $A$ to $B$,
and the authorization secret key $sk_\sigma$ is utilized to decrypt the corresponding ciphertext (refer to section \ref{SubSec:Authorization} for the details).
Given a keyword $X$ (in arbitrary language with arbitrary symbols), secure keyword to ciphertext algorithm (\texttt{K2C}) \cite{YangY_TIFS17} encodes it to a unique element in $Z_N$ and outputs its ciphertext $[\![X]\!]_{pk}$.
Given $[\![X]\!]_{pk_A}$ and $[\![Y]\!]_{pk_B}$, secure addition protocol (\texttt{SAD})
outputs $[\![X+Y]\!]_{pk_\sigma}$, and secure multiplication protocol (\texttt{SMD})
outputs $[\![X\cdot Y]\!]_{pk_\sigma}$.
Secure greater or equal protocol (\texttt{SGE}) outputs $[\![u^{*}]\!]_{pk_\sigma}\leftarrow\texttt{SGE}([\![X]\!]_{pk_A},[\![Y]\!]_{pk_B})$
such that $u^* = 1$ if $X\geq Y$ and $u^* = 0$ if $X<Y$.
Secure less or equal protocol (\texttt{SLE}) outputs $[\![u^{*}]\!]_{pk_\sigma}\leftarrow\texttt{SLE}([\![X]\!]_{pk_A},[\![Y]\!]_{pk_B})$
such that $u^* = 1$ if $X\leq Y$ and $u^* = 0$ if $X>Y$.
Secure less than protocol (\texttt{SLT}) outputs $[\![u^{*}]\!]_{pk_\sigma}\leftarrow\texttt{SLT}([\![X]\!]_{pk_A},[\![Y]\!]_{pk_B})$
such that $u^* = 1$ if $X<Y$ and $u^* = 0$ if $X\geq Y$.
Secure greater than protocol (\texttt{SGT}) outputs $[\![u^{*}]\!]_{pk_\sigma}\leftarrow\texttt{SGT}([\![X]\!]_{pk_A},[\![Y]\!]_{pk_B})$
such that $u^* = 1$ if $X>Y$ and $u^* = 0$ if $X\leq Y$.
Secure equivalent test protocol (\texttt{SET}) outputs $[\![u^{*}]\!]_{pk_\sigma}$
such that $u^* = 1$ if $X=Y$ and $u^* = 0$ if $X\neq Y$.
Secure unequal test protocol (\texttt{SUT}) outputs $[\![u^{*}]\!]_{pk_\sigma}$
such that $u^* = 0$ if $X=Y$ and $u^* = 1$ if $X\neq Y$.
Given $[\![X]\!]_{pk_A}$, $[\![Y_1]\!]_{pk_B}$ and $[\![Y_2]\!]_{pk_B}$,
secure range comparison protocol (\texttt{SRC})
outputs $[\![u^{*}]\!]_{pk_\sigma}\leftarrow\texttt{SRC}([\![X]\!]_{pk_A},[\![Y_1]\!]_{pk_B},[\![Y_2]\!]_{pk_B})$
such that $u^* = 1$ if $Y_1\leq X\leq Y_2$ and $u^* = 0$ otherwise.

\section{Problem Formulation}
\label{Sec:ProblemFormulation}

\subsection{System Model}
\label{SubSec:SystemModel}

\begin{figure}[htp]
\begin{center}
\includegraphics[width=3.5in]{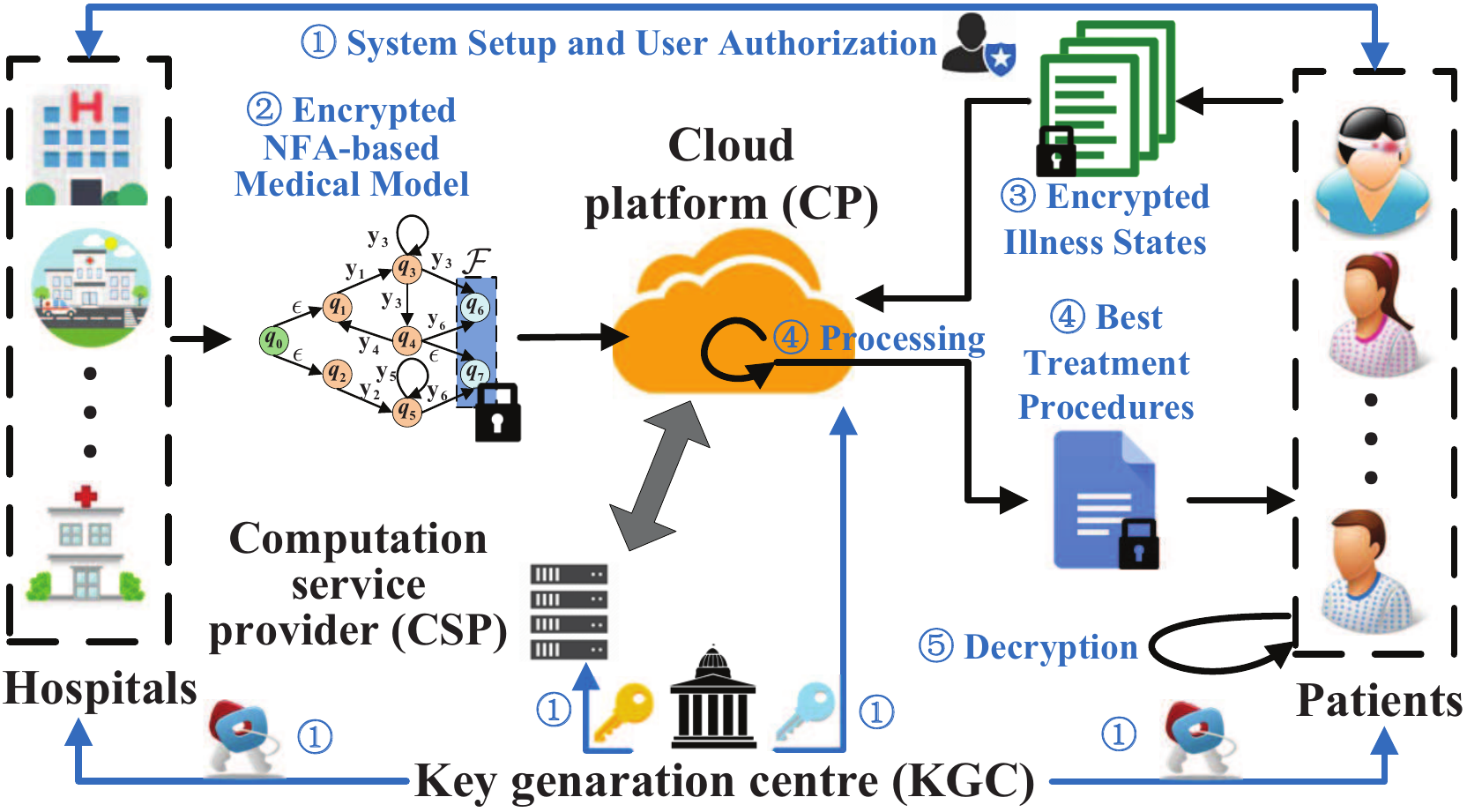}\\
\caption{System Model}\label{FigSystemModel}
\end{center}
\end{figure}

P-Med consists of five entities (Fig. \ref{FigSystemModel}):
key generation center (KGC), cloud platform (CP), computing service provider (CSP),
hospitals and patients.

$\bullet$ \textbf{KGC} is a trusted party, and tasked to distribute the public/secret keys and grant authorizations $($\textcircled{\small{1}}$)$.

$\bullet$ \textbf{Hospital} designs medical models for distinct diseases. Without loss of generality, we consider just one medical model per hospital in our description.
After encryption, a hospital outsources its own encrypted medical model to CP $($\textcircled{\small{2}}$)$.

$\bullet$ \textbf{Patient} is monitored by mIoT.
If patient needs diagnostic and treatment service, the encrypted illness states are sent to CP  $($\textcircled{\small{3}}$)$ to issue a query. After the result is returned, patient recovers it using the secret key $($\textcircled{\small{5}}$)$. 

$\bullet$ \textbf{CP} has powerful storage and computation capability,
   tasked to provide storage service for hospitals and respond on the medical query from the patients. \textbf{CSP} provides online calculation service. Upon receiving a patient's query, CP and CSP interactively execute the outsource computing protocols to find the best encrypted treatment procedures 
   $($\textcircled{\small{4}}$)$. 

\subsection{Attack Model and Security Model}
\label{SubSec:AttackSecurityModel}

In the attack model \cite{LiuX_TDSC16,DoQ15},
CP/CSP/hospital/patient are ``\emph{honest-but-curious}" entities that are
honest to execute the protocols but curious with
other entity's data. An adversary $\cal A^*$ is defined:
1) $\cal A^*$ could \emph{eavesdrop} all communications.
2) $\cal A^*$ could \emph{get} all ciphertext stored in CP,
  and data sent by CSP.
3) $\cal A^*$ could \emph{compromise} CSP to get the data sent by CP.
4) $\cal A^*$ could \emph{compromise} patients (except the
  challenge patient) aiming at getting the challenge patient's plaintext.
But $\cal A^*$ could not compromise
CP and CSP at the same time, nor challenge patient.

We adopt the security model in \cite{Kamara11,YangY_TSC17}.
Consider four parties: patient/hospital (a.k.a "$D_1$"),
CP (a.k.a "$S_1$") and CSP (a.k.a "$S_2$"). We construct three
simulators $(Sim_{D_1},Sim_{S_1},Sim_{S_2})$ against three types
of attackers
$(\mathcal{A}_{D_1},\mathcal{A}_{S_1},\mathcal{A}_{S_2})$ that
corrupt $D_1$, $S_1$ and $S_2$, respectively. These attackers are
non-colluding and semi-honest. Refer to
supplemental material B for the details.

\subsection{Design Goals}
\label{SubSec:DesignGoals}

The core requirement of P-Med is that the servers cannot deduce any useful information from the NFA-based medical model, the patient's illness states, and the treatment recommendation result. We summarize the design goals as below.

\begin{itemize}
	\item \emph{Medical Model and Data Confidentiality.} The servers are not able to recover any useful information from encrypted NFA-based medical model, which includes the encrypted illness states, the encrypted treatment methods and the encrypted transition weights. No useful information should be leaked from the patient's encrypted illness states.
	\item \emph{Treatment Recommendation Confidentiality.} The servers can not derive any useful information from the encrypted treatment recommendation result, which includes the illness state match result, the transitions caused by the patient's illness states, the treatment procedure weights, and the treatment recommendations.
	\item \emph{Soundness.} Soundness includes completeness and correctness: completeness ensures that all the match treatment procedures can be found; correctness guarantees that top-$k$ best procedures are recommended.
\end{itemize}

\section{Design Principle of P-Med}

This section introduces weighted NFA-based medical model and the challenges of P-Med design.
Next, the design principle is explicitly analyzed in plaintext to improve the readability.

\subsection{Weighted NFA-based Medical Model Representation}

NFA-based medical model  \cite{Gambheer16,Alkhaldi18,Ruiz17} is denoted as $\mathbb{M}_0
=(\mathcal{Q},\Sigma,q_0,\mathcal{F},\delta)$: 1) $\mathcal{Q}=(q_0,\cdots,q_{n_1})$ represents the illness state set; 2)  $\Sigma=(y_{\sigma_1},\cdots,y_{\sigma_{n_2}})$ represents the treatment method set; 3) $\delta$ represents the changing of illness states when a particular treatment method is implemented; 4) $\epsilon$ represents that no treatment is implemented; 5) nondeterministic characteristic represents the individual specificity in the treatment. Suppose several patients are in the illness state $q_i$
and treated with the same therapeutic method $y_j$, their state may transit to different illness states due to diverse medical responses.

\begin{figure}[htp]
	\begin{center}
		\includegraphics[width=0.95\linewidth]{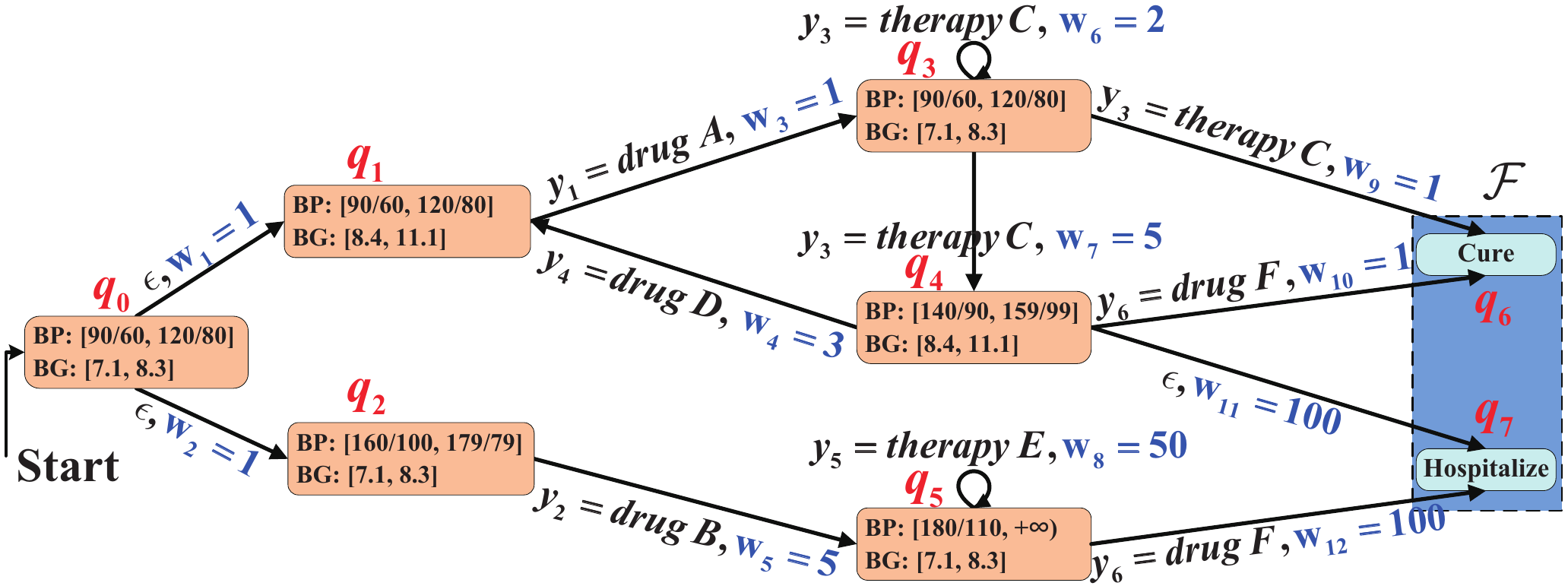}
		\caption{An Example of Weighted NFA-based Medical Model}
		\label{Fig:M-IoT}
	\end{center}
\end{figure}

Fig. \ref{Fig:M-IoT} depicts an example of NFA-based medical model of gestational diabetes \cite{Ruiz17}, in which the 8 states
describe the illness conditions and the 5 input symbols describe the therapies. The initial state is $q_0$ (represents ``gestational diabetes") and the accept states are $(q_6,q_7)$ = (``cure", ``hospitalize"). At the beginning, the initial state $q_0$ transits to $q_1$ or $q_2$ with the empty symbol input $\epsilon$, which indicates that the patient with gestational diabetes may have state $q_1$ (represent ``moderate diabetes") or $q_2$ (represent ``mild diabetes \& moderate hypertension") without any medical intervention. If the patient is in the state $q_3$ and exerted with $y_3$ = ``therapy C", the state may transit to $q_3$, $q_4$ or $q_6$, which differs due to individual differences.
The illness states are measured and represented by the physiological indexes such as blood pressure (BP), blood glucose (BG), which is elaborated in Section \ref{SubSec:SST}.

To provide treatment recommendations to the patients,
the medical model designer sets weight $w$ for each transition in the NFA-based medical model $\mathbb{M}_0$
and the lower transition weight indicates higher recommend level, which is measured by the effectiveness, side effects, cost/performance ratio, etc. The weighted NFA-based medical model is denoted as $\mathbb{M}
=(\mathcal{Q},\Sigma,q_0,\mathcal{F},\delta,\mu)$, where $\mu$ is defined in Subsection \ref{SubSec:FA}. In Fig. \ref{Fig:M-IoT}, the transitions that lead to the accept state $q_6$ = ``cure" have lower weights, which infers preferred treatment methods; on the other hand, the transitions that lead to $q_7$ = ``hospitalize" have obviously higher weights.  

\subsection{The Challenges of Treatment Recommendation}
\label{SubSec:Challenge}

When a patient $B$ submits the recent illness states $\Phi=(\phi_1,\cdots,\phi_m)$ to the server to query treatment recommendation, CP and CSP searches for the successive states in $\mathbb{M}$ that match with $\Phi$ and the result is classified into two situations:
1) $\Phi$ does not appear in $\mathbb{M}$, which indicates that the patient's condition is not included in the medical model. 2) $\Phi$ is found in $\mathbb{M}$ and the match state set is $(q_{\theta_{\bar{j}}},\cdots,q_{\theta_{\bar{j}+m-1}})$. For the former, the servers do not take further computation and output a symbol indicating that no match result is found. For the latter, the servers continue to find the best treatment procedures: start at the state $q_{\theta_{\bar{j}+m-1}}$ to find all the paths that leads to the accept states in $\mathcal{F}$; evaluate each path and calculate a treatment procedure weight to indicate the recommendation level. Suppose the path is $q_{\theta_{\bar{j}+m-1}}\xrightarrow{y_{\theta_{\bar{j}+m}},w_{\theta_{\bar{j}+m}}} q_{\theta_{\bar{j}+m}}\rightarrow\cdots\xrightarrow{y_{\theta_{\tau}},w_{\theta_{\tau}}} q_{\theta_{\tau}}$ with $q_{\theta_{\tau}}\in\mathcal{F}$, the path weight is calculated as $W=\Sigma_{k=\bar{j}+m}^{\tau}{w_{\theta_{k}}}$. Then, the servers rank these weights and recommend the top-$k$ treatment procedures that have the lowest weights.

The above treatment recommendation process is analyzed in the plaintext viewpoint and the process is pretty straightforward. However, it is much more complicated when the hospital's medical model $\mathbb{M}$ and the patient's query $\Phi$ are encrypted. The challenges that are faced by the P-Med design are elaborated below. 1) Since the states in $\mathbb{M}$ and $\Phi$ are encrypted, the servers search for the match states in $\mathbb{M}$ in a blind manner. According to the design goal, the illness state match result should be kept confidential, which indicates that the servers do not know whether the state in $\Phi$ matches with any given state in $\mathbb{M}$. 2) Since the state match results are unknown, the servers are uncertain about whether $\Phi$ appears in $\mathbb{M}$. They cannot distinguish two situations nor take different approaches to do the following calculations. 3) If $\Phi$ indeed appears in $\mathbb{M}$, the server can not determine which illness state sets in $\mathbb{M}$ match with $\Phi$, nor decide which state in $\mathbb{M}$ is the right start point to search for the treatment procedures. 4) Since the NFA-based medical model could be quite complicated and contain many different paths leading to the accept states, it is a challenge for the servers to find all the treatment procedures for the patient to satisfy the completeness requirement in the design goal, especially when the medical model contains loops and self-loops. 5) According to the design goal, the weights of the match treatment procedures are kept secret, and the servers need to rank the weight in a  privacy-preserving way.

\subsection{Design Principles}
\label{SubSec:Princeple}

To deal with the above challenges, P-Med should be designed from the ciphertext viewpoint and follow the following principles. 1) On the premise of the confidentiality of medical model, medical data and treatment result, the design goal of completeness should be realized by the traverse of all the paths from the initial state to the accept states. 2) Since the state match result is unknown to the server, the path weight has to be computed for each traversed paths. A single unified path weight calculation algorithm should be designed, which is applicable to both of the situations: $\Phi$ appears in the path (named matched path), and $\Phi$ does not appear in the path (named unmatched path). To realize the design goal of correctness, the weights of the matched paths should be absolutely lower than that of the unmatched paths. And the matched paths leading to the accept state with good therapeutic effect should result in lower path weights. 3) Privacy-preserving treatment procedure selection algorithm needs to be designed to rank the encrypted weights and select the top-$k$ best treatment procedures. To realize the treatment recommendation confidentiality, it is required that
the servers cannot distinguish which treatment procedures are recommended.

For ease of understanding, we use the plaintext form of Fig. \ref{Fig:M-IoT}
to illustrate how the design principles work.

(1) According to principle 1, we design a traverse algorithm to traverse all the paths that start from the initial state $q_0$ and end at an accept state. To avoid endless loops, it is necessary to set the maximum visit time $MVisit$ that each state
can be included in a path,
and the maximum state number $MState$ contained in a path. The $i$-th traversed path is represented as
$q_0
\xrightarrow{y_{i,\theta_1},w_{i,\theta_1}}q_{i,\theta_{1}}
\xrightarrow{y_{i,\theta_2},w_{i,\theta_2}}q_{i,\theta_{2}}
\rightarrow\cdots\xrightarrow{y_{i,\theta_{\tau_i}},w_{i,\theta_{\tau_i}}} q_{i,\theta_{\tau_i}}$
with $q_{\theta_{i,\tau_i}}\in\mathcal{F}$, and denoted as the $i$-th treatment procedure $\mathcal{TP}_i=(\mathcal{Q}_i,\mathcal{Y}_i,\mathcal{W}_i)$, where the illness state set $\mathcal{Q}_i=(q_0,q_{i,\theta_{1}},\cdots, q_{i,\theta_{\tau_i}})$, 
the treatment method set $\mathcal{Y}_i=(y_{i,\theta_{1}},\cdots, y_{i,\theta_{\tau_i}})$,
the transition weight set 
$\mathcal{W}_i=(w_{i,\theta_{1}},\cdots, w_{i,\theta_{\tau_i}})$. If $n$ paths are found in NFA-based medical model $\mathbb{M}$, the traversed treatment procedure set is denoted as $\mathbb{TP}=(\mathcal{TP}_1,\cdots,\mathcal{TP}_n)$. The concrete construction is shown in secure treatment procedure traverse algorithm (\texttt{TPT}) in Section \ref{SubSec:TPT}.

(2) According to principle 2, we design a unified path weight calculation algorithm that is suitable for both of the matched and unmatched situations. If $\Phi=(\phi_1,\cdots,\phi_m)$ does not match with any successive $m$ states in $\mathcal{TP}_i$, the treatment procedure weight is set to $W_i=MWeight$, where 
$MWeight$ is the maximum weight pre-defined by a hospital for any unmatched path. $MWeight$ should be defined absolutely larger than all the weights of the matched paths. If $\Phi=(\phi_1,\cdots,\phi_m)$ match with $(q_{i, \theta_{\bar{j}}},\cdots,q_{i,\theta_{\bar{j}+m-1}})$ for the first time in $\mathcal{TP}_i$, the weight is the sum of the transition weights from $q_{i,\theta_{\bar{j}+m-1}}$ to the accept state $q_{i,\theta_{\tau}}$ and calculated as $W_i=\Sigma_{k=\bar{j}+m}^{\tau_i}{w_{i,\theta_{k}}}$. The transitions that lead to better treatment accept result are set to have relatively lower weight. Then, we have $W_i=MWeight$ for the unmatched paths and $W_i=\Sigma_{k=\bar{j}+m}^{\tau_i}{w_{i,\theta_{k}}}$ for the matched paths.
After the calculation, the weighted treatment procedure is denoted as $\mathcal{WTP}_i=(\mathcal{Q}_i,\mathcal{Y}_i,W_i)$,
which substitutes the transition weight set 
$\mathcal{W}_i=(w_{i,\theta_{1}},\cdots, w_{i,\theta_{\tau_i}})$
in $\mathcal{TP}_i$ with the weight $W_i$ in $\mathcal{WTP}_i$. The set is denoted as  $\mathbb{WTP}=(\mathcal{WTP}_1,\cdots,\mathcal{WTP}_n)$.
The concrete construction is shown in secure treatment procedure weight calculation algorithm (\texttt{TPW}) in Section \ref{SubSec:TPW}.

(3) According to principle 3, we design a best treatment procedure selection algorithm with privacy protection mechanism. To prevent the server from discovering which treatment procedures are recommended to the patient, it is necessary to unify the lengths of the treatment procedures in $\mathbb{WTP}$. Without this processing, the server could get useful information about the result according to the length of the recommended result. Suppose the treatment procedure $\mathcal{WTP}_j$ with the longest path in $\mathbb{WTP}$ has $MState$ illness states, i.e., $|\mathcal{Q}_j|=MState,|\mathcal{Y}_j|=MState-1$. Then, we expand $\mathcal{WTP}_i$ into expanded treatment procedure $\mathcal{ETP}_i$ by appending dummy symbols (e.g., $\bot$) after the illness states and treatment methods such that $|\mathcal{Q}_i|=MState,|\mathcal{Y}_i|=MState-1$,
for $1\leq i\leq n$. The weight values $W_i$ are the same in $\mathcal{WTP}_i$ and $\mathcal{ETP}_i$. 
The set is denoted as $\mathbb{ETP}=(\mathcal{ETP}_1,\cdots,\mathcal{ETP}_n)$.
It is noted that these dummy symbols are indistinguishable from the normal data after the encryption in the privacy-preserving algorithm. 

The treatment procedure selection process consists of three protocols to fulfill the task. \textbf{S}ecure \textbf{min}imum selection protocol (\texttt{SMin}) selects the best treatment procedure from two procedures: given the ciphertexts of $(\mathcal{ETP}_1, \mathcal{ETP}_2)$, it outputs the ciphertext of $\mathcal{ETP}_{Min}$ such that $W_{Min}=min(W_1,W_2)$. 
\textbf{S}ecure \textbf{min}imum selection from $\boldsymbol{n}$ treatment procedures protocol (\texttt{SMin$_n$}) leverages \texttt{SMin} as sub-protocol: given the ciphertexts of $(\mathcal{ETP}_1, \cdots,\mathcal{ETP}_n)$, it outputs the ciphertext of $\mathcal{ETP}_{Min}$ such that $W_{Min}=min(W_1,\cdots,W_n)$. 
Secure top-$k$ \textbf{b}est treatment \textbf{p}rocedure \textbf{s}election protocol (\texttt{BPS-$k$}) leverages \texttt{SMin$_n$} as sub-protocol: given the ciphertexts of $(\mathcal{ETP}_1, \cdots,\mathcal{ETP}_n)$, it outputs the ciphertexts of $(\mathcal{ETP}_{Min_1},\cdots,\mathcal{ETP}_{Min_k})$ such that $(W_{Min_1},\cdots,W_{Min_k})$ are the top-$k$ lowest weights. The concrete construction of these protocols is shown in Section \ref{SubSec:BPS}.

\subsection{Example}
A toy example is used to illustrate how the design principles work in P-Med. Suppose $MState=4$ and three paths are traversed: 
1) $q_0
\xrightarrow{y_1,\textcolor[rgb]{0,0,1}{w_1}}q_1$;
2) $q_0
\xrightarrow{y_2,\textcolor[rgb]{0,0,1}{w_2}}q_2
\xrightarrow{y_3,\textcolor[rgb]{0,0,1}{w_3}}q_1$;
3) $q_0
\xrightarrow{y_2,\textcolor[rgb]{0,0,1}{w_2}}q_2
\xrightarrow{y_4,\textcolor[rgb]{0,0,1}{w_4}}q_3
\xrightarrow{y_5,\textcolor[rgb]{0,0,1}{w_5}}q_4$.
Then we have $\mathcal{TP}_1=((q_0,q_1),(y_1),(w_1))$,
$\mathcal{TP}_2=((q_0,q_2,q_1),(y_2,y_3),(w_2,w_3))$,
$\mathcal{TP}_3=((q_0,q_2,q_3,q_4),(y_2,y_4,y_5),(w_2,w_4,w_5))$.
After the weight calculation, we obtain
$\mathcal{WTP}_1=((q_0,q_1),(y_1),W_1)$,
$\mathcal{WTP}_2=((q_0,q_2,q_1),(y_2,y_3),W_2)$,
$\mathcal{WTP}_3=((q_0,q_2,q_3,$ $q_4),(y_2,y_4,y_5),W_3)$.
Then, expand treatment procedures into

\noindent 
$\mathcal{ETP}_1=((q_0,q_1,\bot,\bot),(y_1,\bot,\bot),W_1)$,

\noindent
$\mathcal{ETP}_2=((q_0,q_2,q_1,\bot),(y_2,y_3,\bot),W_2)$,

\noindent
$\mathcal{ETP}_3=((q_0,q_2,q_3,q_4),(y_2,y_4,y_5),W_3)$.

Next, we take the weighted medical model $\mathbb{M}$ in Fig. \ref{Fig:M-IoT} as an example to illustrate the framework of P-Med.
Firstly, set the parameters $MVisit =2,MState=8$. The traverse algorithm finds all the paths that start from $q_0$ and end at $q_6$ or $q_7$. The traversed treatment procedures are  $\mathbb{TP}=(\mathcal{TP}_1,\cdots,\mathcal{TP}_{11})$:

\noindent 1. $q_0
\xrightarrow{\epsilon,\textcolor[rgb]{0,0,1}{w_1}}q_1
\xrightarrow{y_1,\textcolor[rgb]{0,0,1}{w_3}}q_3
\xrightarrow{y_3,\textcolor[rgb]{0,0,1}{w_6}}q_3
\xrightarrow{y_3,\textcolor[rgb]{0,0,1}{w_7}}q_4
\xrightarrow{y_6,\textcolor[rgb]{0,0,1}{w_{10}}}q_6$,

\noindent 2. $q_0
\xrightarrow{\epsilon,\textcolor[rgb]{0,0,1}{w_1}}q_1
\xrightarrow{y_1,\textcolor[rgb]{0,0,1}{w_3}}q_3
\xrightarrow{y_3,\textcolor[rgb]{0,0,1}{w_6}}q_3
\xrightarrow{y_3,\textcolor[rgb]{0,0,1}{w_7}}q_4
\xrightarrow{\epsilon,\textcolor[rgb]{0,0,1}{w_{11}}}q_7$,

\noindent 3. $q_0
\xrightarrow{\epsilon,\textcolor[rgb]{0,0,1}{w_1}}q_1
\xrightarrow{y_1,\textcolor[rgb]{0,0,1}{w_3}}q_3
\xrightarrow{y_3,\textcolor[rgb]{0,0,1}{w_6}}q_3
\xrightarrow{y_3,\textcolor[rgb]{0,0,1}{w_9}}q_6$, 

\noindent 4. $q_0
\xrightarrow{\epsilon,\textcolor[rgb]{0,0,1}{w_1}}
\textcolor[rgb]{1,0,0}{q_1}
\xrightarrow{y_1,\textcolor[rgb]{0,0,1}{w_3}}
\textcolor[rgb]{1,0,0}{q_3}
\xrightarrow{y_3,\textcolor[rgb]{0,0,1}{w_7}}
\textcolor[rgb]{1,0,0}{q_4}
\xrightarrow{y_4,\textcolor[rgb]{0,0,1}{w_4}}
\textcolor[RGB]{148,0,211}{q_1}
\xrightarrow{y_1,\textcolor[rgb]{0,0,1}{w_3}}
\textcolor[RGB]{148,0,211}{q_3}
\xrightarrow{y_3,\textcolor[rgb]{0,0,1}{w_7}}
\textcolor[RGB]{148,0,211}{q_4}
\xrightarrow{y_6,\textcolor[rgb]{0,0,1}{w_{10}}}q_6$, 

\noindent 5. $q_0
\xrightarrow{\epsilon,\textcolor[rgb]{0,0,1}{w_1}}
\textcolor[rgb]{1,0,0}{q_1}
\xrightarrow{y_1,\textcolor[rgb]{0,0,1}{w_3}}
\textcolor[rgb]{1,0,0}{q_3}
\xrightarrow{y_3,\textcolor[rgb]{0,0,1}{w_7}}
\textcolor[rgb]{1,0,0}{q_4}
\xrightarrow{y_4,\textcolor[rgb]{0,0,1}{w_4}}
\textcolor[RGB]{148,0,211}{q_1}
\xrightarrow{y_1,\textcolor[rgb]{0,0,1}{w_3}}
\textcolor[RGB]{148,0,211}{q_3}
\xrightarrow{y_3,\textcolor[rgb]{0,0,1}{w_7}}
\textcolor[RGB]{148,0,211}{q_4}
\xrightarrow{\epsilon,\textcolor[rgb]{0,0,1}{w_{11}}}q_7$, 

\noindent 6. $q_0
\xrightarrow{\epsilon,\textcolor[rgb]{0,0,1}{w_1}}
\textcolor[rgb]{1,0,0}{q_1}
\xrightarrow{y_1,\textcolor[rgb]{0,0,1}{w_3}}
\textcolor[rgb]{1,0,0}{q_3}
\xrightarrow{y_3,\textcolor[rgb]{0,0,1}{w_7}}
\textcolor[rgb]{1,0,0}{q_4}
\xrightarrow{y_4,\textcolor[rgb]{0,0,1}{w_4}}q_1
\xrightarrow{y_1,\textcolor[rgb]{0,0,1}{w_3}}q_3
\xrightarrow{y_3,\textcolor[rgb]{0,0,1}{w_9}}q_6$, 

\noindent 7. $q_0
\xrightarrow{\epsilon,\textcolor[rgb]{0,0,1}{w_1}}
\textcolor[rgb]{1,0,0}{q_1}
\xrightarrow{y_1,\textcolor[rgb]{0,0,1}{w_3}}
\textcolor[rgb]{1,0,0}{q_3}
\xrightarrow{y_3,\textcolor[rgb]{0,0,1}{w_7}}
\textcolor[rgb]{1,0,0}{q_4}
\xrightarrow{y_6,\textcolor[rgb]{0,0,1}{w_{10}}}q_6$, 

\noindent 8. $q_0
\xrightarrow{\epsilon,\textcolor[rgb]{0,0,1}{w_1}}
\textcolor[rgb]{1,0,0}{q_1}
\xrightarrow{y_1,\textcolor[rgb]{0,0,1}{w_3}}
\textcolor[rgb]{1,0,0}{q_3}
\xrightarrow{y_3,\textcolor[rgb]{0,0,1}{w_7}}
\textcolor[rgb]{1,0,0}{q_4}
\xrightarrow{\epsilon,\textcolor[rgb]{0,0,1}{w_{11}}}q_7$, 

\noindent 9. $q_0
\xrightarrow{\epsilon,\textcolor[rgb]{0,0,1}{w_1}}q_1
\xrightarrow{y_1,\textcolor[rgb]{0,0,1}{w_3}}q_3
\xrightarrow{y_3,\textcolor[rgb]{0,0,1}{w_9}}q_6$, 

\noindent 10. $q_0
\xrightarrow{\epsilon,\textcolor[rgb]{0,0,1}{w_2}}q_2
\xrightarrow{y_2,\textcolor[rgb]{0,0,1}{w_5}}q_5
\xrightarrow{y_5,\textcolor[rgb]{0,0,1}{w_8}}q_5
\xrightarrow{y_6,\textcolor[rgb]{0,0,1}{w_{12}}}q_7$, 

\noindent 11. $q_0
\xrightarrow{\epsilon,\textcolor[rgb]{0,0,1}{w_2}}q_2
\xrightarrow{y_2,\textcolor[rgb]{0,0,1}{w_5}}q_5
\xrightarrow{y_6,\textcolor[rgb]{0,0,1}{w_{12}}}q_7$.

Then, set the parameter $MWeight=10000$
for weight calculation algorithm. Suppose patient's illness state set $\Phi=(\phi_1,\phi_2,\phi_3)$ matches with $(q_1, q_3, q_4)$ in $\mathbb{M}$.
The weight calculation algorithm try to find the match state set in each treatment procedure. In above paths, the match state set $(q_1, q_3, q_4)$ that appears for the first time are labeled with red; $(q_1, q_3, q_4)$ that appears for the second time are labeled with purple.
According to design principle 2, we have $W_1=W_2=W_3=W_9=W_{10}=W_{11}=MWeight=10000$ for unmatched paths;
for matched paths, we have $W_4=w_4+w_3+w_7+w_{10}=10$, $W_5=w_4+w_3+w_7+w_{11}=109$,
$W_6=w_4+w_3+w_9=5$,
$W_7=w_{10}=1$, $W_8=w_{11}=100$.
Lastly, set the parameter $k=3$ for selection algorithm to get the top-3 best treatment procedures: 7-th, 6-th and 4-th treatment procedures with the lowest weights 1, 5, 10, respectively.

\section{Basic Component of P-Med}
This section introduces how to distribute keys, grant user authentication, encrypt medical model and patient's query, represent illness state and make state match test for P-Med.

\subsection{Key Distribution and User Authorization}
\label{SubSec:Authorization}

KGC runs the $KeyGen$ algorithm in PCTD
to generate the public parameter $PP=(g,N)$
and master secret key $SK=\lambda$ for the system,
and $SKeyS$ algorithm to generate the partial strong keys $SK_1=\lambda_1$ and $SK_2=\lambda_2$ for CP and CSP, respectively.
KGC generates the secret/public key pair $sk_{A}=a,pk_{A}=g^{a}$ for hospital $A$,
and $sk_{B}=b,pk_{B}=g^{b}$ for patient $B$, where $a, b$ are randomly selected from $Z_N$.

If a patient $B$ wishes to request service from hospital $A$, $A$ defines a valid service time period in the form of
$SP=$ ``20190101-20191231". Then,
KGC generates a certificate number $CN$,
and a certificate $\texttt{CER}_{A,B}$:
$\langle cer=(CN,A,B,SP,pk_\sigma), Sig(cer,SK)\rangle$, 
where $pk_\sigma=g^{sk_\sigma}$, $sk_\sigma\in_R Z_N$, and  $sk_\sigma$ is confidentially sent to $B$.

\subsection{Encryption of Medical Model and Query}

\begin{figure}[htp]
	\begin{center}
		\includegraphics[width=3.5in]{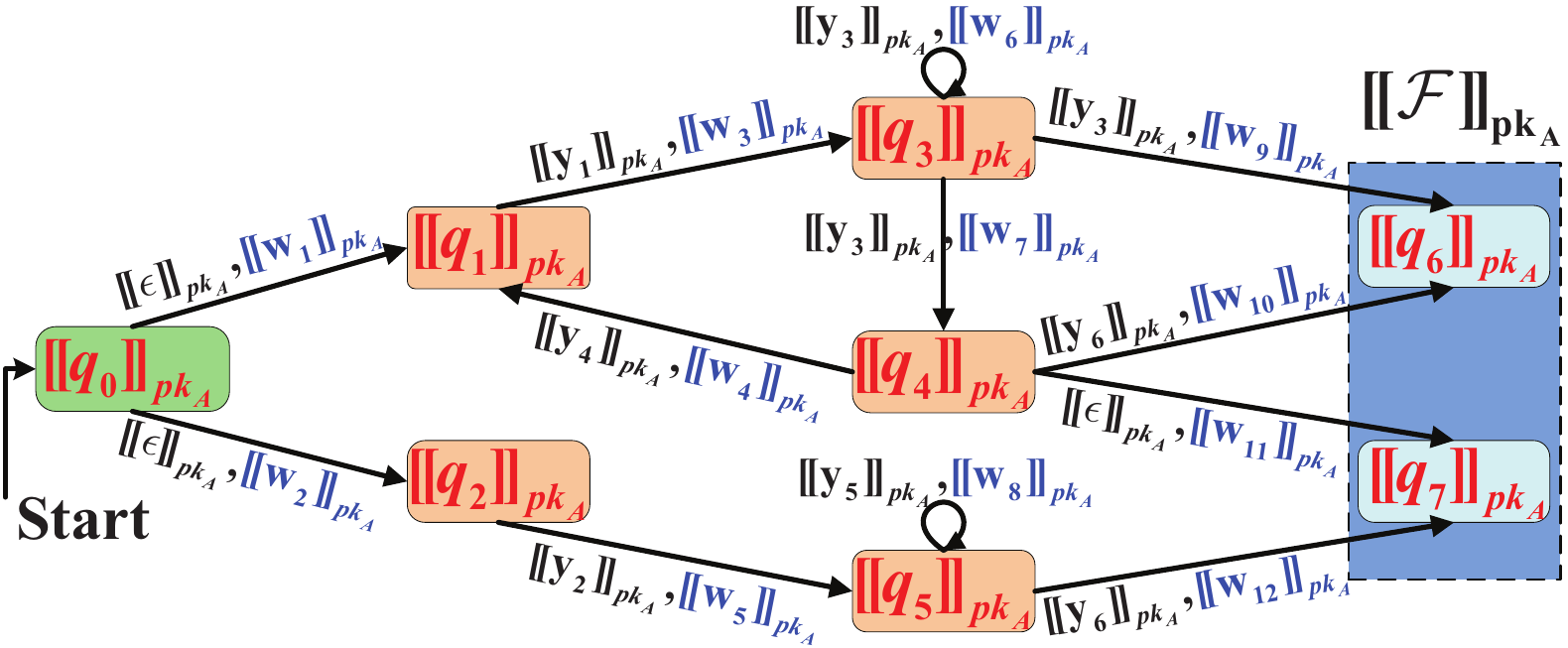}\\
		\caption{Encryption of Weighted NFA}
		\label{Fig:Hospital}
	\end{center}
\end{figure}

Hospital $A$ encrypts the weighted NFA-based medical model to
$[\![\mathbb{M}]\!]_{pk_A}=([\![\mathcal{Q}]\!]_{pk_A},[\![\Sigma]\!]_{pk_A},[\![q_0]\!]_{pk_A},[\![\mathcal{F}]\!]_{pk_A},[\![\delta]\!]_{pk_A},$ $[\![\mu]\!]_{pk_A})$,
where the encrypted states $[\![\mathcal{Q}]\!]=([\![q_0]\!],\cdots,[\![q_{n_1}]\!])$, the encrypted treatment methods
$[\![\Sigma]\!]=([\![y_{\sigma_1}]\!],\cdots,[\![y_{\sigma_{n_2}}]\!])$, the encrypted accept states
$[\![\mathcal{F}]\!]=([\![q_{\varrho_1}]\!],\cdots,[\![q_{\varrho_{n_3}}]\!])$, the encrypted transition weights
$[\![\mathcal{W}]\!]=([\![w_1]\!],\cdots,[\![w_{n_5}]\!])$, and the empty symbol $\epsilon$ is encrypted to $[\![\epsilon]\!]$.\footnote{The encryption key $pk_A$ is omitted here to simplify the expression.} 
For example, the weighted NFA-based medical model shown in Fig. \ref{Fig:M-IoT} is encrypted to the one show in Fig. \ref{Fig:Hospital}.
The encrypted transition function $[\![\delta]\!]_{pk_A}$ and encrypted transition weight function $[\![\mu]\!]_{pk_A}$ of Fig. \ref{Fig:Hospital} is represented by a weighted state transition table shown in Table \ref{Tab:STT},
which is a two-dimensional table with one dimension representing the current states
and the other dimension the next states.
The row/column intersection contains the input symbol that leads the current state transiting to the next state, and the transition weight (in blue characters).

\renewcommand\arraystretch{1.2}
\begin{table}[htbp]
\caption{Weighted State Transition Table}\label{Tab:STT} \centering
\tiny
\begin{tabular}
{I@{}c|@{}c@{}I@{}c@{}|@{}c@{}|@{}c@{}|@{}c@{}|@{}c@{}|@{}c@{}|@{}c@{}|@{}I}
\thickhline
    \multicolumn{2}{IcI}{\multirow{2}{*}{\rotatebox{0}{\textbf{Input}}}} &\multicolumn{7}{cI}{\textbf{Next State}}\\
  \cline{3-9}  \multicolumn{2}{IcI}{} & $\boldsymbol{[\![q_1]\!]}$ & $\boldsymbol{[\![q_2]\!]}$ & $\boldsymbol{[\![q_3]\!]}$ & $\boldsymbol{[\![q_4]\!]}$  & $\boldsymbol{[\![q_5]\!]}$
  & $\boldsymbol{[\![q_6]\!]}$ & $\boldsymbol{[\![q_7]\!]}$\\
  \thickhline
   \multirow{6}{*}{\rotatebox{90}{\textbf{Current State}}}
   &\multirow{1}{*}{$\boldsymbol{[\![q_0]\!]}$}     & $[\![\epsilon]\!]$, \textcolor[rgb]{0,0,1}{$[\![w_1]\!]$}
   & $[\![\epsilon]\!],\textcolor[rgb]{0,0,1}{[\![w_2]\!]}$    & \multirow{1}{*}{$\bot$}   & \multirow{1}{*}{$\bot$}  & \multirow{1}{*}{$\bot$}    & \multirow{1}{*}{$\bot$}    & \multirow{1}{*}{$\bot$}     \\
  \cline{2-9}   &\multirow{1}{*}{$\boldsymbol{[\![q_1]\!]}$}      & \multirow{1}{*}{$\bot$}    & \multirow{1}{*}{$\bot$}
  & $[\![y_1]\!]$, \textcolor[rgb]{0,0,1}{$[\![w_3]\!]$}   & \multirow{1}{*}{$\bot$}   & \multirow{1}{*}{$\bot$}   & \multirow{1}{*}{$\bot$}    & \multirow{1}{*}{$\bot$}    \\
  \cline{2-9}   & \multirow{1}{*}{$\boldsymbol{[\![q_2]\!]}$}     & \multirow{1}{*}{$\bot$}    & \multirow{1}{*}{$\bot$}    & \multirow{1}{*}{$\bot$}    & \multirow{1}{*}{$\bot$}
  & $[\![y_2]\!]$, \textcolor[rgb]{0,0,1}{$[\![w_5]\!]$}   & \multirow{1}{*}{$\bot$}  & \multirow{1}{*}{$\bot$}   \\
  \cline{2-9}   & $\boldsymbol{[\![q_3]\!]}$   & $\bot$    & $\bot$
  & $[\![y_3]\!]$, \textcolor[rgb]{0,0,1}{$[\![w_6]\!]$}
  &  $[\![y_3]\!]$, \textcolor[rgb]{0,0,1}{$[\![w_7]\!]$}  & $\bot$
  & $[\![y_3]\!]$, \textcolor[rgb]{0,0,1}{$[\![w_9]\!]$}   & $\bot$    \\
  \cline{2-9}   & $\boldsymbol{[\![q_4]\!]}$
  & $[\![y_4]\!]$,\textcolor[rgb]{0,0,1}{$[\![w_4]\!]$}    & $\bot$    & $\bot$   & $\bot$    & $\bot$
  & $[\![y_6]\!]$, \textcolor[rgb]{0,0,1}{$[\![w_{10}]\!]$}
  & $[\![\epsilon]\!]$, \textcolor[rgb]{0,0,1}{$[\![w_{11}]\!]$}   \\
  \cline{2-9}   & \multirow{1}{*}{$\boldsymbol{[\![q_5]\!]}$}    & \multirow{1}{*}{$\bot$}    & \multirow{1}{*}{$\bot$}   & \multirow{1}{*}{$\bot$}    & \multirow{1}{*}{$\bot$}
  & $[\![y_5]\!]$, \textcolor[rgb]{0,0,1}{$[\![w_8]\!]$}    & \multirow{1}{*}{$\bot$}
  & $[\![y_6]\!]$ ,\textcolor[rgb]{0,0,1}{$[\![w_{12}]\!]$} \\
  \thickhline
\end{tabular}
\end{table}

When patient $B$ queries the telemedical service,
the illness states $\Phi=(\phi_1,\cdots,\phi_m)$ in the last few days are encrypted into
$[\![\Phi]\!]_{pk_B}=([\![\phi_1]\!]_{pk_B},\cdots,[\![\phi_m]\!]_{pk_B})$ and sent to CP, which is used for diagnosis and treatment recommendation.

\subsection{Illness State Representation and Match Test}
\label{SubSec:SST}

\begin{figure}[htp]
	\begin{center}
		\includegraphics[width=3.2in]{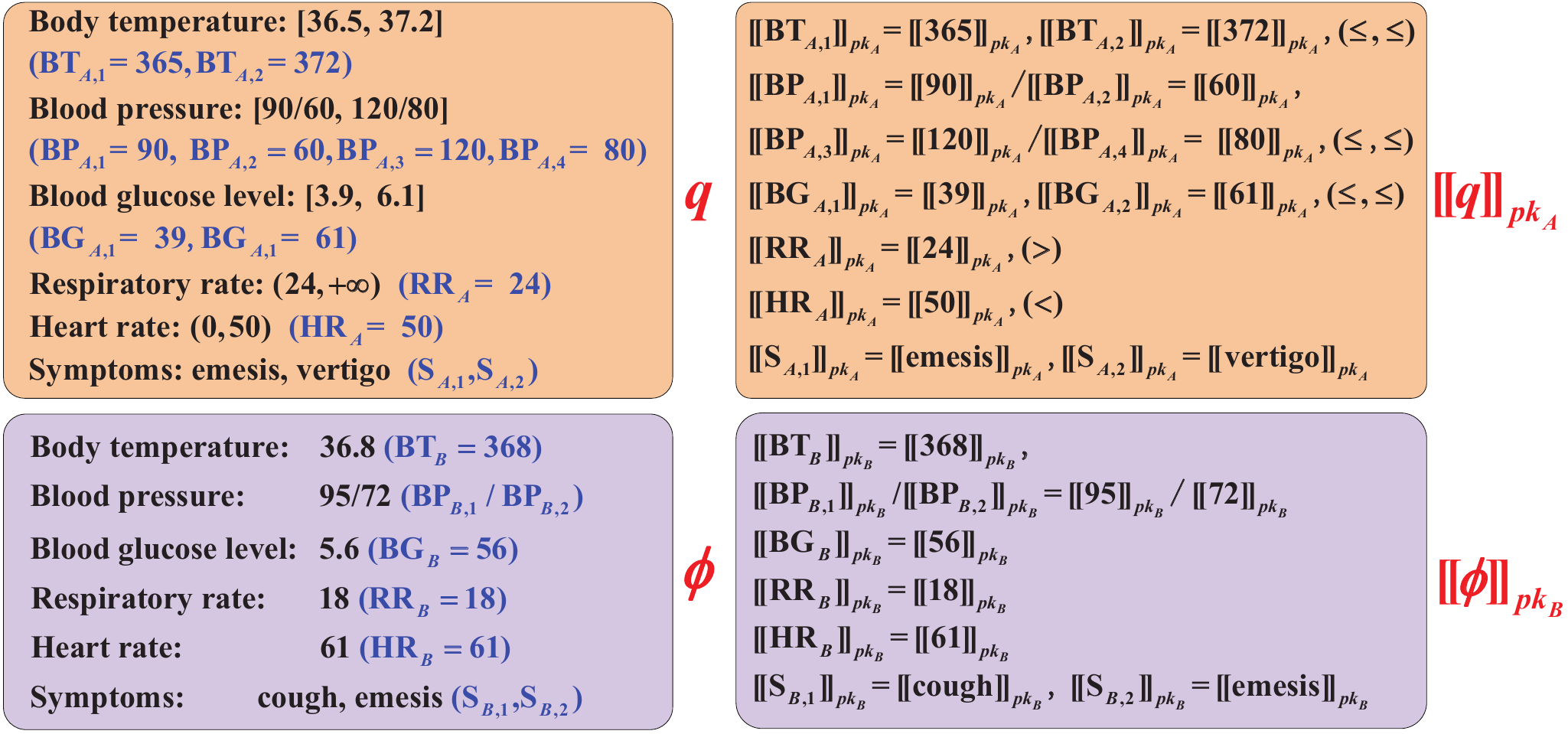}
		\caption{State Encryption Example}
		\label{Fig:SSM}
	\end{center}
\end{figure}

\emph{Illness State Representation}. In the healthcare domain, the illness state can be expressed by symptoms and a set of physiological index, where the former can be described by the patient and the latter can be monitored by the mIoT.
P-Med adopts this method, and Fig. \ref{Fig:SSM}
shows an example for the representation of $q$, $\phi$ and the encryption of them.
Five vital signs of human body are body temperature (BT), blood pressure (BP), blood glucose level (BG),
respiratory rate (RR) and heart rate (HR), which have
frequently-used units $^{\circ}$C, mmHg, mmol/L, breaths per minute and beats per minute, respectively.
In Fig. \ref{Fig:SSM}, the illness state $q$ in medical model utilizes intervals to describe the five vital signs and several medical terminologies
(in lexicographical order) to describe the symptoms. The illness state $\phi$ of patient $B$ is represented by the concrete physiological index rather than interval. If the value of the physiological index is a decimal, an integer (10 or 100) should be multiplied to the value
such that the decimal is mapped to $Z_N$. The multiplication operation should be uniform between the hospital $A$ and patient $B$.
The symptoms and the treatment methods are encrypted utilizing \texttt{K2C}.

\emph{State Match Test}.
\textbf{S}ecure illness \textbf{s}tate \textbf{m}atch protocol (\texttt{SSM}) takes $[\![q]\!]_{pk_{A}}$, $[\![\phi]\!]_{pk_B}$ as input, and outputs
$[\![u^*]\!]_{pk_\sigma}$, where $u^*=1$ indicates that $q$ and $\phi$ match; otherwise, $u^*=0$.
Since it is impossible to exhaustively enumerate all the illness states of various diseases, we utilize
the example in Fig. \ref{Fig:SSM} to illustrate how to design \texttt{SSM} shown in Algorithm \ref{Algo:SSM}.

\begin{algorithm}
	{\small{\KwIn{$[\![q]\!]_{pk_A},[\![\phi]\!]_{pk_B}$.}
			\KwOut{$[\![u^*]\!]_{pk_\sigma}$.}
			initializes $[\![u^*]\!]_{pk_\sigma}=[\![1]\!]_{pk_\sigma}$\;
			CP and CSP jointly calculate
			$[\![u_1]\!]_{pk_\sigma}\leftarrow\texttt{SRC}([\![BT_B]\!]_{pk_B},[\![BT_{A,1}]\!]_{pk_A},[\![BT_{A,2}]\!]_{pk_A})$\;
			$[\![u_{2,1}]\!]_{pk_\sigma}\leftarrow\texttt{SRC}([\![BP_{B,1}]\!]_{pk_B},[\![BP_{A,1}]\!]_{pk_A},[\![BP_{A,3}]\!]_{pk_A})$\;
			$[\![u_{2,2}]\!]_{pk_\sigma}\leftarrow\texttt{SRC}([\![BP_{B,2}]\!]_{pk_B},[\![BP_{A,2}]\!]_{pk_A},[\![BP_{A,4}]\!]_{pk_A})$\;
			$[\![u_2]\!]_{pk_\sigma}\leftarrow\texttt{SMD}([\![u_{2,1}]\!]_{pk_\sigma},[\![u_{2,2}]\!]_{pk_\sigma})$\;
			$[\![u_3]\!]_{pk_\sigma}\leftarrow\texttt{SRC}([\![BG_B]\!]_{pk_B},[\![BG_{A,1}]\!]_{pk_A},[\![BG_{A,2}]\!]_{pk_A})$\;
			$[\![u_4]\!]_{pk_\sigma}\leftarrow\texttt{SGT}([\![RR_B]\!]_{pk_B},[\![RR_A]\!]_{pk_A})$\;
			$[\![u_5]\!]_{pk_\sigma}\leftarrow\texttt{SLT}([\![HR_B]\!]_{pk_B},[\![HR_A]\!]_{pk_A})$\;
			$[\![u_{6,1}]\!]_{pk_\sigma}\leftarrow\texttt{SET}([\![S_{B,1}]\!]_{pk_B},[\![S_{A,1}]\!]_{pk_A})$\;
			$[\![u_{6,2}]\!]_{pk_\sigma}\leftarrow\texttt{SET}([\![S_{B,2}]\!]_{pk_B},[\![S_{A,2}]\!]_{pk_A})$\;
			$[\![u_6]\!]_{pk_\sigma}\leftarrow\texttt{SMD}([\![u_{6,1}]\!]_{pk_\sigma},[\![u_{6,2}]\!]_{pk_\sigma})$\;
			
			\For { $i= 1$ to $6$}
			{
				$[\![u^*]\!]_{pk_\sigma}\leftarrow\texttt{SMD}([\![u^*]\!]_{pk_\sigma},[\![u_i]\!]_{pk_\sigma})$\;
			}
			
			\textbf{Return} $[\![u^*]\!]_{pk_\sigma}$.
			\caption{{\sc Secure illness state Match Protocol
					(\texttt{SSM})}}
			\label{Algo:SSM}}}
\end{algorithm}

In Algorithm \ref{Algo:SSM}, line 2 tests whether the patient $B$'s body temperature $BT_B$ is in the range $[BT_{A,1},BT_{A,2}]$,
and we have $u_1=1$ if it holds (otherwise, $u_1=0$). Line 3-5 tests whether $B$'s blood pressure $BP_{B,1}/BP_{B,2}$ is in the
range $[BP_{A,1}/BP_{A,2},BP_{A,3}/BP_{A,4}]$, and we have $u_2=1$ if it holds (otherwise, $u_2=0$).
Line 6 tests whether $B$'s blood glucose level $BG_B$ is in the range $[BG_{A,1},BG_{A,2}]$, and we have $u_3=1$ if it holds (otherwise, $u_3=0$). Line 7 tests whether $B$'s respiratory rate $RR_B>RR_A$, and we have $u_4=1$ if it holds (otherwise, $u_4=0$).
Line 8 tests whether $B$'s heart rate $HR_B<HR_A$, and we have $u_5=1$ if it holds (otherwise, $u_5=0$).
Line 9-11 tests whether $B$'s symptoms $(S_{B,1},S_{B,2})$ match with $(S_{A,1},S_{A,2})$ in $q$,
and we have $u_6=1$ if it holds (otherwise, $u_6=0$). Line 12-13 tests whether $B$'s illness state $\phi$
matches with $q$, and we have $u^*=1$ if it holds (otherwise, $u^*=0$).
Algorithm \ref{Algo:SSM} is elaborated in
supplemental material C-1.

\section{Proposed P-Med Framework}
This section introduces the system overview and constructs the concrete privacy-preserving algorithms (protocols) according to the design principles proposed in Section \ref{SubSec:Princeple}.

\subsection{System Overview}
\label{SubSec:Overview}

\begin{figure}[htp]
	\begin{center}
		\includegraphics[width=3.4in]{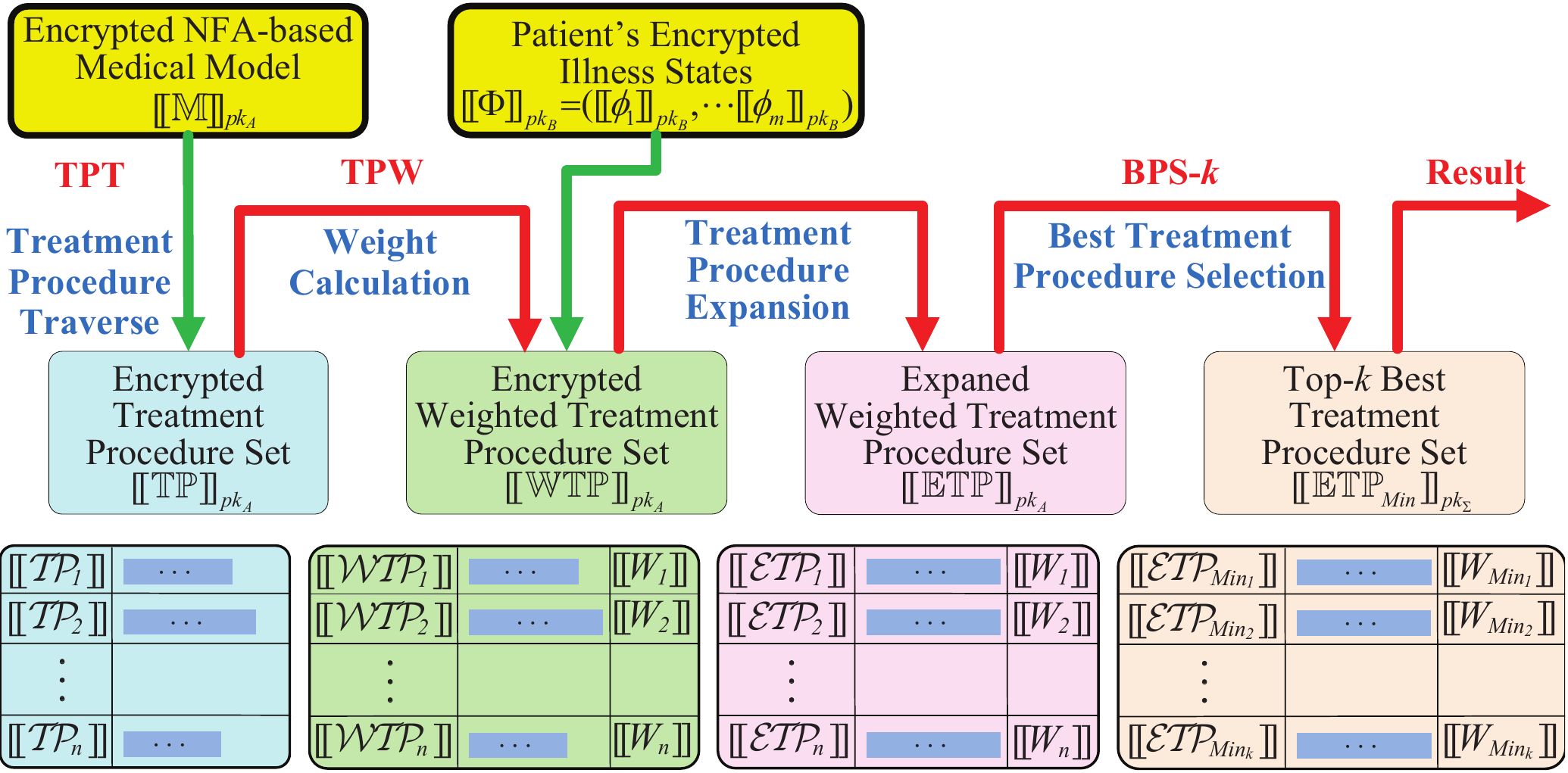}\\
		\caption{P-Med Framework}
		\label{Fig:Framework}
	\end{center}
\end{figure}

P-Med consists of the following four phases (Fig. \ref{Fig:Framework}).

\emph{Treatment Procedure Traverse}. According to design principle 1, CP traverses all the treatment procedures in the medical model $[\![\mathbb{M}]\!]_{pk_{A}}$
and obtains the encrypted treatment procedure set
$[\![\mathbb{TP}]\!]_{pk_A}$ with elements $([\![\mathcal{TP}_1]\!]_{pk_A},\cdots,[\![\mathcal{TP}_n]\!]_{pk_A})$, where 
$[\![\mathcal{TP}_i]\!]_{pk_A}=([\![\mathcal{Q}_i]\!]_{pk_{A}}, [\![\mathcal{Y}_i]\!]_{pk_{A}}, [\![\mathcal{W}_i]\!]_{pk_{A}})$ contains
the encrypted sets of illness states, treatment methods and transition weights, respectively.

\emph{Treatment Procedure Weight Calculation}. According to design principle 2, CP and CSP calculates on $[\![\Phi]\!]_{pk_B}$ and $[\![\mathcal{TP}_i]\!]_{pk_A}$ to get the treatment procedure weight $[\![W_i]\!]_{pk_\sigma}$ and obtain the \textbf{w}eighted \textbf{t}reatment \textbf{p}rocedures set $[\![\mathbb{WTP}]\!]_{pk_A}=([\![\mathcal{WTP}_1]\!]_{pk_A},\cdots,[\![\mathcal{WTP}_n]\!]_{pk_A})$, where
$[\![\mathcal{WTP}_i]\!]_{pk_A}=([\![\mathcal{Q}_i]\!]_{pk_{A}}, [\![\mathcal{Y}_i]\!]_{pk_{A}}, [\![W_i]\!]_{pk_\sigma})$.

\emph{Treatment Procedure Expansion}. Since the elements in $[\![\mathbb{WTP}]\!]_{pk_A}$ may contain different numbers of encrypted illness states and treatment methods, CP and CSP may utilize the length to distinguish the results of P-Med. According to design principle 3, it is necessary to uniform the length by appending encrypted dummy symbols to get \textbf{e}xpended weighted \textbf{t}reatment \textbf{p}rocedure set $[\![\mathbb{ETP}]\!]_{pk_A}$.

\emph{Best Treatment Procedure Selection}. According to design principle 3, secure top-$k$ \textbf{b}est treatment \textbf{p}rocedure \textbf{s}election protocol (\texttt{BPS-$k$})
calculates the top-$k$ most recommended treatment procedures (with the top-$k$ lowest weights). \texttt{BPS-$k$} makes use of
the \textbf{s}ecure \textbf{min}imum selection subprotocols \texttt{SMin}/\texttt{SMin}$_n$ to select the treatment procedure with the lowest weight
from two/$n$ treatment procedures, and \texttt{SMin} is a subprotocol of \texttt{SMin}$_n$.

\subsection{Secure Treatment Procedure Traverse}
\label{SubSec:TPT}

\begin{algorithm}
{\small{\KwIn{$[\![\mathbb{M}]\!]_{pk_{A}}$, $MVisit$, $MState$.}
\KwOut{$[\![\mathbb{TP}]\!]_{pk_A}=([\![\mathcal{TP}_1]\!]_{pk_A},\cdots,[\![\mathcal{TP}_n]\!]_{pk_A})$.}

set the arrays $value(\cdot,\cdot)$, $weight(\cdot,\cdot)$ according to the state transition table of the encrypted NFA $[\![\mathbb{M}]\!]_{pk_{A}}$\;
set the stacks $Q,Y,W$ to be empty and set $n=0$\;

\For { $i= 0$ to $n_1$}
{
    set $count_i = 0$\;
}

\For { $k= 0$ to $MVisit$}
{
    \For { $i= 0$ to $n_1$}
    {
        \For { $j= 0$ to $n_1$}
        {
            set $visit_{k,i,j}=0$\;
        }
    }
}

$Q$.push ($[\![q_0]\!]_{pk_{A}}$), $count_0=count_0+1$\;

\While { $Q\neq\emptyset$}
{
    set $\alpha$ = $Q.peak.element$, $\beta = -1$\;
    \For { $i=1$ to $n_1$}
    {
        \If { $(value_{\alpha,i}\neq\bot)\ \&\ (visit_{count[\alpha],\alpha,i}=0)$}
        {

            set $\beta=i$, $visit_{count[\alpha],\alpha,\beta}=1$\;
        }
    }

    \If {$\beta = -1$}
    {

        \For { $j= 0$ to $n_1$}
        {
            set $visit_{count[\alpha],\alpha,j}=0$\;
        }

        $Q$.pop\;
        $count_{\alpha}=count_{\alpha}-1$\;

        \If { $Y\neq\emptyset$}
        {
            $Y$.pop, $W$.pop\;
        }
    }

    \ElseIf {$(\beta \neq -1)\ \&\ (count_{\beta}<MVisit)$}
    {
        $Y$.push ($value_{\alpha,\beta}$), $W$.push ($weight_{\alpha,\beta}$),
        $Q$.push ($[\![q_{\beta}]\!]_{pk_{A}}$), $count_{\beta}=count_{\beta}+1$\;
    }

    \If { $Q\neq\emptyset$}
    {
        set $\alpha'$ = $Q.peak.element$\;

        \If {$([\![q_{\alpha'}]\!]_{pk_{A}}\in [\![\mathcal{F}]\!]_{pk_{A}})$ }
        {
            $n=n+1$,
            $[\![\mathcal{Q}_{n}]\!]_{pk_{A}} = Q$, $[\![\mathcal{Y}_{n}]\!]_{pk_{A}} = Y$, $[\![\mathcal{W}_{n}]\!]_{pk_{A}} = W$\;
            $Q$.pop, $Y$.pop, $W$.pop,
            $count_{\alpha'}=count_{\alpha'}-1$\;
        }
        \ElseIf {$([\![q_{\alpha'}]\!]_{pk_{A}}\notin [\![\mathcal{F}]\!]_{pk_{A}})\ \&\ (Q.size=MState)$ }
        {
            $Q$.pop, $Y$.pop, $W$.pop,
            $count_{\alpha'}=count_{\alpha'}-1$\;
        }
    }
}
set $[\![\mathcal{TP}_i]\!]_{pk_A}=([\![\mathcal{Q}_i]\!]_{pk_{A}}, [\![\mathcal{Y}_i]\!]_{pk_{A}}, [\![\mathcal{W}_i]\!]_{pk_{A}})$ ($1\leq i\leq n$)\;
\textbf{Return} $[\![\mathbb{TP}]\!]_{pk_A}=([\![\mathcal{TP}_1]\!]_{pk_A},\cdots,[\![\mathcal{TP}_n]\!]_{pk_A})$.
\caption{{\sc Secure Treatment Procedure Traverse Algorithm
(\texttt{TPT})}}
\label{Algo:TPT}}}
\end{algorithm}

According to design principle 1, secure treatment procedure traverse algorithm (\texttt{TPT}) finds all the treatment procedures in $[\![\mathbb{M}]\!]_{pk_{A}}$,
which comes from $[\![q_0]\!]_{pk_{A}}$ to $[\![\mathcal{F}]\!]_{pk_{A}}$.
In computation theory's view, \texttt{TPT} finds the regular language of the NFA $[\![\mathbb{M}]\!]_{pk_{A}}$.
In graph theory's view, \texttt{TPT} finds all the paths from $[\![q_0]\!]_{pk_{A}}$ to $[\![\mathcal{F}]\!]_{pk_{A}}$ in the graph of $[\![\mathbb{M}]\!]_{pk_{A}}$,
which may contain loops and self-loops.
To prevent endless loops, hospital $A$ needs to designate the maximum visit time $MVisit$ that each state
can be included in a treatment procedure,
and the maximum state number $MState$ that each treatment procedure contains.

Given $[\![\mathbb{M}]\!]_{pk_{A}},MVisit,MState$ as input,
\texttt{TPT} (Algorithm \ref{Algo:TPT}) outputs
$[\![\mathbb{TP}]\!]_{pk_A}=([\![\mathcal{TP}_1]\!]_{pk_A},\cdots,[\![\mathcal{TP}_n]\!]_{pk_A})$, where $[\![\mathcal{TP}_i]\!]_{pk_A}=([\![\mathcal{Q}_i]\!]_{pk_{A}}, [\![\mathcal{Y}_i]\!]_{pk_{A}}, [\![\mathcal{W}_i]\!]_{pk_{A}})$,
illness states
$[\![\mathcal{Q}_i]\!]_{pk_{A}}=$ $([\![q_0]\!]_{pk_{A}}, [\![q_{i,\theta_1}]\!]_{pk_{A}},\cdots,[\![q_{i,\theta_{\tau_i}}]\!]_{pk_{A}})$,
treatment methods
$[\![\mathcal{Y}_i]\!]_{pk_{A}}=([\![y_{i,\theta_1}]\!]_{pk_{A}},\cdots,[\![y_{i,\theta_{\tau_i}}]\!]_{pk_{A}})$
\footnote{To facilitate the expression, the encrypted empty string $[\![\epsilon]\!]$ is also denoted as a symbol in $([\![y_{i,\theta_1}]\!]_{pk_{A}},\cdots,[\![y_{i,\theta_{\tau_i}}]\!]_{pk_{A}})$.},
and transition weights
$[\![\mathcal{W}_i]\!]_{pk_{A}}=([\![w_{i,\theta_1}]\!]_{pk_{A}},\cdots,[\![w_{i,\theta_{\tau_i}}]\!]_{pk_{A}})$.
The elements in $[\![\mathcal{Q}_i]\!]_{pk_{A}}$, $[\![\mathcal{Y}_i]\!]_{pk_{A}}$ and $[\![\mathcal{W}_i]\!]_{pk_{A}}$
are arranged in accordance to the path in the directed graph of $[\![\mathbb{M}]\!]_{pk_{A}}$.
For a particular $[\![\mathbb{M}]\!]_{pk_{A}}$,
CP runs \texttt{TPT} only once and stores $[\![\mathbb{TP}]\!]_{pk_A}$ in the cloud.
The basic idea of \texttt{TPT} is summarized as the following steps.

1) Push $[\![q_0]\!]_{pk_{A}}$ into stack $Q$.

2) Set the encrypted illness state on the top of the stack $Q$ as $\alpha$.
  Check whether there exists any illness state satisfies the following requirements:
  it is connected with the illness state $\alpha$ and has not been visited starting from the encrypted illness state $\alpha$.

3) If such illness state in step 2 exists in the graph, push it into $Q$;
otherwise, pop $\alpha$ from $Q$.

4) If the top element in $Q$ does not belong to $[\![\mathcal{F}]\!]_{pk_{A}}$ and $Q$ achieves the maximum illness state number $MState$,
  pop it from $Q$; otherwise, \texttt{TPT} algorithm successfully finds out a treatment procedure from $[\![q_0]\!]_{pk_{A}}$ to $[\![F]\!]_{pk_{A}}$
  The found treatment procedure is recorded in $\mathcal{TP}_i=\{[\![\mathcal{Q}_i]\!]_{pk_{A}}, [\![\mathcal{Y}_i]\!]_{pk_{A}}, [\![\mathcal{W}_i]\!]_{pk_{A}}\}$ ($i\in[1,n]$),
  and then pop the top element in $Q$.

5) Repeat the steps 2-4 until the stack $Q$ is empty.

The notations in the \texttt{TPT} are introduced below.

$\bullet$ $\boldsymbol{count(\cdot)}$. The one-dimensional array $count(\cdot)$ has $n_1+1$ elements,
  and the element $count_i$ counts the number of encrypted illness state
  $[\![q_i]\!]_{pk_{A}}$ in $Q$, for $0\leq i\leq n_1$.

$\bullet$ $\boldsymbol{value(\cdot,\cdot),~weight(\cdot,\cdot)}$.
  The two-dimensional arrays $value(\cdot,\cdot),~weight(\cdot,\cdot)$ have $n_1+1$ rows and $n_1+1$ columns,
  and are initialized according to the state transition table of
  $[\![\mathbb{M}]\!]_{pk_{A}}$.
  The label of the first (resp. second) dimension represents the label of the current (resp. next) state.
  The element in $value(\cdot,\cdot)$ (resp. $weight(\cdot,\cdot)$) represents the encrypted treatment method
  (resp. encrypted transition weight) from the current state to the next one.

$\bullet$ $\boldsymbol{visit(\cdot,\cdot,\cdot)}$. The three-dimensional array $visit(\cdot,\cdot,\cdot)$
  is an array of matrices and it has $(MVisit+1)$ matrices,
  where each matrices has $n_1+1$ rows and $n_1+1$ columns.
  Since each state $[\![q_i]\!]_{pk_{A}}$ may appear at most $MVisit$ times
  in a treatment procedure, and $count_i$ counts its occurrence number in $Q$.
  For $[\![q_i]\!]_{pk_{A}}$ that appears the $count_i$-th time in $Q$,
  the element $visit_{count[i],i,j}=1$ indicates that transition from $[\![q_i]\!]_{pk_{A}}$
  to $[\![q_j]\!]_{pk_{A}}$ is visited, and $visit_{count[i],i,j}=0$ indicates that transition is not visited.

\texttt{TPT} is elaborated in supplemental material C-2.

\subsection{Secure Treatment Procedure Weight Calculation}
\label{SubSec:TPW}

\begin{algorithm}
{\small{\KwIn{$MWeight$, $[\![\Phi]\!]_{pk_B}=([\![\phi_1]\!]_{pk_B},\cdots,[\![\phi_m]\!]_{pk_B})$, $[\![\mathbb{TP}]\!]_{pk_A}=([\![\mathcal{TP}_1]\!]_{pk_A},\cdots,[\![\mathcal{TP}_n]\!]_{pk_A})$.}
\KwOut{$[\![\mathbb{WTP}]\!]_{pk_{A}}=([\![\mathcal{WTP}_1]\!]_{pk_A},\cdots,[\![\mathcal{WTP}_n]\!]_{pk_A})$.}

\For { $i= 1$ to $n$}
{
    $[\![W_i]\!]_{pk_\sigma}=[\![0]\!]_{pk_\sigma}$, $[\![v_i]\!]_{pk_\sigma}=[\![1]\!]_{pk_\sigma}$, $[\![v_i']\!]_{pk_\sigma}=[\![0]\!]_{pk_\sigma}$\;

    \If {$\theta_{\tau_i}-1\geq m$}
    {   
	    \For { $j= 0$ to $\theta_{\tau_i}-m-1$}
	    {
	        $[\![s_j]\!]_{pk_\sigma}=[\![0]\!]_{pk_\sigma}$, $[\![s_j']\!]_{pk_\sigma}=[\![0]\!]_{pk_\sigma}$, $[\![s_j'']\!]_{pk_\sigma}=[\![0]\!]_{pk_\sigma}$\;
	        $[\![a_j]\!]_{pk_\sigma}=[\![0]\!]_{pk_\sigma}$, $[\![a_j']\!]_{pk_\sigma}=[\![0]\!]_{pk_\sigma}$, $[\![a_j'']\!]_{pk_\sigma}=[\![1]\!]_{pk_\sigma}$\;
	        \For { $k= 1$ to $m$}
	        {
	            $[\![u_k]\!]_{pk_\sigma}\leftarrow\texttt{SSM}([\![\phi_k]\!]_{pk_B},[\![q_{i,\theta_{j+(k-1)}}]\!]_{pk_A})$\;
	            $[\![a_j]\!]_{pk_\sigma}=[\![a_j]\!]_{pk_\sigma}\cdot[\![u_k]\!]_{pk_\sigma}$\;
	        }
	        $[\![s_j']\!]_{pk_\sigma}\leftarrow\texttt{SET}([\![a_j]\!]_{pk_\sigma},[\![m]\!]_{pk_\sigma})$\;
	        $[\![a_j'']\!]_{pk_\sigma}\leftarrow\texttt{SET}([\![v_i']\!]_{pk_\sigma},[\![0]\!]_{pk_\sigma})$\;
	        $[\![v_i']\!]_{pk_\sigma}=[\![v_i']\!]_{pk_\sigma}\cdot[\![s_j']\!]_{pk_\sigma}$\;
	
	        \For { $k= j+m$ to $\tau_i$}
	        {
	            $[\![s_j'']\!]_{pk_\sigma}\leftarrow\texttt{SAD}([\![s_j'']\!]_{pk_\sigma},[\![w_{i,\theta_k}]\!]_{pk_A})$\;
	        }
	        $[\![a_j']\!]_{pk_\sigma}\leftarrow\texttt{SMD}([\![s_j']\!]_{pk_\sigma},[\![s_j'']\!]_{pk_\sigma})$\;
	        $[\![s_j]\!]_{pk_\sigma}\leftarrow\texttt{SMD}([\![a_j']\!]_{pk_\sigma},[\![a_j'']\!]_{pk_\sigma})$\;
	        $[\![W_i]\!]_{pk_\sigma}=[\![W_i]\!]_{pk_\sigma}\cdot[\![s_j]\!]_{pk_\sigma}$\;
	    }
	
	    $[\![v_i]\!]_{pk_\sigma}\leftarrow\texttt{SET}([\![v_i']\!]_{pk_\sigma},[\![0]\!]_{pk_\sigma})$\;
	    $[\![W_i]\!]_{pk_\sigma}=[\![W_i]\!]_{pk_\sigma}\cdot([\![v_i]\!]_{pk_\sigma})^{MWeight}$\;
	    set $[\![\mathcal{WTP}_i]\!]_{pk_A}=([\![\mathcal{Q}_i]\!]_{pk_{A}}, [\![\mathcal{Y}_i]\!]_{pk_{A}}, [\![W_i]\!]_{pk_\sigma})$\;
	}
	\Else
	{
		$[\![W_i]\!]_{pk_\sigma}=[\![MWeight]\!]_{pk_\sigma}$\;
	}
}
\textbf{Return} $[\![\mathbb{WTP}]\!]_{pk_{A}}=([\![\mathcal{WTP}_1]\!]_{pk_A},\cdots,[\![\mathcal{WTP}_n]\!]_{pk_A})$.
\caption{{\sc Secure Treatment Procedure Weight Calculation Protocol
(\texttt{TPW})}}
\label{Algo:TPW}}}
\end{algorithm}

According to design principle 2, secure treatment procedure weight calculation protocol (\texttt{TPW}) takes $MWeight$, $[\![\Phi]\!]_{pk_B}$, $[\![\mathbb{TP}]\!]_{pk_A}$ as input,
and outputs the encrypted weighted treatment procedures
$[\![\mathbb{WTP}]\!]_{pk_A}=([\![\mathcal{WTP}_1]\!]_{pk_A},\cdots,[\![\mathcal{WTP}_n]\!]_{pk_A})$ with 
$[\![\mathcal{WTP}_i]\!]_{pk_A}=([\![\mathcal{Q}_i]\!]_{pk_{A}}, [\![\mathcal{Y}_i]\!]_{pk_{A}}, [\![W_i]\!]_{pk_\sigma}),$
where the elements $[\![\mathcal{Q}_i]\!]_{pk_{A}}, [\![\mathcal{Y}_i]\!]_{pk_{A}}$ in $[\![\mathcal{WTP}_i]\!]_{pk_A}$ are the same
as that in $[\![\mathcal{TP}_i]\!]_{pk_A}$, and $[\![W_i]\!]_{pk_\sigma}$ is the calculated treatment procedure weight.
\texttt{TPW} is shown in Algorithm \ref{Algo:TPW} and elaborated in supplemental material C-3.

The result of \texttt{TPW} is classified into the following scenarios.
(1) If patient $B$'s illness condition set $[\![\Phi]\!]_{pk_B}$ does not appear in the treatment procedure $[\![\mathcal{TP}_i]\!]_{pk_A}$,
we have $[\![W_i]\!]_{pk_\sigma}=[\![MWeight]\!]_{pk_\sigma}$.
(2) If $[\![\Phi]\!]_{pk_B}$ appears at least one time \footnote{$[\![\Phi]\!]_{pk_B}$ may appear more than one time in $[\![\mathcal{TP}_i]\!]_{pk_A}$
due to that loop may exist.} in $[\![\mathcal{TP}_i]\!]_{pk_A}$, we have $[\![W_i]\!]_{pk_\sigma}=[\![\Sigma_{k=\bar{j}+m}^{\tau_i}{w_{i,\theta_{k}}}]\!]_{pk_\sigma}$,
and $([\![q_{i,\theta_{\bar{j}}}]\!]_{pk_{A}},\cdots,[\![q_{i,\theta_{\bar{j}+m-1}}]\!]_{pk_{A}})$ is the first-match state set.

\subsection{Secure Treatment Procedure Expansion and Selection}
\label{SubSec:BPS}

\emph{Secure Treatment Procedure Expansion}. To uniform the lengths of weighted treatment procedures,
$[\![\mathbb{WTP}]\!]_{pk_A}$ is converted to $[\![\mathbb{ETP}]\!]_{pk_A}$ by appending the encrypted dummy symbols (such as $[\![\bot]\!]_{pk_A}$) such that the element $[\![\mathcal{ETP}_i]\!]_{pk_A}=([\![\mathcal{Q}_i]\!]_{pk_{A}}, [\![\mathcal{Y}_i]\!]_{pk_{A}}, [\![W_i]\!]_{pk_\sigma})$ in $[\![\mathbb{ETP}]\!]_{pk_A}$ has 
$|[\![\mathcal{Q}_i]\!]_{pk_{A}}|=MState$ and $|[\![\mathcal{Y}_i]\!]_{pk_{A}}|=MState-1$.

\emph{Secure Best Treatment Procedure Selection}. According to design principle 3, the selection task is fulfilled by three protocols: \textbf{s}ecure \textbf{min}imum selection protocol (\texttt{SMin}), \textbf{s}ecure \textbf{min}imum selection from $n$ treatment procedures protocol (\texttt{SMin$_n$}) and
secure top-$k$ \textbf{b}est treatment \textbf{p}rocedure \textbf{s}election protocol (\texttt{BPS-$k$}). \texttt{SMin}
selects the best treatment procedure from two procedures; \texttt{SMin$_n$} leverages \texttt{SMin} as sub-protocol
to select the best one from $n$ treatment procedures;
\texttt{BPS-$k$} leverages \texttt{SMin$_n$} as sub-protocol to select the top-$k$ most recommended treatment procedures (with the lowest weights)
in a privacy-preserving way.

\subsubsection{Secure Minimum Selection Protocol (\texttt{SMin})}
\label{SubSec:SMin}

On input $[\![\mathcal{ETP}_1]\!]_{pk_{A}}$ and $[\![\mathcal{ETP}_2]\!]_{pk_{A}}$, \texttt{SMin} outputs $[\![\mathcal{ETP}_{Min}]\!]_{pk_\sigma}$
such that $W_{Min}=min(W_1,W_2)$,
and $[\![\mathcal{Q}_{Min}]\!]_{pk_\sigma}, [\![\mathcal{Y}_{Min}]\!]_{pk_\sigma}$ are the corresponding
encrypted illness states and treatment methods, respectively. It is required that CP and CSP can not distinguish
$[\![\mathcal{ETP}_{Min}]\!]_{pk_\sigma}$ comes from $[\![\mathcal{ETP}_1]\!]_{pk_{A}}$ or $[\![\mathcal{ETP}_2]\!]_{pk_{A}}$.

\textbf{Step 1}: CP calculates $[\![W_1']\!]_{pk_\sigma}=[\![W_1]\!]_{pk_\sigma}^2\cdot[\![1]\!]_{pk_\sigma}$,
$[\![W_2']\!]_{pk_\sigma}=[\![W_2]\!]_{pk_\sigma}^2$, and flips a random coin $s\in\{0,1\}$.
CP randomly selects $r_0',r_0,r_1,r_{2,i},r_{3,i}\in Z_N$ ($1\leq i\leq MState-1$) such that $\mathcal{L}(N)/8<\mathcal{L}(r_0')<\mathcal{L}(N)/4-1$
and $\mathcal{L}(r_0)<\mathcal{L}(N)/8$.
CP calculates
\begin{align*}
l_0&=([\![W_{2-s}']\!]_{pk_\sigma})^{r_0'}\cdot([\![W_{s+1}']\!]_{pk_\sigma})^{N-r_0'}\cdot[\![r_0]\!]_{pk_\sigma},\\
l_1&=[\![W_{s+1}]\!]_{pk_\sigma}\cdot([\![W_{2-s}]\!]_{pk_\sigma})^{N-1}\cdot[\![r_1]\!]_{pk_\sigma},\\
l_{2,i}&\leftarrow\texttt{SAD}([\![q_{s+1,i}]\!]_{pk_{A}}\cdot([\![q_{2-s,i}]\!]_{pk_{A}})^{N-1},[\![r_{2,i}]\!]_{pk_\sigma}),\\
l_{3,i}&\leftarrow\texttt{SAD}([\![y_{s+1,i}]\!]_{pk_{A}}\cdot([\![y_{2-s,i}]\!]_{pk_{A}})^{N-1},[\![r_{3,i}]\!]_{pk_\sigma}).
\end{align*}
CP computes $l_0'=\texttt{PD1}_{SK_1}(l_0)$ and sends $(l_0',l_0,l_1,l_{2,i},$ $l_{3,i})$ to CSP ($1\leq i\leq MState-1$).

\textbf{Step 2}: CSP calculates $l_0''=\texttt{PD2}_{SK_2}(l_0,l_0')$.
If $\mathcal{L}(l_0'')>\mathcal{L}(N)/2$, CSP sets $t=0$ and calculates
$l_4=[\![0]\!]_{pk_\sigma},~l_{5,i}=[\![0]\!]_{pk_\sigma},~l_{6,i}=[\![0]\!]_{pk_\sigma}$.
If $\mathcal{L}(l_0'')<\mathcal{L}(N)/2$, CSP sets $t=1$ and calculates
$l_4=\texttt{CR}(l_1),~l_{5,i}=\texttt{CR}(l_{2,i}),~l_{6,i}=\texttt{CR}(l_{3,i})$.
CSP sends $([\![t]\!]_{pk_\sigma},l_4,l_{5,i},l_{6,i})$ to CP.

\textbf{Step 3}: CP calculates
\begin{align*}
[\![W_{Min}]\!]_{pk_\sigma}&=[\![W_{2-s}]\!]_{pk_\sigma}\cdot l_4\cdot([\![t]\!]_{pk_\sigma})^{N-r_1},\\
[\![q_{Min,i}]\!]_{pk_\sigma}&\leftarrow\texttt{SAD}([\![q_{2-s,i}]\!]_{pk_{A}},l_{5,i})\cdot([\![t]\!]_{pk_\sigma})^{N-r_{2,i}},\\
[\![y_{Min,i}]\!]_{pk_\sigma}&\leftarrow\texttt{SAD}([\![y_{2-s,i}]\!]_{pk_{A}},l_{6,i})\cdot([\![t]\!]_{pk_\sigma})^{N-r_{3,i}},\\
[\![q_0]\!]_{pk_\sigma}&\leftarrow\texttt{SAD}([\![q_0]\!]_{pk_A}, [\![0]\!]_{pk_\sigma}),
\end{align*}
and sets $[\![\mathcal{ETP}_{Min}]\!]=([\![\mathcal{Q}_{Min}]\!], [\![\mathcal{Y}_{Min}]\!], [\![W_{Min}]\!])$,
where
$[\![\mathcal{Q}_{Min}]\!]=([\![q_0]\!], [\![q_{Min,1}]\!],\cdots,[\![q_{Min,MState-1}]\!])$,
$[\![\mathcal{Y}_{Min}]\!]=([\![y_{Min,1}]\!],\cdots,[\![y_{Min,MState-1}]\!])$.\footnote{The encryption key $pk_\sigma$ is omitted here to simplify the expression.}

The detail of \texttt{SMin} is depicted in Fig. \ref{Fig:WorkflowSMin}, and the correctness is elaborated in
supplemental material C-4.

\subsubsection{Secure Minimum Treatment Selection from $n$ (\texttt{SMin$_n$})}
\label{SubSec:SMinN}

On input $[\![\mathbb{ETP}]\!]_{pk_A}=([\![\mathcal{ETP}_1]\!]_{pk_{A}},\cdots, [\![\mathcal{ETP}_n]\!]_{pk_{A}})$,
\texttt{SMin$_n$}
outputs $[\![\mathcal{ETP}_{Min}]\!]_{pk_\sigma}$
such that $W_{Min}=min(W_1,\cdots,W_n)$,
and $[\![\mathcal{Q}_{Min}]\!]_{pk_\sigma}, [\![\mathcal{Y}_{Min}]\!]_{pk_\sigma}$ are the corresponding
encrypted illness states and treatment methods, respectively. It is required that CP and CSP cannot distinguish
$[\![\mathcal{ETP}_{Min}]\!]_{pk_\sigma}$ comes from which element in $[\![\mathbb{ETP}]\!]_{pk_A}$.

\begin{figure}[htp]
	\begin{center}
		\includegraphics[width=3.2in]{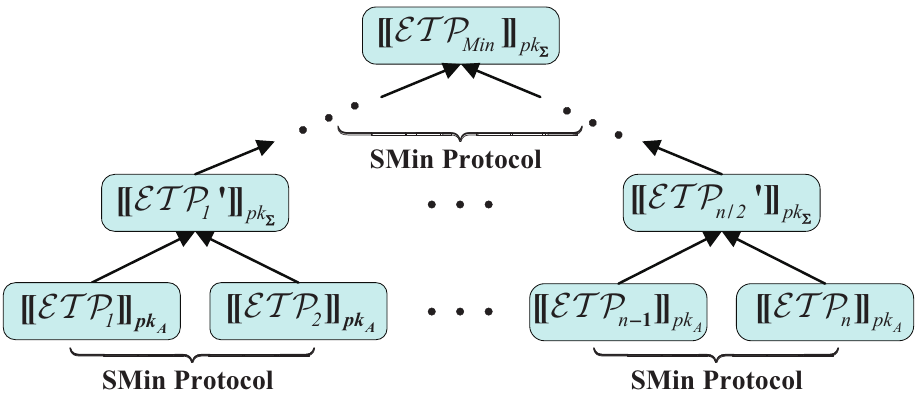}
		\caption{Running Procedure of \texttt{SMin$_n$}}
		\label{Fig:SMinN}
	\end{center}
\end{figure}

Fig. \ref{Fig:SMinN} shows the procedure to get $[\![\mathcal{ETP}_{Min}]\!]_{pk_\sigma}$
using \texttt{SMin} as sub-protocol. The best treatment procedure is selected
between two adjacent procedures in each layer,
which executes continuously until it reaches the top.

\subsubsection{Secure Top-$k$ Treatment Selection (\texttt{BPS-$k$})}
\label{SubSec:BPSK}

Taken as input $[\![\mathbb{ETP}]\!]_{pk_A}$,
\texttt{BPS-$k$} (shown in Algorithm \ref{Algo:Top-K}) outputs $[\![\mathbb{ETP}_{Min}]\!]_{pk_\sigma}=([\![\mathcal{ETP}_{Min_1}]\!]_{pk_\sigma},\cdots,[\![\mathcal{ETP}_{Min_k}]\!]_{pk_\sigma})$
such that
$W_{Min_1},\cdots,W_{Min_k}$ are the top-$k$ lowest weights,
and $[\![\mathcal{Q}_{Min_i}]\!]_{pk_\sigma}, [\![\mathcal{Y}_{Min_i}]\!]_{pk_\sigma}$ are the corresponding
encrypted illness states and treatment methods for $1\leq i \leq k$. It is required that CP and CSP cannot distinguish
the elements in $[\![\mathbb{ETP}_{Min}]\!]_{pk_\sigma}$ comes from which elements in $[\![\mathbb{ETP}]\!]_{pk_A}$.
The basic idea of \texttt{BPS-$k$}
is to find the treatment procedure with the lowest weight in each round.
Then, multiply the corresponding weight with $MWeight$, and set the other weights unchanged. After $k$ rounds,
top-$k$ best treatment procedures are found.

The correctness of \texttt{BPS-$k$} is elaborated in
supplemental material C-5, and a toy example is given in Fig. \ref{Fig:WorkflowBPSK} to illustrate the workflow.

\begin{algorithm}
{\small{\KwIn{$[\![\mathbb{ETP}]\!]_{pk_A}$.}
\KwOut{$[\![\mathbb{ETP}_{Min}]\!]_{pk_\sigma}$.}

Set $S=[\![\mathbb{ETP}]\!]_{pk_A}$\;

\For {$i= 1$ to $k$}
{CP and CSP jointly calculate
 $[\![\mathcal{ETP}_{Min_i}]\!]_{pk_\sigma}\leftarrow\texttt{SMin}_n(S)$\;

 \For{$j=1$ to $n$}
 {CP randomly selects $r_j \in Z_N$ and computes
$l_j=([\![W_{Min_i}]\!]_{pk_\sigma})^{r_j}\cdot([\![W_{j}]\!]_{pk_\sigma})^{N-r_j}$,
$l_j'=\texttt{PD1}_{SK_1}(l_j)$, where $\mathcal{L}(r_j)<\mathcal{L}(N)/4-1$\;}

Permute $(l_j,l_j')$ using permutation function $\pi_i$ and get
$(l_{\pi_i(j)},l_{\pi_i(j)}')$ for $1\leq j\leq n$, which are sent to CSP\;

CSP computes
$l_{\pi_i(j)}''=\texttt{PD2}_{SK_2}(l_{\pi_i(j)},l_{\pi_i(j)}')$\;
If $l_{\pi_i(j)}''=0$, set $A_{\pi_i(j)}= [\![MWeight]\!]_{pk_\sigma}$; otherwise, $A_{\pi_i(j)}= [\![1]\!]_{pk_\sigma}$.
 CSP sends $A_{\pi_i(j)}$ to CP, $1\leq j\leq n$\;
 CP obtains $(A_1,\cdots,A_n)$ by using permutation $\pi^{-1}_i$\;
 Refresh $([\![W_1]\!]_{pk_\sigma},\cdots,[\![W_n]\!]_{pk_\sigma})$ in $S$ by computing
 $[\![W_j]\!]_{pk_\sigma}\leftarrow\texttt{SMD}([\![W_j]\!]_{pk_\sigma}, A_j )$, $1\leq j\leq n$\;
 }
 \textbf{Return} $[\![\mathbb{ETP}_{Min}]\!]_{pk_\sigma}$.
\caption{{\sc  Secure Top-$k$ Best Treatment Procedures Selection (\texttt{BPS-$k$})}}
\label{Algo:Top-K}
 }}
\end{algorithm}

\section{Security and Design Goal Analysis of P-Med}
This section proves the security of the subprotocols and the security of P-Med. It is also analyzed that the design goal of P-Med is achieved.

\subsection{Security of Subprotocols}
\label{SubSec:SecProtocol}

\begin{theorem}
\label{Theorem:TPT}
\texttt{TPT} is secure against semi-honest
(non-colluding) attackers $\mathcal{A}=(\mathcal{A}_{D_1},\mathcal{A}_{S_1},\mathcal{A}_{S_2})$
and the adversary $\mathcal{A}^*$ defined in the attack model.
\end{theorem}

\emph{\textbf{Proof}}.
All data calculated in \texttt{TPT} are encrypted using PCTD encryption,
and not decrypted by CP or CSP in the processing process.
Due to the security of PCTD in \cite{Bresson03},
\texttt{TPT} is secure against attackers $\mathcal{A}, \mathcal{A}^*$.
\hfill$\square$

\begin{theorem}
\label{Theorem:SSM}
\texttt{SSM} is secure against semi-honest
(non-colluding) attackers $\mathcal{A}=(\mathcal{A}_{D_1},\mathcal{A}_{S_1},\mathcal{A}_{S_2})$
and the adversary $\mathcal{A}^*$ defined in the attack model.
\end{theorem}

\emph{\textbf{Proof}}.
\texttt{SSM} calls \texttt{SRC}, \texttt{SMD}, \texttt{SLT}, \texttt{SGT} and \texttt{SET} as subprotocol, which are proved secure in \cite{YangY_TIFS17,LiuX_TIFS16,YangY_INS18}.
All data calculated in \texttt{SSM} are encrypted using PCTD.
Thus, \texttt{SSM} is secure against the attackers $\mathcal{A}, \mathcal{A}^*$.
\hfill$\square$

\begin{theorem}
\label{Theorem:TPW}
\texttt{TPW} is secure against semi-honest
(non-colluding) attackers $\mathcal{A}=(\mathcal{A}_{D_1},\mathcal{A}_{S_1},\mathcal{A}_{S_2})$
and the adversary $\mathcal{A}^*$ defined in the attack model.
\end{theorem}

\emph{\textbf{Proof}}.
\texttt{TPW} calls \texttt{SSM}, \texttt{SET}, \texttt{SAD}
and \texttt{SMD} as subprotocol, which are proved in the above theorems and \cite{YangY_TIFS17,LiuX_TIFS16}.
All data calculated in \texttt{TPW} are encrypted using PCTD.
Thus, \texttt{TPW} is secure against attackers $\mathcal{A}, \mathcal{A}^*$.
\hfill$\square$

\begin{theorem}
\label{Theorem:SMin} \texttt{SMin} is secure against semi-honest
(non-colluding) attackers $\mathcal{A}=(\mathcal{A}_{D_1},\mathcal{A}_{S_1},\mathcal{A}_{S_2})$
and the adversary $\mathcal{A}^*$ defined in the attack model.
\end{theorem}

\emph{\textbf{Proof}}. Please see supplemental materials D-1.

\begin{theorem}
\label{Theorem:SMinN} $\texttt{SMin}_n$ is secure against semi-honest (non-colluding)
attackers
$\mathcal{A}=(\mathcal{A}_{D_1},\mathcal{A}_{S_1},\mathcal{A}_{S_2})$
and the adversary $\mathcal{A}^*$ defined in the attack model.
\end{theorem}

\emph{\textbf{Proof}}. $\texttt{SMin}_n$ protocol calls
\texttt{SMin} as subprotocol,
which is proved secure in Theorem \ref{Theorem:SMin}.
Thus, $\texttt{SMin}_n$ is also secure against
attackers $\mathcal{A},\mathcal{A}^*$.
\hfill$\square$

\begin{theorem}
\label{Theorem:BPSK} \texttt{BPS-$k$}
 is secure against
(non-colluding) attackers $\mathcal{A}=(\mathcal{A}_{D_1},\mathcal{A}_{S_1},\mathcal{A}_{S_2})$
and the adversary $\mathcal{A}^*$ defined in the attack model.
\end{theorem}

\emph{\textbf{Proof}}. Please see supplemental materials D-2.

%

\subsection{Security of P-Med Framework}
\label{SubSec:SecPMed}

Here, we utilize the attack model to demonstrate that P-Med can resist the adversary $\mathcal{A}^*$.
1) If $\mathcal A^*$ eavesdrops on the transmission between the hospital
and CP, and on the transmission between the challenge patient and CP,
all data transmitted in these two links are obtained by $\mathcal A^*$,
which includes encrypted NFA-based medical model, encrypted illness states and encrypted treatment result.
Moreover, the intermediate calculated ciphertext
(obtained by executing \texttt{SSM}, \texttt{TPW}, \texttt{SMin},\texttt{SMin}$_n$ and \texttt{BPS-$k$} protocols)
that are transmitted between CP and CSP can also be eavesdropped by
$\mathcal A^*$. However, since these data are encrypted by PCTD before transmission,
$\mathcal A^*$ cannot derive the medical model without knowing hospital's secret key
nor decrypt challenge patient's ciphertext without knowing the challenge patient's secret key.
2-3) Assume $\mathcal A^*$ compromise CP (or CSP) to obtain the partial
strong private key $\lambda_1$ (or $\lambda_2$), but $\mathcal
A^*$ cannot compromise CP and CSP concurrently. $\mathcal A^*$
could not get the strong secret key $\lambda$ since it is randomly
split into two parts using \texttt{SKeyS} algorithm
of PCTD. Even when CSP is compromised and the intermediate result in the protocols of P-Med are obtained,
$\mathcal A^*$ cannot get any useful information since the ``blinding" method
\cite{PeterA13} is applied to conceal the
plaintext: a random number is added to or multiplied with the
plaintext before transmitted to CSP. 4) If $\mathcal A^*$ obtains the secret keys of the
  patients (except the challenge patient), $\mathcal A^*$ cannot decrypt the challenge patient's ciphertext since the secret keys of different patients are irrelevant.
  
\subsection{Design Goal Analysis}
\label{SubSec:GoalAnal}

P-Med has achieved the design goals that are proposed in Section \ref{SubSec:DesignGoals}, which is analyzed below.

\emph{Medical Model and Data Confidentiality}. According to Section \ref{SubSec:SecPMed}, although an adversary $\mathcal A^*$ eavesdrops all communications, $\mathcal A^*$ still can not obtain the medical model and patient's medical data, which guarantees the confidentiality.

\emph{Treatment Recommendation Confidentiality}. According to Section \ref{SubSec:SecPMed}, although an attacker $\mathcal A^*$ compromises CP or CSP, $\mathcal A^*$ still can not get: the illness state match result in \texttt{SSM}/\texttt{TPW}, the state transitions process in \texttt{TPT}/\texttt{TPW}, the treatment procedure weights in \texttt{TPW}/\texttt{SMin}/\texttt{SMin$_n$}/\texttt{BPS-$k$} and the treatment recommendation result in \texttt{BPS-$k$}.

\emph{Soundness}. P-Med realizes completeness for finding out all the match treatment procedures since all the procedures are traversed in \texttt{TPT}, weighted in \texttt{TPW} and ranked in \texttt{BPS-$k$}. P-Med is correct to return the top-$k$ best procedures, which is guaranteed by the correctness of \texttt{BPS-$k$} and analyzed in supplemental material C-5.

\section{Secure Error-tolerant Gene Matching based on NFA}
\label{SubSec:PGene}

DFA and NFA can be leveraged as the modeling tool for many other applications: pattern matching \cite{Bremler13}, search on genome \cite{Troncoso07,Sasakawa14}, internet protocol parsing \cite{Graham14}, deep packet inspection \cite{Bando12,LiuC11}, 
regular expressions search \cite{YangY18TCC,Bando12}, etc. Here, we focus on \textbf{p}rivacy-preserving NFA based error-tolerant \textbf{gene} matching (P-Gene).

Human Desoxyribo-Nucleic Acid (DNA) is a double helix structured molecule to carry genetic instructions for growth and development. Since genes may mutate and cause disease, DNA diagnosis is an exciting new frontier to discover predisposition to various diseases for medical diagnosis. However, the disclosure of patient's DNA may pose significant risks to individual privacy, and incur genome-based discrimination. Due to business or safety concerns (such as bio-terrorism), the exact DNA pattern of pathogene should also be kept confidential for healthcare providers. Therefore, the privacy of both sides should be kept rigorously. In this section, we leverage Ukkonen NFA model to construct a privacy-preserving error-tolerant gene matching scheme for mutant gene detection or virus genome detection.

The Ukkonen NFA \cite{Ukkonen85,Troncoso07} is a type of NFA arranged in grid, where each row $i$ denotes the number of errors tolerated. Suppose an Ukkonen NFA contains $(\mu+1)\times(m+1)$ states, where $\mu$ is the error-tolerance degree and $m$ is the number of symbols contained in a pattern. We represent the state at row $i$ and column $j$ as $q_{i,j}$, where $0\leq i\leq\mu$ and  $0\leq j\leq m$. The state $q_{0,0}$ is the initial state, and the states in the rightmost column are accepting states, i.e., $\mathcal{F}=(q_{0,m},\cdots,q_{\mu,m})$. The arrows labeled as $\Sigma$ represent transitions induced by any symbol in $\Sigma$. The transitions in Ukkonen NFA are categorized into the following types.

\begin{itemize}
	\item Horizontal arrows from $q_{i,j}$ to $q_{i,j+1}$ denote matching a symbol in the pattern, which are called $h$-trans (horizontal transitions).
	
	\item Vertical arrows from $q_{i,j}$ to $q_{i+1,j}$ denote inserting a symbol in the pattern, which are called $v$-trans (vertical transitions).
	
	\item Solid diagonal arrows from $q_{i,j}$ to $q_{i+1,j+1}$ denote replacing a symbol in the pattern by an element from $\Sigma$, which are called $\Sigma$-$d$-trans ($\Sigma$ induced diagonal transitions).
	
	\item Dashed diagonal arrows from $q_{i,j}$ to $q_{i+1,j+1}$ denote deleting a symbol in the pattern, which are called $\varepsilon$-$d$-trans ($\varepsilon$ induced diagonal transitions).
	
	\item The $\Sigma$-$d$-trans and $\varepsilon$-$d$-trans are called $d$-trans (diagonal transitions), which can be induced by any symbol in $\Sigma$ or $\varepsilon$. The $d$-trans can always be triggered with or without symbol input, which is a unconditional transition.
\end{itemize}

\begin{figure}[htp]
	\begin{center}
		\includegraphics[width=3.5in]{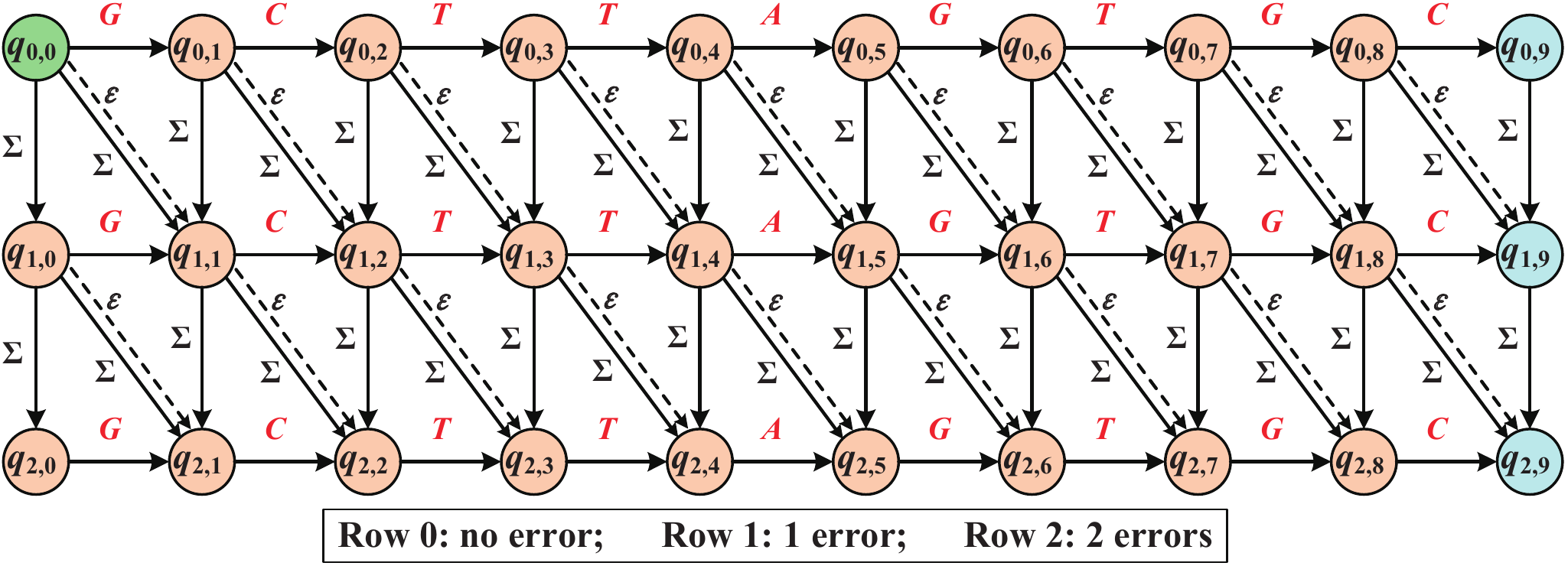}\\
		\caption{Ukkonen DFA of F8 Gene Sequence Fragment}
		\label{Fig:Ukkonen}
	\end{center}
\end{figure}

In humans, factor VIII (an essential blood-clotting protein) is encoded by the F8 gene, and defects in F8 gene result in hemophilia A (a recessive X-linked coagulation disorder). According to NCBI database\footnote{NCBI: National Center for Biotechnology Information of U.S.\\NCBI reference sequence of F8 gene: NM$\_$000132.3.\\https://www.ncbi.nlm.nih.gov/nuccore/NM$\_$000132.}, human F8 gene sequence contains
``GCT TAG TGC TGA GCA CAT CCA GTG GGT AAA GTT CCT TAA AAT GCT CTG CAA AGA AAT ...". 
We take the first 9 nucleobases ``GCT TAG TGC" in F8 gene as a DNA pattern to construct its error tolerant Ukkonen NFA model in Fig. \ref{Fig:Ukkonen}, where the symbol set $\Sigma=\{A,C,G,T\}$ and the error-tolerance degree $\mu=2$. 

Suppose the DNA pattern is $\Psi=\{\psi_1,\cdots,\psi_m\}$ and its error tolerant model is represented by Ukkonen NFA, and the searched symbol sequence is $\Phi=\{\phi_1,\cdots,\phi_n\}$. If $q_{i,m}$ is the accepted state in the matching algorithm, it indicates that the Hamming distance (edit distance) between $\Psi$ and $\Phi$ is $i$. To construct a privacy-preserving error-tolerant gene matching protocol (P-Gene) based on Ukkonen NFA, we define an encrypted transition matrix $\mathbb{E}$ to represent the state transition diagram. The element $E_{q_{i,j}\rightarrow q_{i',j'}}=[\![e_{q_{i,j}\rightarrow q_{i',j'}}]\!]_{pk_A}$ in $\mathbb{E}$ denotes the state transition between $q_{i,j}$ and $q_{i',j'}$, which is assigned different values in diverse situations (shown in the following equation).
$$e_{q_{i,j}\rightarrow q_{i',j'}}=\left\{
\begin{aligned}
&\psi_{j'},~if~(i'=i)\&(j'=j+1),\\
&0,~if~(i'=i+1)\&(j'=j~or~ j+1),\\
&1,~otherwise.
\end{aligned}
\right.$$
\begin{itemize}
	\item It equals $[\![\psi_i]\!]_{pk_A}$ for the $h$-trans ($q_{i,j}\rightarrow q_{i,j+1}$), which indicates the symbol $\psi_i$ induced the transition.
	\item It equals $[\![0]\!]_{pk_A}$ for the $v$-trans ($q_{i,j}\rightarrow q_{i+1,j}$) and the $d$-trans ($q_{i,j}\rightarrow q_{i+1,j+1}$).
	\item It equals $[\![1]\!]_{pk_A}$ for the other situations, which indicates no transition is possible.
\end{itemize}

To record the state transition induced by a sequence of symbols, we define an encrypted active state matrix $$\mathbb{S}=\{S_{i,j}=[\![s_{i,j}]\!]_{pk_\sigma}|0\leq i\leq\mu,0\leq j\leq m\}.$$ The element $S_{i,j}$ equals $[\![0]\!]_{pk_\sigma}$ indicating that the state $q_{i,j}$ is activated; otherwise, it equals $[\![1]\!]_{pk_\sigma}$. 

The secure error-resistant DNA match protocol is proposed in Algorithm \ref{Algo:DNAMatch}. The input includes encrypted DNA pattern $[\![\Psi]\!]_{pk_A}=([\![\psi_1]\!]_{pk_A},\cdots,[\![\psi_{m}]\!]_{pk_A})$,	
encrypted DNA sequence $[\![\Phi]\!]_{pk_B}=([\![\phi_1]\!]_{pk_B},\cdots,[\![\phi_{n}]\!]_{pk_B})$, encrypted transition matrix $\mathbb{E}$ (constructed from $[\![\Psi]\!]_{pk_A}$) and encrypted active state matrix $\mathbb{S}$, where the elements in  $\mathbb{S}$ are initialized to be $[\![0]\!]_{pk_\sigma}$ (for $S_{0,0}$) and $[\![1]\!]_{pk_\sigma}$ (otherwise). The protocol outputs $[\![\mathcal{F}_S]\!]_{pk_{\sigma}}=(S_{0,m},\cdots,S_{\mu,m})$, which are the encrypted activation states of $\mathcal{F}=(q_{0,m},\cdots,q_{\mu,m})$. If $s_{i,m}=0$ for some accept state $q_{i,m}\in\mathcal{F}$ ($0\leq i\leq\mu$), it indicates $[\![\Phi]\!]_{pk_B}$ is accepted by the model.

Each encrypted symbol $[\![\phi_k]\!]_{pk_B}$ ($1\leq k\leq n$) in DNA sequence $[\![\Phi]\!]_{pk_B}$ may induce the state transition and activation (in line 1) in the following ways.

\begin{itemize}
	\item For row $i=0$ and column $j=1\cdots m$, the state $q_{0,j}$ is activated (i.e., $s_{0,j}=0$) if $(s_{0,j-1}=0)\&(\phi_k=\psi_j)$, which is executed in line 2-4. 
	
	\item For row $i=1,\cdots,\mu$ and column $j=1\cdots m$, the state $q_{i,j}$ is activated (i.e., $s_{i,j}=0$) if $s_{i-1,j-1}=0$ or $s_{i-1,j}=0$ or $(s_{i,j-1}=0)\&(\phi_k=\psi_j)$, which is calculated in line 5-11.
\end{itemize}

\begin{algorithm}
	{\small{
			\KwIn{
				DNA pattern $[\![\Psi]\!]_{pk_A}=([\![\psi_1]\!]_{pk_A},\cdots,[\![\psi_{m}]\!]_{pk_A})$,	
				DNA sequence $[\![\Phi]\!]_{pk_B}=([\![\phi_1]\!]_{pk_B},\cdots,[\![\phi_{n}]\!]_{pk_B})$,
				transition matrix $\mathbb{E}$,
				active state matrix $\mathbb{S}$.}
			\KwOut{$[\![\mathcal{F}_S]\!]_{pk_{\sigma}}$.}

			\For { $k= 1$ to $n$}
			{
				\For { $j= 1$ to $m$}
				{	
					$B_0\leftarrow\texttt{SUT}(\phi_k,E_{q_{0,j-1}\rightarrow q_{0,j}})$\;
					$S_{0,j}\leftarrow\texttt{SAD}(S_{0,j-1},B_0)$\;
				}
				\For { $i= 1$ to $\mu$}
				{	
					\For { $j= 1$ to $m$}
					{
						$B_0\leftarrow\texttt{SUT}(\phi_k,E_{q_{i,j-1}\rightarrow q_{i,j}})$\;
						$B_1\leftarrow\texttt{SAD}(S_{i,j-1},B_0)$\;
						$B_2\leftarrow\texttt{SMD}(S_{i-1,j-1},S_{i-1,j})$\;
						$B_3\leftarrow\texttt{SMD}(S_{i,j-1},B_2)$\;
						$S_{i,j}\leftarrow\texttt{SMD}(B_1,B_3)$\;
					}					
				}
			}
			
			Set $[\![\mathcal{F}_S]\!]_{pk_{\sigma}}=(S_{0,m},\cdots,S_{\mu,m})$\;
			\textbf{Return} $[\![\mathcal{F}_S]\!]_{pk_{\sigma}}$.
			\caption{{\sc Privacy-preserving Error-resistant Gene Match Protocol (P-Gene)}}
			\label{Algo:DNAMatch}}}
\end{algorithm}

\begin{figure}[htp]
	\begin{center}
		\includegraphics[width=3.5in]{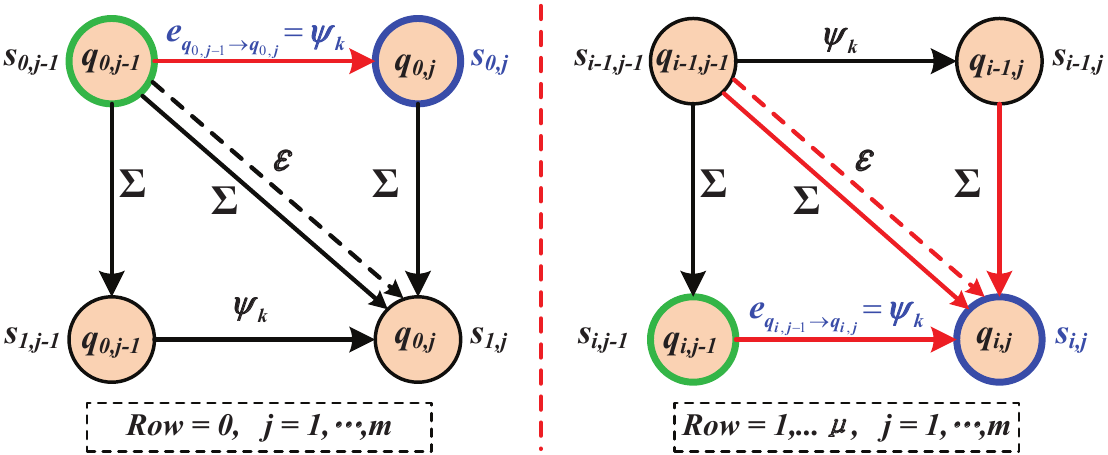}\\
		\caption{Illustration of Gene Match Protocol}
		\label{Fig:DNAProtocol}
	\end{center}
\end{figure}

Figure \ref{Fig:DNAProtocol} illustrates the state activation process for the two conditions. In P-Gene, the secure unequal test protocol $\texttt{SUT}([\![X]\!]_{pk_A},[\![Y]\!]_{pk_B})$ equals $[\![1]\!]_{pk_\sigma}$ for $X\neq Y$, and equals
$[\![0]\!]_{pk_\sigma}$ for $X=Y$.

\begin{figure}[htp]
	\begin{center}
		\includegraphics[width=3.5in]{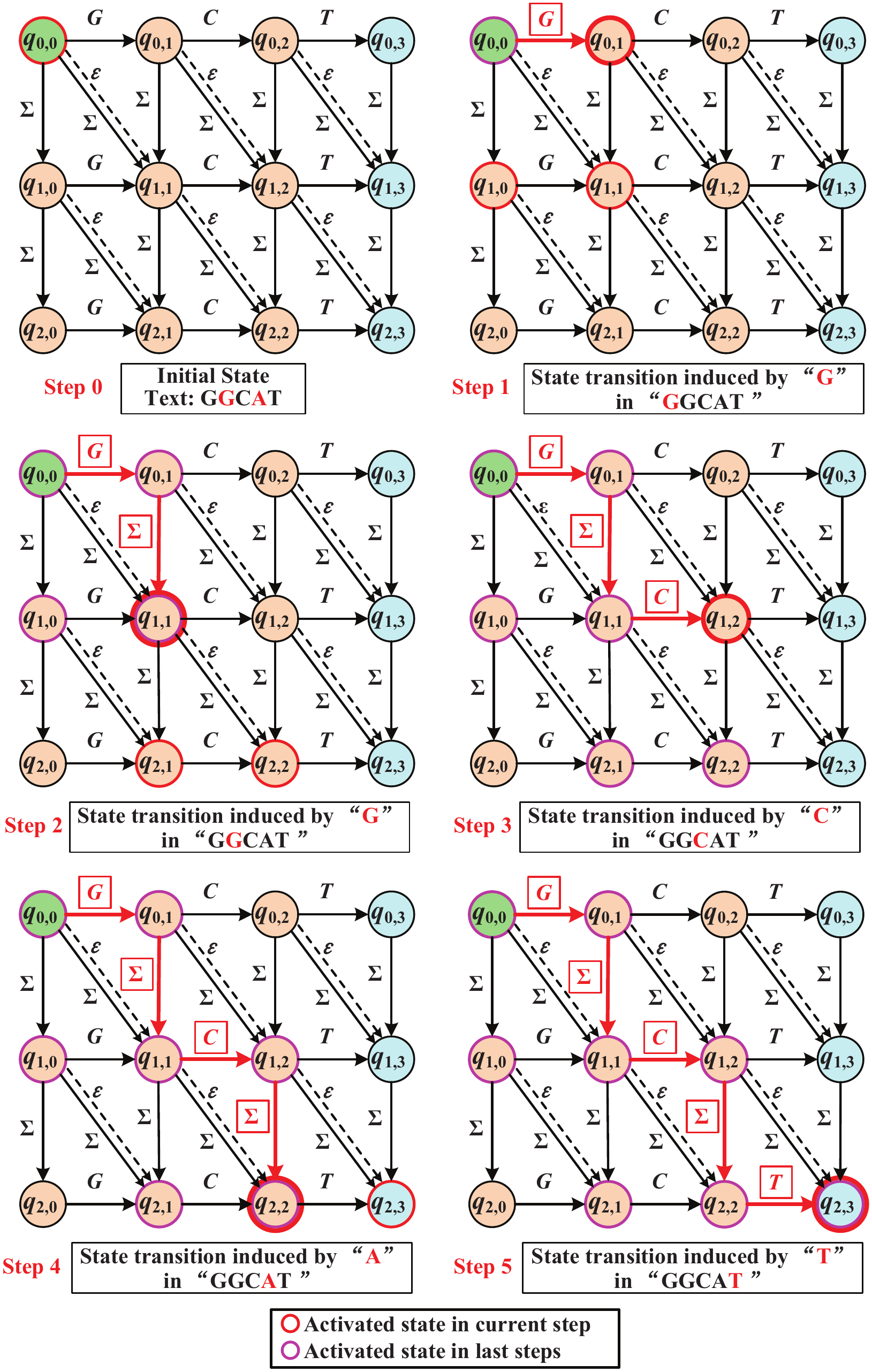}\\
		\caption{State Transition for Gene Matching}
		\label{Fig:UkkonenTrans}
	\end{center}
\end{figure}

In Fig. \ref{Fig:UkkonenTrans}, we take DNA pattern $\Psi=(G,C,T)$ to construct the Ukkonen DFA model with $\mu=2$. The DNA sequence for testing is $\Phi=(G,$\textcolor[rgb]{1,0,0}{\textit{G}}$,C,$\textcolor[rgb]{1,0,0}{\textit{A}}$,T)$, which has two errors compared with $\Psi$. The steps in Fig. \ref{Fig:UkkonenTrans} show the activation process of states induced by inputting the symbols $\phi_k$ ($1\leq k\leq n$) in $\Phi$. The red circles indicate the activated states in current step, and the purple circles indicate the activated states in the last steps. The red arrows represent the effective activated transitions induced by the sequence $\Phi=(G,$\textcolor[rgb]{1,0,0}{\textit{G}}$,C,$\textcolor[rgb]{1,0,0}{\textit{A}}$,T)$.

\section{Performance Analysis}
\label{Sec:Performance}
This subsection presents the experiment and theoretical analysis of the performance of P-Med. A comparison is made among P-Med and other related schemes. The P-Gene protocol is tested using real dataset in NCBI in Section \ref{Sec:Expriment}.

\subsection{Experiment Analysis}
\label{Sec:Expriment}
We evaluate the impact factors of the performance of P-Med on a PC with Intel(R) Core(TM) i9-7920X CPU@2.9GHz, 32 GB RAM,
and a custom simulator built in Java. To achieve 80-bit security level \cite{NIST}, it is necessary to choose the parameter $\mathcal{L}(N)=1024$.
In the programming, thread pool technology is utilized to optimize the
parallel operation.

\renewcommand\arraystretch{1.2}
\begin{table}[htbp]
\caption{Performance of \texttt{TPT}}
\label{Table:TPT-time} \centering
\newcommand{\tabincell}[2]{\begin{tabular}{@{}#1@{}}#2\end{tabular}}
\begin{tabular}{Ic|c|c|cI}
  \thickhline
  $\textbf{VertNum}$ & $\boldsymbol{MVisit}$ & $\boldsymbol{MState}$ &\textbf{Time (ms)} \\
  \thickhline						
  \multirow{3}{*}{\tabincell{c}{10}} &1 &  10  & 0.036 \\
  \cline{2-4}						
   &2 &  10  &  0.046  \\
  \cline{2-4}						
   &3 &  10  &  0.064   \\
  \hline
  \multirow{3}{*}{\tabincell{c}{20}} &1 &  20  & 0.132  \\
  \cline{2-4}						
   &2 &  20  &  0.823  \\
  \cline{2-4}						
   &3 &  20  &  2.543 \\
  \hline
  \multirow{3}{*}{\tabincell{c}{30}} &1 &  30  & 0.649 \\
  \cline{2-4}						
   &2 &  30  &  23.324 \\
  \cline{2-4}						
   &3 &  30  &  114.008 \\
  \hline
  \multirow{3}{*}{\tabincell{c}{40}} &1 &  40  & 7.328  \\
  \cline{2-4}						
   &2 &  40  &  4,857 \\
  \cline{2-4}						
   &3 &  40  &  33,844  \\
  \hline
  \multirow{3}{*}{\tabincell{c}{50}} &1 &  50  & 16.433 \\
  \cline{2-4}						
   &2 &  50  &  19,783 \\
  \cline{2-4}						
   &3 &  50  &  226,898 \\
  \thickhline
\end{tabular}
\end{table}

\renewcommand\arraystretch{1.2}
\begin{table}[htbp]
\caption{Performance of \texttt{SMin} ($\mathcal{L}(N)=1024$)}
\label{Table:SMin} \centering
\newcommand{\tabincell}[2]{\begin{tabular}{@{}#1@{}}#2\end{tabular}}
\begin{tabular}{Ic|c|cI}
  \thickhline
  $\boldsymbol{MState}$  & \textbf{Computation (s)} & \textbf{Communication (MB)} \\
  \thickhline						
  10 &  4.418	& 0.363 \\
  \hline
  20 &  8.866	& 0.721  \\
  \hline
  30 &  14.267	& 1.079  \\
  \hline
  40 &  19.276	& 1.437 \\
  \hline
  50 &  23.347	& 1.795 \\
  \thickhline
\end{tabular}
\end{table}

The performance of \texttt{TPT} is irrelevant with the bit length of the crypto parameter $N$,
since the encrypted illness states and treatment methods are simply deemed as vertexes and edges in a graphically represented medical model $[\![\mathbb{M}]\!]_{pk_A}$, respectively. Due to that \texttt{TPT} is independently executed by CP, there is no communication overhead in \texttt{TPT}.
There are three factors that impact the computation overhead of \texttt{TPT}, namely: 1) the complexity of the NFA-based medical model $[\![\mathbb{M}]\!]_{pk_A}$,
2) the maximum visit time $MVisit$, and 3) the maximum illness state number $MState$.

The complexity of the medical model can be measured by the parameters in the NFA-based graph of $[\![\mathbb{M}]\!]_{pk_A}$, i.e.,
the total number of the vertexes ($VertNum$), the number of accept states, indegree and outdegree of each vertex, the number of loops and self-loops.
Table \ref{Table:TPT-time} shows the performance of \texttt{TPT}, where the computation cost increases
with $VertNum$, $MVisit$, $MState$. In the test, we randomly generate 1000 medical model graphs for
each pair of parameters to get the average \texttt{TPT} execution time, where each generated graph
has one initial state, two accept states, two indegrees and two outdegrees for each vertex, two loops
and two self-loops. According to the experimental test, it consumes 226,898 ms ($\approx$ 3.782 minutes) to traverse
all the treatment procedures when the medical model contains 50 vertexes (i.e. illness states).
Since \texttt{TPT} is executed only once by CP and the result is stored in the cloud, the computation cost is appropriate for real application.

\begin{figure}[htbp]
\centering
    \subfigure[\scriptsize Computation Cost]
    {
        \includegraphics[width=0.22\textwidth]{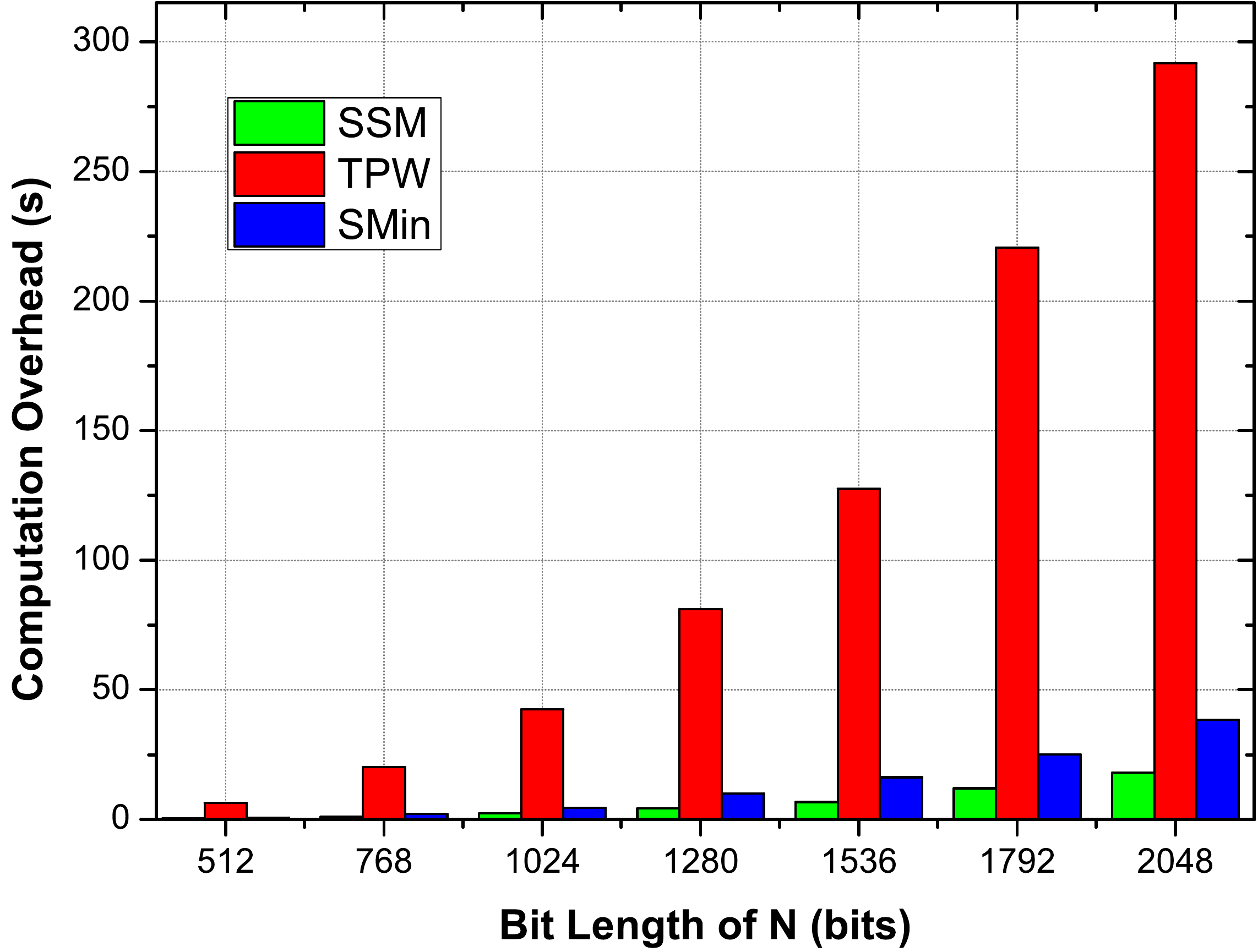}
        \label{SubFig:SSM-TPW-SMin-cal}
    }
    \subfigure[\scriptsize Communication Cost]
    {
        \includegraphics[width=0.22\textwidth]{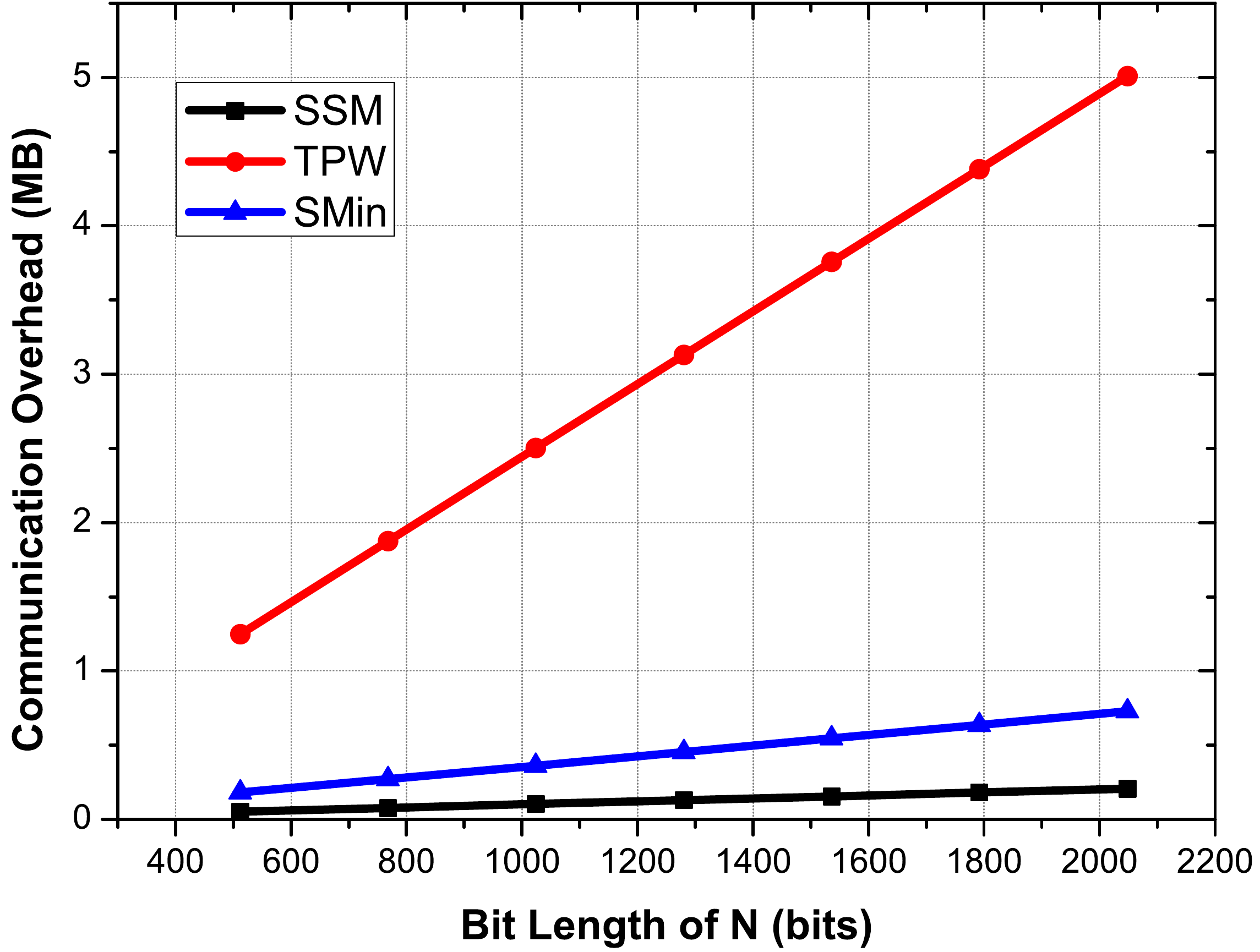}
        \label{SubFig:SSM-TPW-SMin-com}
    }\\
    \subfigure[\scriptsize Computation Cost (\texttt{TPW})]
    {
        \includegraphics[width=0.22\textwidth]{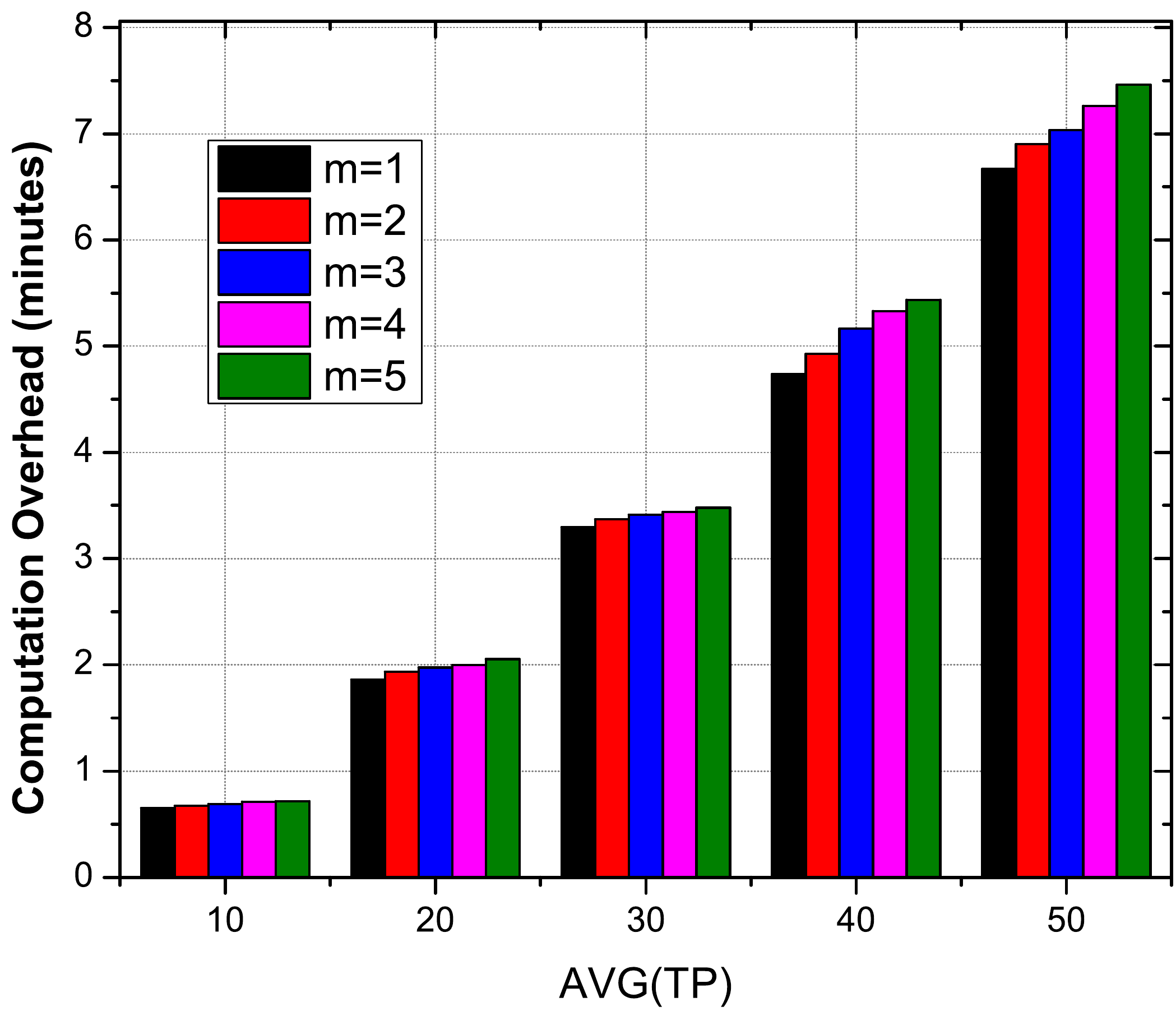}
        \label{SubFig:TPW-cal}
    }
    \subfigure[\scriptsize Communication Cost (\texttt{TPW})]
    {
        \includegraphics[width=0.22\textwidth]{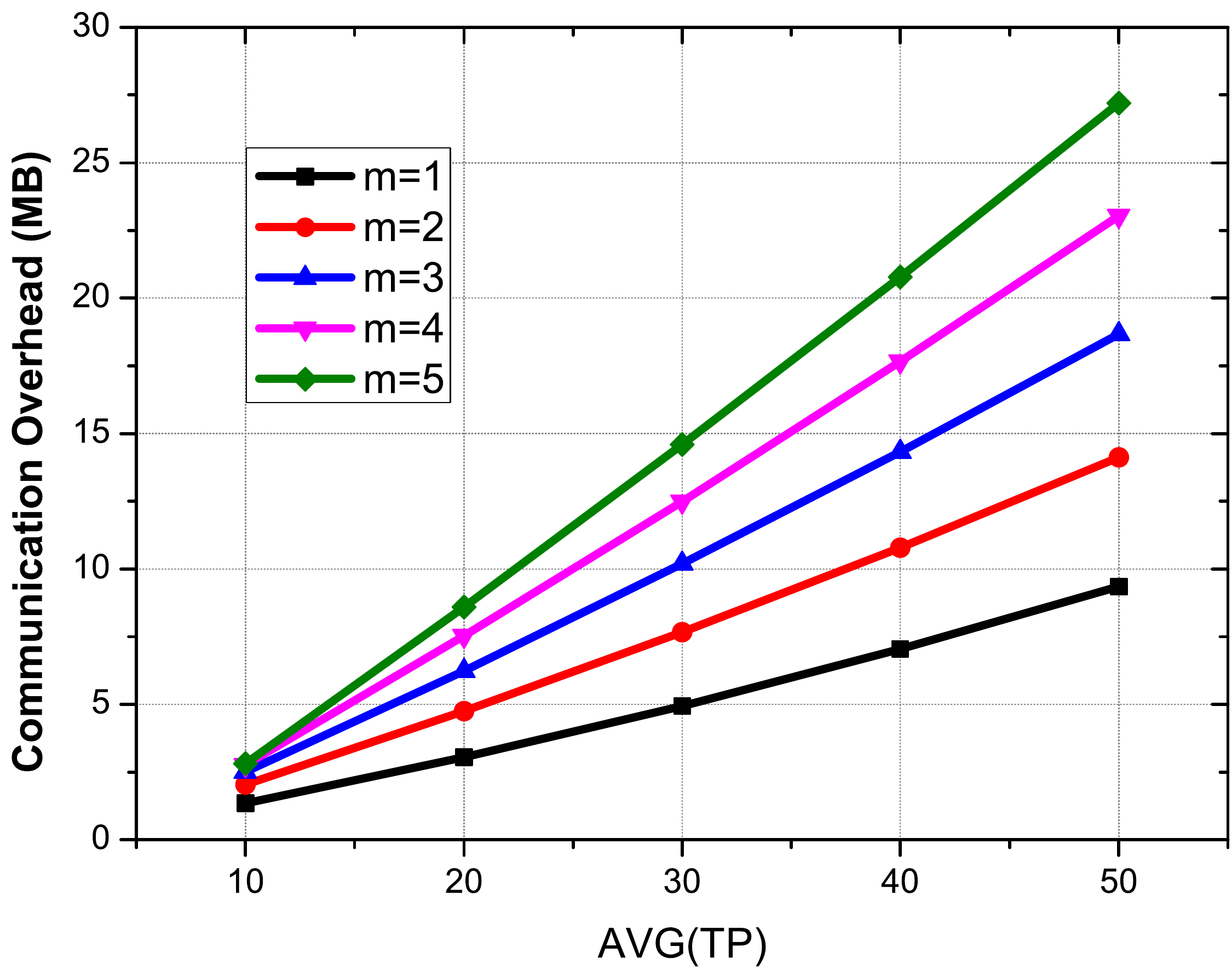}
        \label{SubFig:TPW-com}
    }\\
    \subfigure[\scriptsize Computation Cost (\texttt{SMin$_n$})]
    {
        \includegraphics[width=0.22\textwidth]{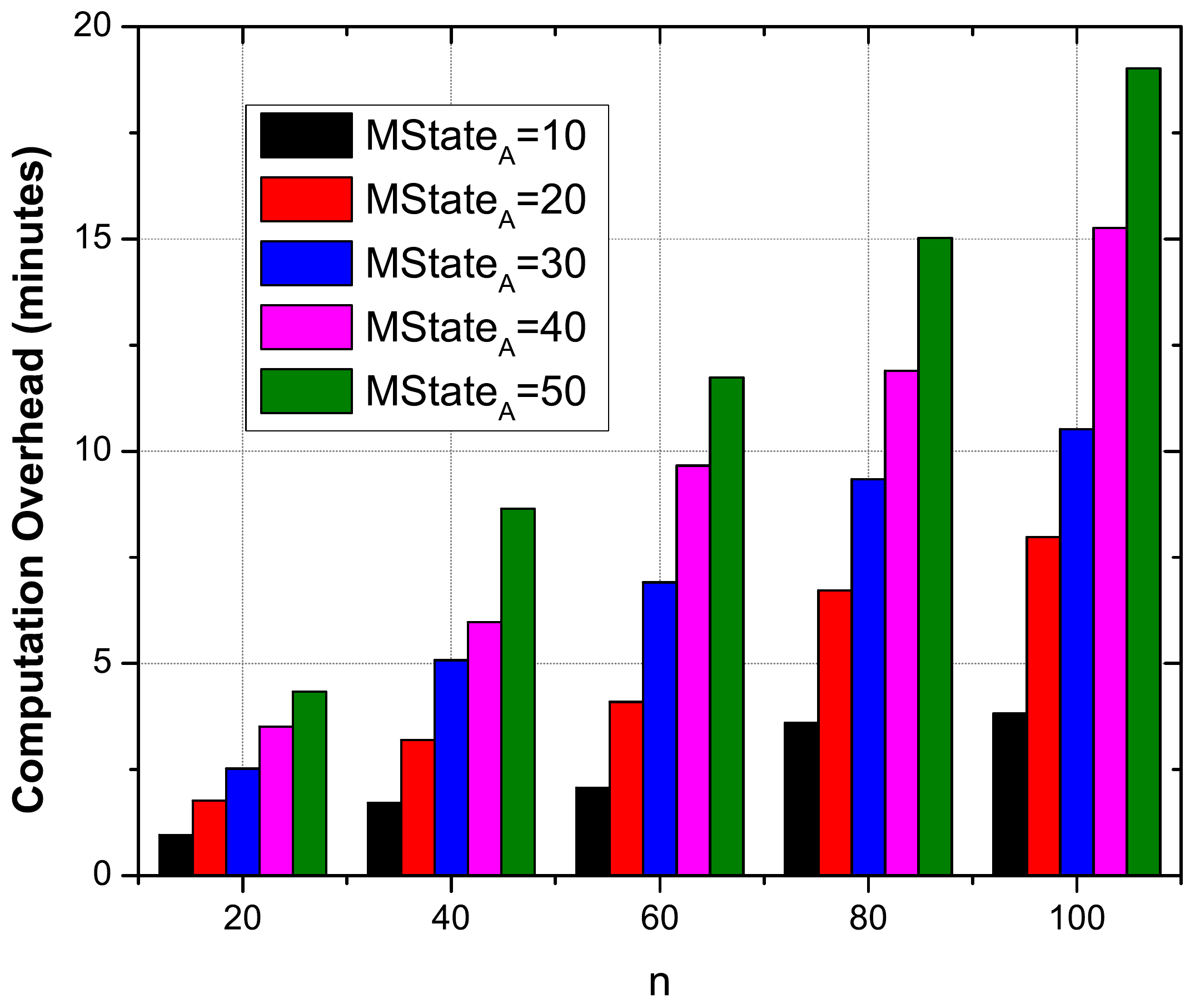}
        \label{SubFig:SMinN-cal}
    }
    \subfigure[\scriptsize Communication Cost (\texttt{SMin$_n$})]
    {
        \includegraphics[width=0.22\textwidth]{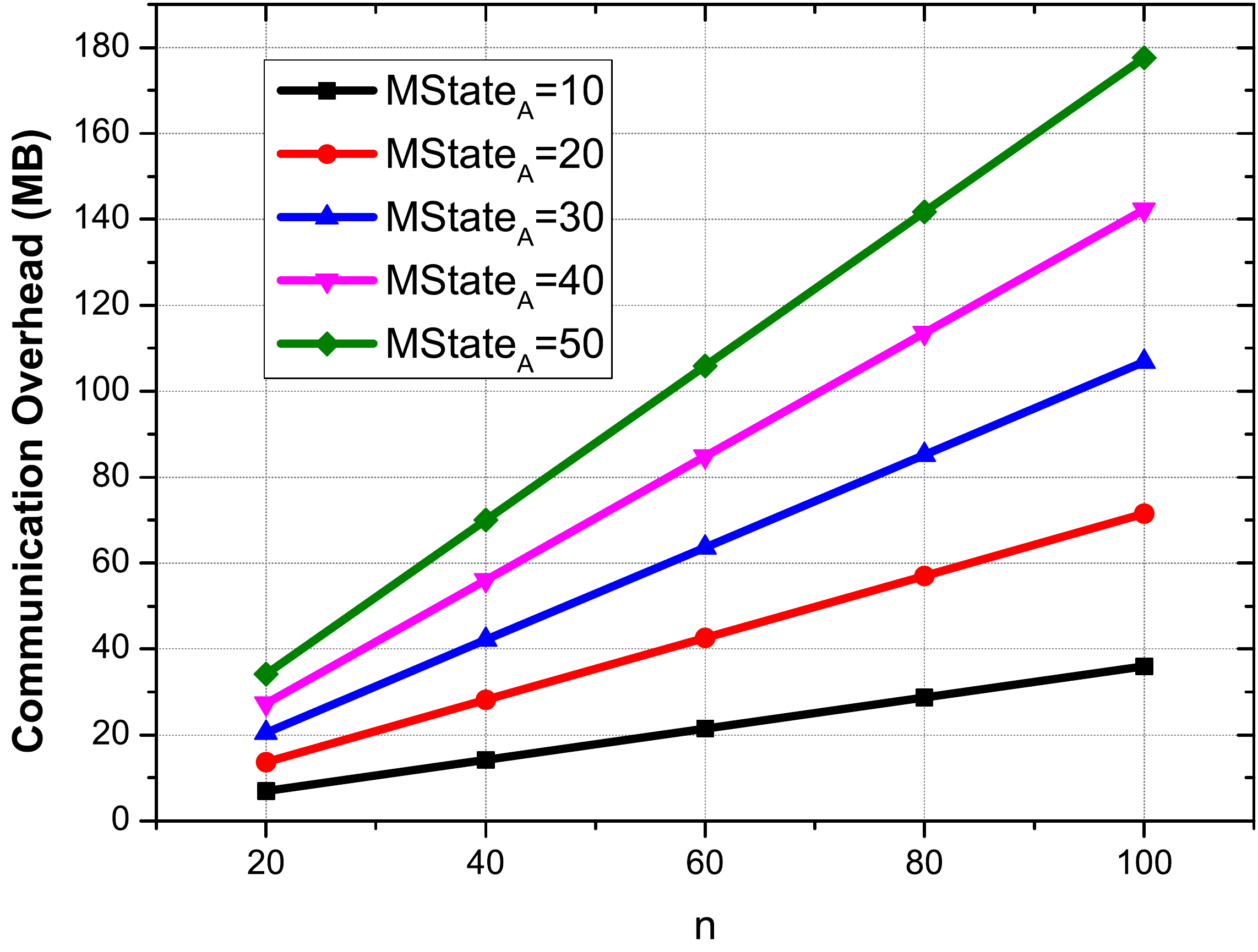}
        \label{SubFig:SMinN-com}
    }\\
    \subfigure[\scriptsize Computation Cost (\texttt{BPS$_k$})]
    {
        \includegraphics[width=0.22\textwidth]{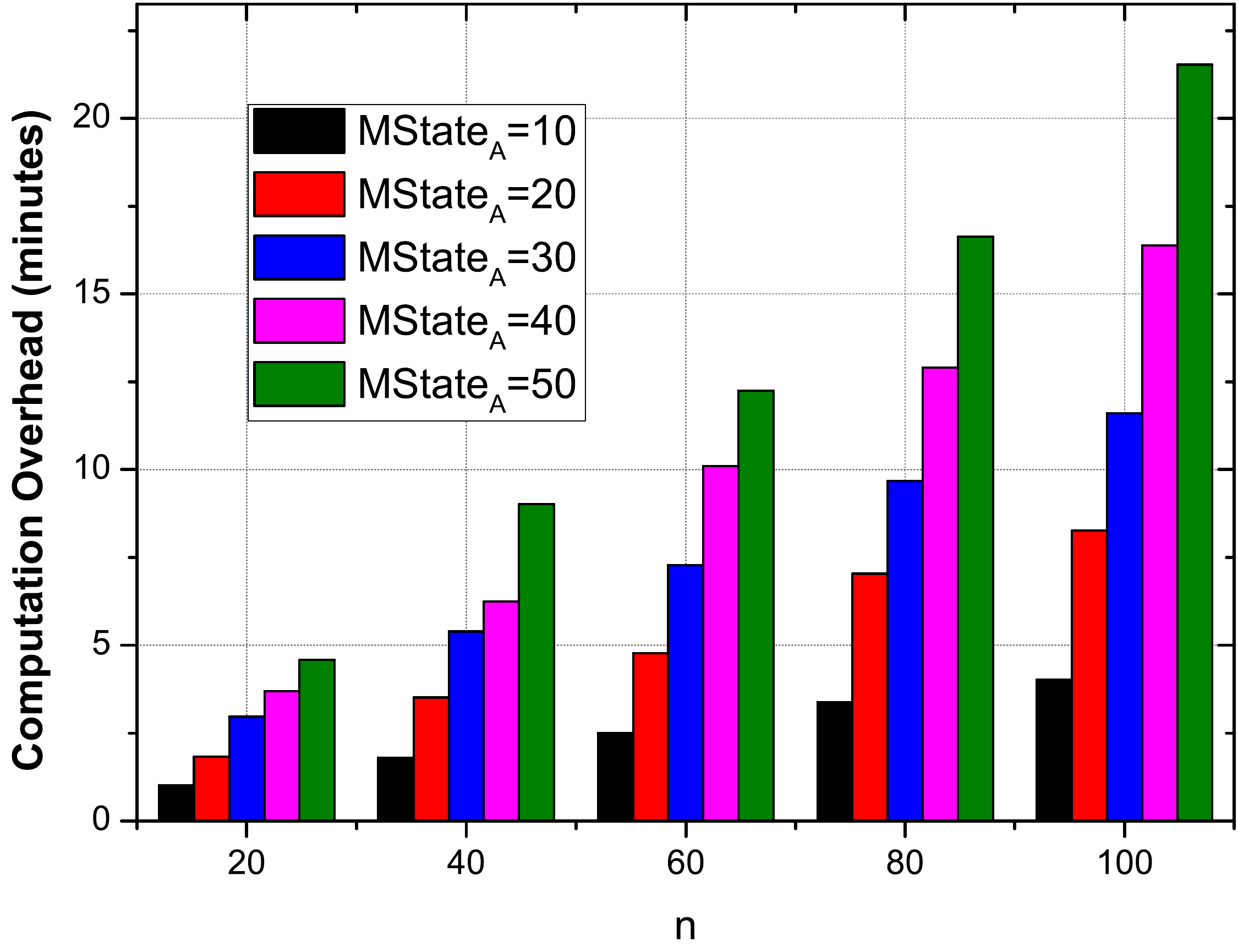}
        \label{SubFig:BPSK-cal}
    }
    \subfigure[\scriptsize Communication Cost (\texttt{BPS$_k$})]
    {
        \includegraphics[width=0.22\textwidth]{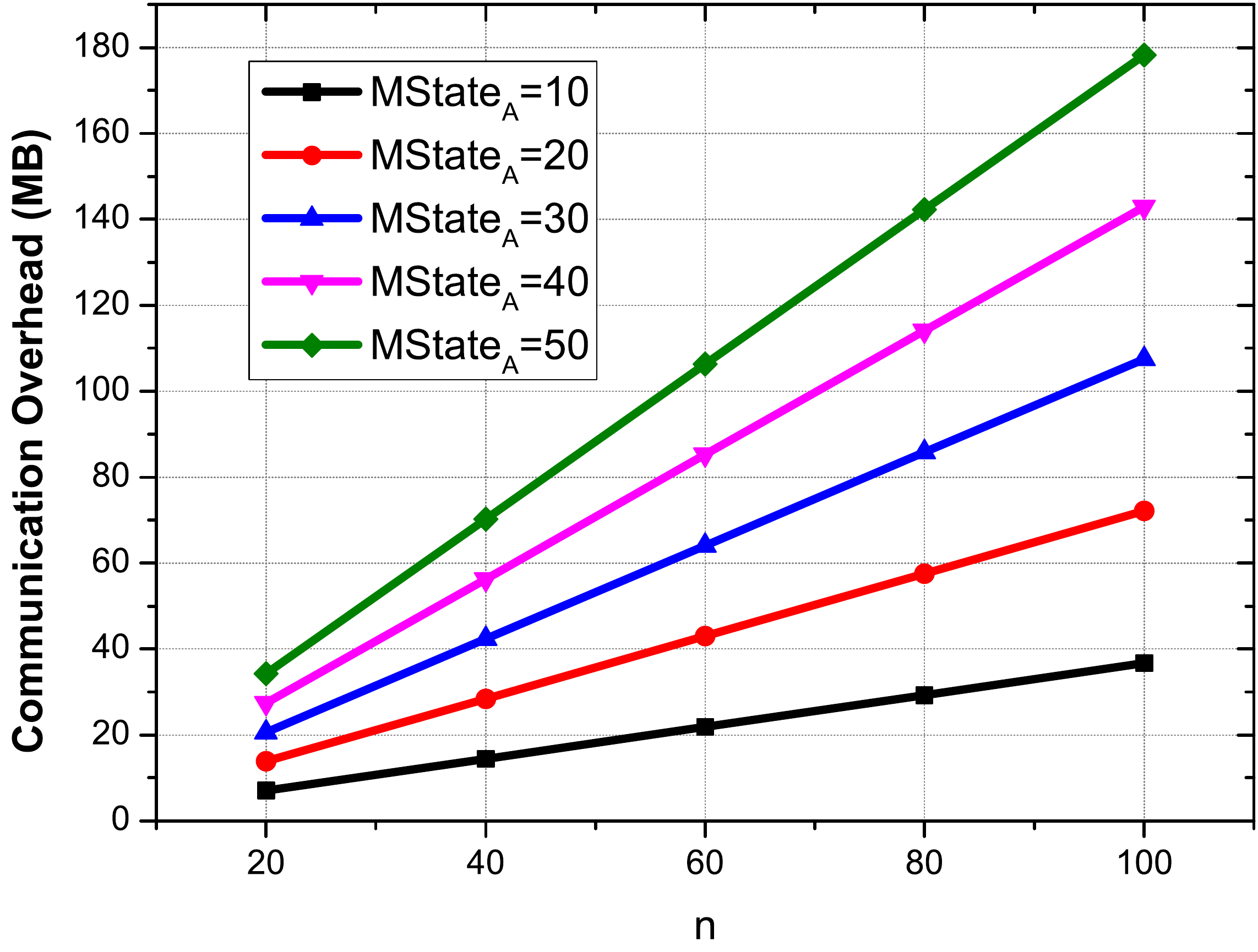}
        \label{SubFig:BPSK-com}
    }\\
    \subfigure[\scriptsize Computation Cost (P-Med)]
    {
        \includegraphics[width=0.22\textwidth]{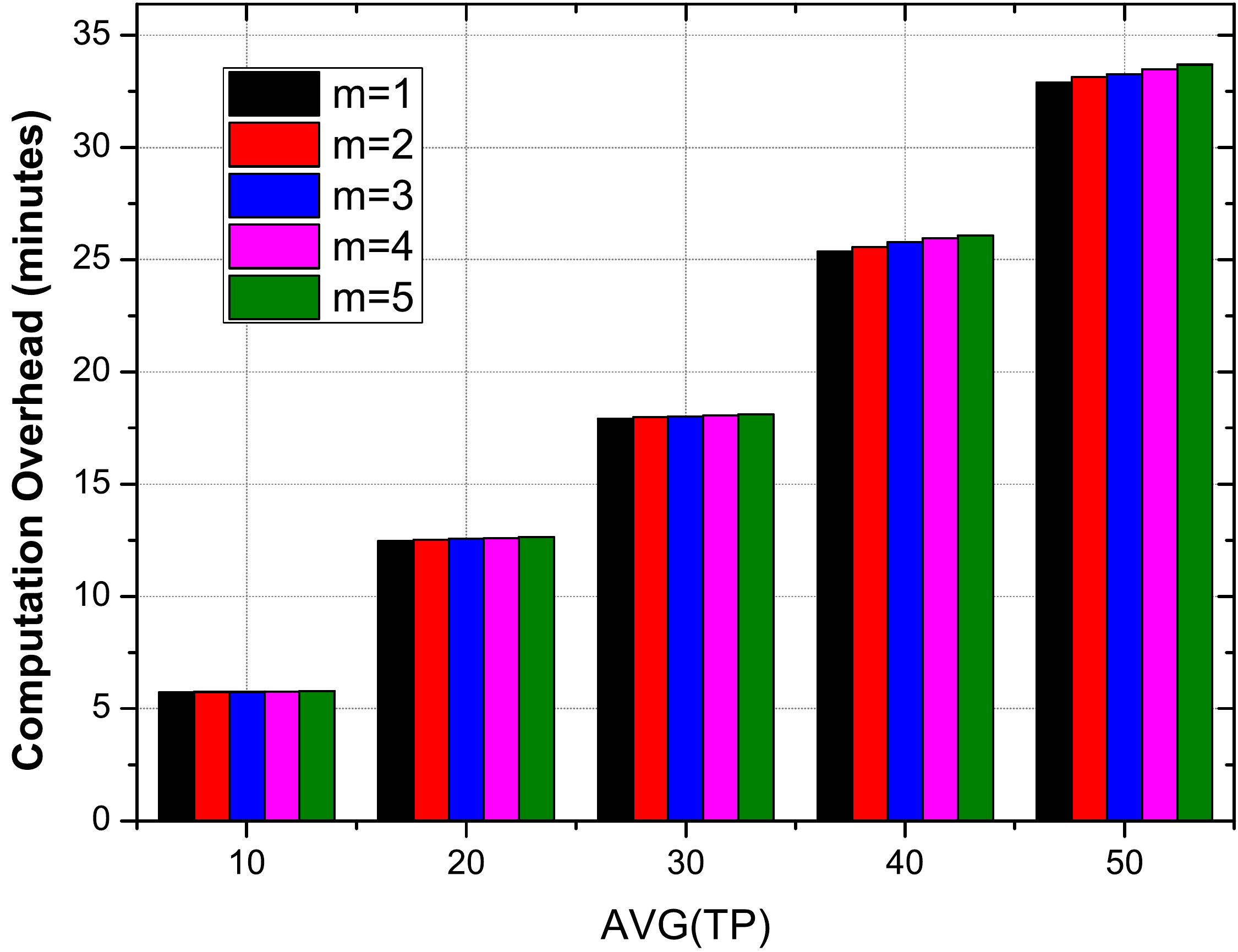}
        \label{SubFig:100-cal}
    }
    \subfigure[\scriptsize Communication Cost (P-Med)]
    {
        \includegraphics[width=0.22\textwidth]{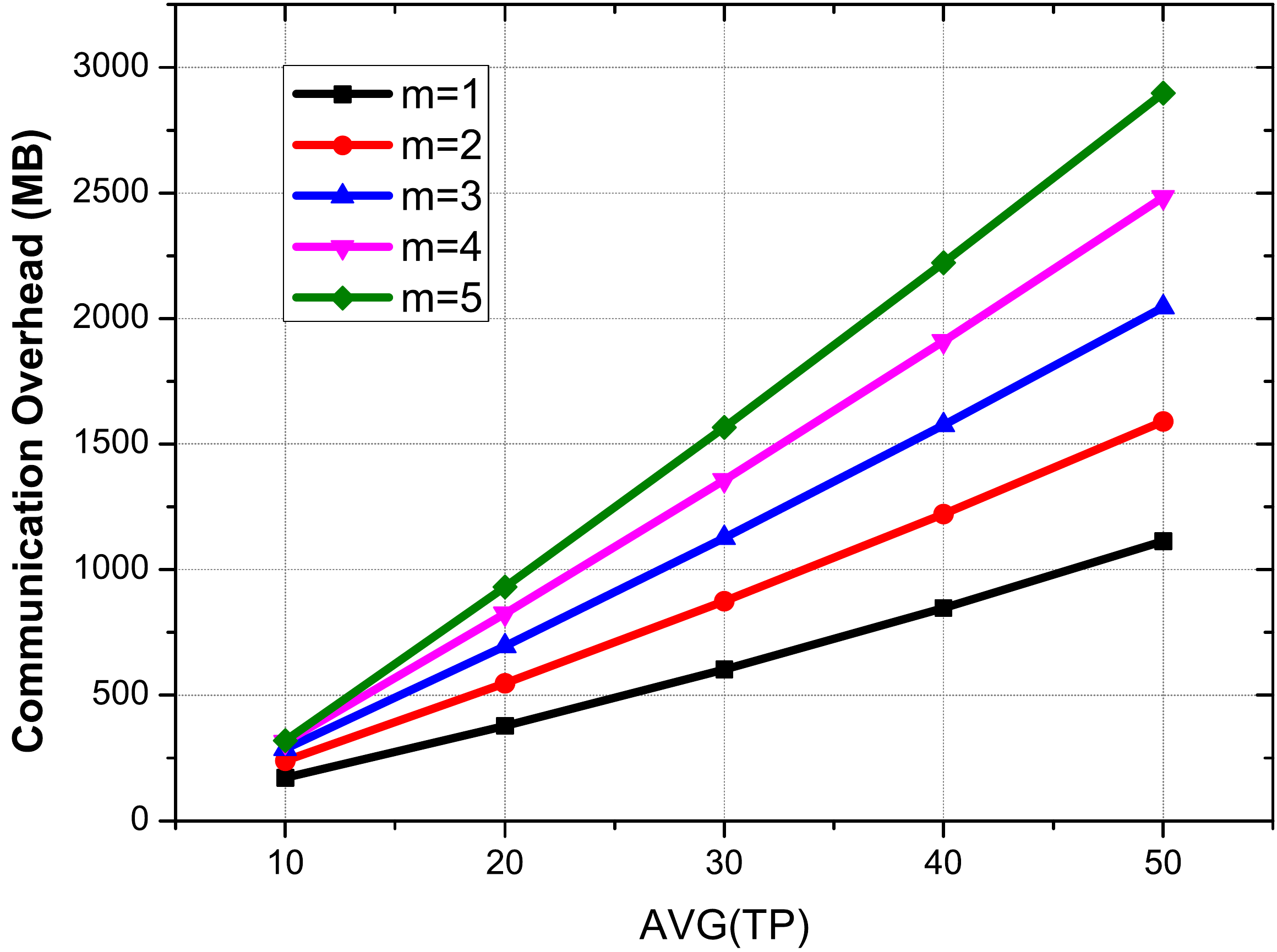}
        \label{SubFig:100-com}
    }\\
    \subfigure[\scriptsize Computation Cost (P-Gene)]
	{
		\includegraphics[width=0.22\textwidth]{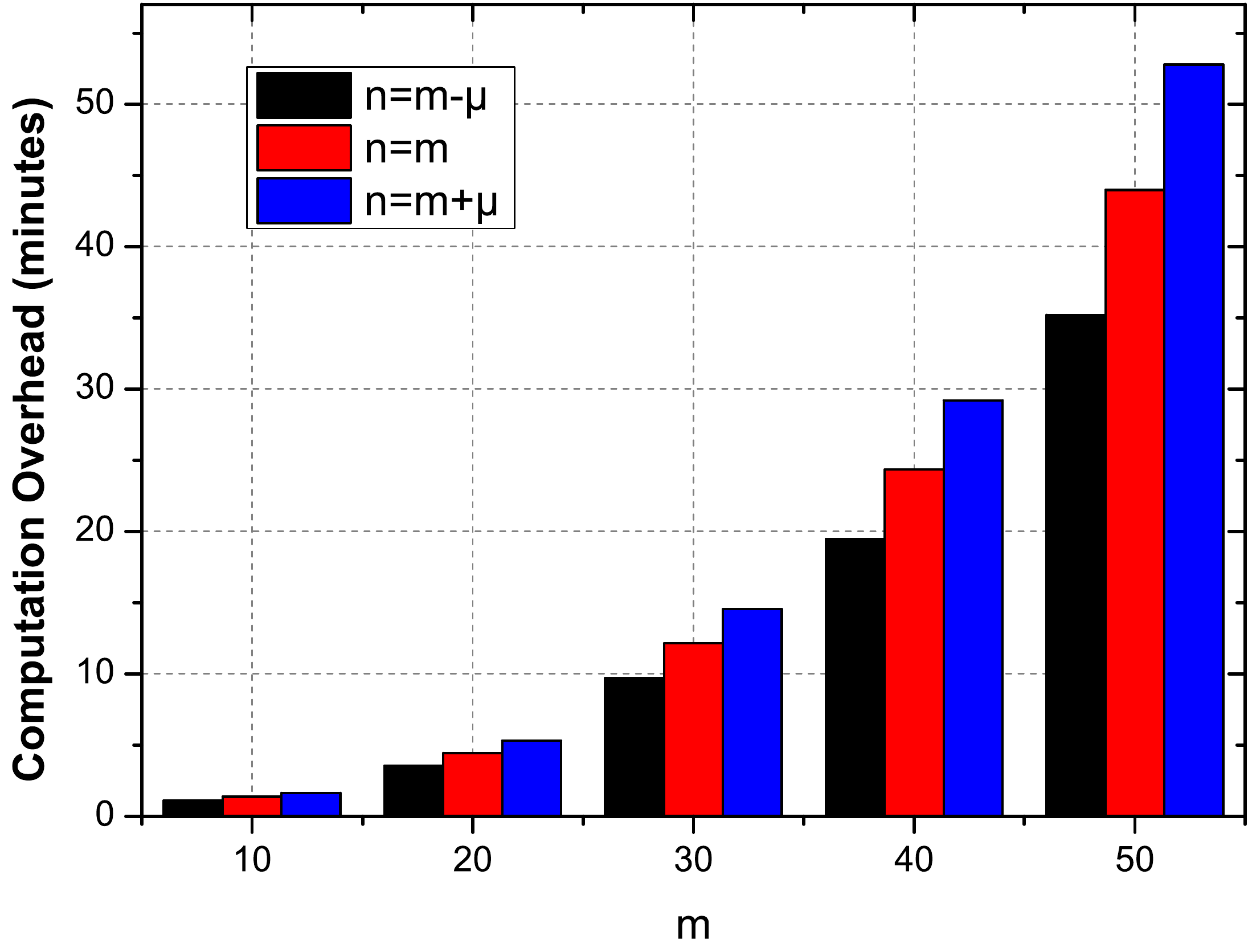}
		\label{SubFig:PGene-cal}
	}
	\subfigure[\scriptsize Communication Cost (P-Gene)]
	{
		\includegraphics[width=0.22\textwidth]{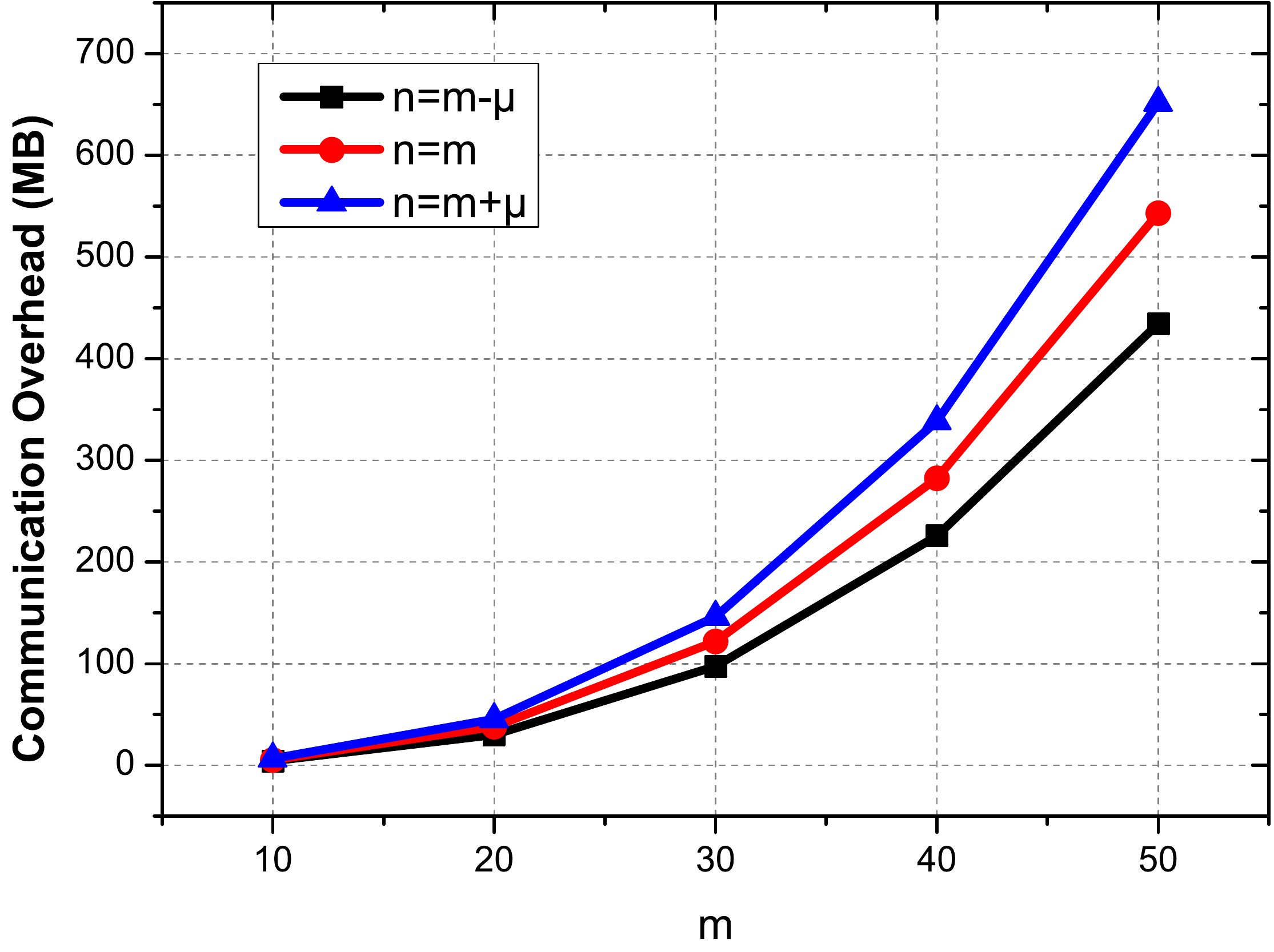}
		\label{SubFig:PGene-com}
	}
\caption{Evaluation of P-Med and P-Gene}
\label{Fig:Per}
\end{figure}

The computation and communication costs of \texttt{SSM}, \texttt{TPW}, \texttt{SMin} \texttt{SMin$_n$} and \texttt{BPS$_k$} increase with $\mathcal{L}(N)$,
due to that PCTD needs more time (and space) to calculate on (and store) the encrypted data.
Fig. \ref{SubFig:SSM-TPW-SMin-cal}-\ref{SubFig:SSM-TPW-SMin-com}
show the performance of \texttt{SSM}, \texttt{TPW}, \texttt{SMin} with
$MState=10$, $m=3$, $n=1$.
When $\mathcal{L}(N)=1024$, it takes 2.343/42.427/4.418 seconds (0.104/2.503/0.363 MB)
to run \texttt{SSM}, \texttt{TPW}, \texttt{SMin}.
Fig. \ref{SubFig:TPW-cal}-\ref{SubFig:TPW-com} show that the performance of \texttt{TPW} also increases with $m$ and the average illness state number $\mathcal{AVG}(\mathbb{TP})$ of $\mathbb{TP}$, where $\mathcal{AVG}(\mathbb{TP})=(\sum_{i=1}^{n}{\theta_{\tau_i}})/n$ and $\theta_{\tau_i}$
is the illness states number in $\mathcal{TP}_i$.
It costs 7.461 minutes (27.201 MB) for \texttt{TPW} to calculate a treatment procedure weight
when $\mathcal{AVG}(\mathbb{TP})=50$, $m=5$ and $\mathcal{L}(N)=1024$.
The performance of \texttt{SMin} (Table \ref{Table:SMin}) increases with $MState$, and it costs 23.347 seconds (1.795 MB) to run \texttt{SMin} when $MState=50$, $\mathcal{L}(N)=1024$.
From Fig. \ref{SubFig:SMinN-cal}-\ref{SubFig:BPSK-com}, we observe that the performance of
\texttt{SMin$_n$} and \texttt{BPS$_k$} increases with $n$
and $MState$. When $n=100$, $MState=50$, $\mathcal{L}(N)=1024$,
it takes 19.018/21.531 minutes (177.625/178.266 MB) to run \texttt{SMin$_n$}/\texttt{BPS$_k$}.

In the analysis, we test on diverse medical models with $n=40,60,80,100$. Since the performance of P-Med increases with $m$ and $\mathcal{AVG}(\mathbb{TP})$,
we select $\mathcal{AVG}(\mathbb{TP})=10,20,30,40,50$ and $m=1,2,3,4,5$ to evaluate  P-Med for each medical model, respectively.
Fig. \ref{SubFig:100-cal}-Fig. \ref{SubFig:100-com} show the computation/communication cost of P-Med ($n=100$),
and it takes 33.689 minutes (2898 MB) to get the encrypted best treatment procedure in P-Med when $\mathcal{AVG}(\mathbb{TP})=50, m=5$.

Then, we utilize the F8 gene sequence in NCBI database to test the performance of P-Gene protocol, and the computation/communication overheads are shown in Fig. \ref{SubFig:PGene-cal}-\ref{SubFig:PGene-com}.  Suppose a Ukkonen NFA model is constructed from a DNA pattern with length $m$ and error-tolerant degree $\mu$. 
The length $n$ of DNA sequence that can be accepted by this model should be in the range $m-\mu\leq n\leq m+\mu$. It is obvious that the cost of P-Gene is affected by $m$, $n$ and $\mu$. 
In the experiment, we set $(m,\mu)\in\{(10,2),(20,4),(30,6),(40,8),(50,10)\}$, and select $n\in\{m-\mu,m,m+\mu\}$ for diverse values of $(m,\mu)$. For $m=50$ and $\mu=10$, the experiment result shows that it takes 35.197/43.996/52.795 minutes (424.619/543.274/651.929 MB) to get the error-tolerant DNA matching result when $n=40,50,60$, respectively.  
The detailed experiment data of Fig. \ref{Fig:Per} is given in Table \ref{Table:SSM-TPW-SMin}-\ref{Table:P-Gene}
in supplemental materials E.

\subsection{Theoretical Analysis}
\emph{Computation Overhead}:
Assume that one exponentiation calculation with an exponent of length $\mathcal{L}(N)$
requires approximately 1.5$\mathcal{L}(N)$ multiplication calculations \cite{Knuth14} (e.g. the computation
of $g^a$ with $\mathcal{L}(a)=\mathcal{L}(N)$ requires 1.5$\mathcal{L}(N)$ multiplications, and denoted as 1.5$\mathcal{L}(N)$ muls).
Compared with exponentiation calculation, the computation overhead of addition and multiplication are negligible.
In PCTD, the operations \texttt{Enc}, \texttt{WDec}, \texttt{SDec}, \texttt{PD1}, \texttt{PD2} and \texttt{CR}
need 4.5$\mathcal{L}(N)$, 1.5$\mathcal{L}(N)$, 1.5$\mathcal{L}(N)$, 4.5$\mathcal{L}(N)$, 4.5$\mathcal{L}(N)$ and 3$\mathcal{L}(N)$ muls, respectively.
An exponentiation computation with a PCTD ciphertext as base number and an exponent of length $\mathcal{L}(N)$
is approximately 3$\mathcal{L}(N)$ muls. Basic protocols \texttt{SAD}, \texttt{SMD}, \texttt{SGE}, \texttt{SLE}, \texttt{SLT}, \texttt{SGT}, \texttt{SET} and \texttt{SRC} need 48$\mathcal{L}(N)$, 85.5$\mathcal{L}(N)$, 36$\mathcal{L}(N)$, 36$\mathcal{L}(N)$, 36$\mathcal{L}(N)$, 36$\mathcal{L}(N)$, 57$\mathcal{L}(N)$ and 157.5$\mathcal{L}(N)$ muls, respectively. In P-Med, the computation overhead of \texttt{TPT} is influenced by the complexity
of the medical model and the parameters $MVisit$, $MState$, which is irrelevant with the exponent operation and bit length of $N$.
\texttt{SSM} needs $\mathcal{O}(\mathcal{L}(N))$ muls and most of the computations can be concurrently calculated in practice.
It takes $\mathcal{O}(n\cdot \mathcal{AVG}(\mathbb{TP})\cdot\mathcal{L}(N))$ muls to run \texttt{TPW}, $\mathcal{O}(MState\cdot\mathcal{L}(N))$ muls to run \texttt{SMin},
$\mathcal{O}(MState\cdot\lceil log_2 n\rceil\cdot\mathcal{L}(N))$ muls to run \texttt{SMin$_n$},
and $\mathcal{O}(MState\cdot\lceil log_2 n\rceil\cdot k\cdot\mathcal{L}(N))$ muls to run \texttt{BPS-$k$}.
Finally, it takes $\mathcal{O}((n\cdot \mathcal{AVG}(\mathbb{TP})+MState\cdot\lceil log_2 n\rceil\cdot k)\cdot\mathcal{L}(N))$ muls
for P-Med to select the top-$k$ best treatment procedures.

\emph{Communication Overhead}:
In PCTD, the transmission overhead of a ciphertext $[\![m]\!]_{pk}$ and partial decrypted ciphertext $C_1^{(1)}$ are 4$\mathcal{L}(N)$
and 2$\mathcal{L}(N)$ bits, respectively. Basic protocols \texttt{SAD}, \texttt{SMD}, \texttt{SGE}, \texttt{SLE}, \texttt{SLT}, \texttt{SGT}, \texttt{SET} and \texttt{SRC} cost 16$\mathcal{L}(N)$, 36$\mathcal{L}(N)$, 10$\mathcal{L}(N)$, 10$\mathcal{L}(N)$, 10$\mathcal{L}(N)$, 10$\mathcal{L}(N)$, 20$\mathcal{L}(N)$ and 56$\mathcal{L}(N)$ bits, respectively. In P-Med, it costs $\mathcal{O}(n\cdot \mathcal{AVG}(\mathbb{TP})\cdot\mathcal{L}(N))$ bits to run \texttt{TPW}, $\mathcal{O}(MState\cdot\mathcal{L}(N))$ bits to run \texttt{SMin}, $\mathcal{O}(MState\cdot\lceil log_2 n\rceil\cdot\mathcal{L}(N))$ bits to run \texttt{SMin$_n$},
and $\mathcal{O}(MState\cdot\lceil log_2 n\rceil\cdot k\cdot\mathcal{L}(N))$ bits to run \texttt{BPS-$k$}.
Finally, it takes $\mathcal{O}((n\cdot \mathcal{AVG}(\mathbb{TP})+MState\cdot\lceil log_2 n\rceil\cdot k)\cdot\mathcal{L}(N))$ bits
for P-Med to select the top-$k$ best treatment procedures.

\subsection{Comparative Analysis}

A comparison (Table \ref{Tab:Comparison}) is made among P-Med and finite automata based diagnosis schemes \cite{Sasakawa14,Lewis10},
IoT based diagnosis schemes \cite{Verma17,Kumar18} and machine learning (ML) based diagnosis schemes \cite{GaussKer14,SVM17,PPDP18}.
DFA is used in scheme \cite{Lewis10} to make medical diagnosis without privacy protection,
while the scheme in \cite{Sasakawa14} and P-Med are constructed based on NFA using homomorphic encryption.
In \cite{Sasakawa14}, NFA describes the linear structured DNA string and leaks the input symbol type in the test algorithm;
in P-Med, NFA describes a complex nonlinear cyclic/acyclic medical model and information leakage is avoided.
The IoT-based disease diagnosis systems in \cite{Verma17,Kumar18} do not have any security protection.

\renewcommand\arraystretch{1.2}
\begin{table}[htbp]
\caption{Comparison}
\label{Tab:Comparison}
\centering
\begin{tabular}{cccccccccc}
 \hline
 & P-Med &\cite{Lewis10} &  \cite{Sasakawa14} &  \cite{Verma17} &  \cite{Kumar18} & \cite{GaussKer14} & \cite{SVM17}  & \cite{PPDP18} \\
\hline
F1 & \checkmark & \checkmark & \checkmark & $\times$   & $\times$   & $\times$   & $\times$   & $\times$   \\
F2 & \checkmark & $\times$   & $\times$   & \checkmark & \checkmark & $\times$   & $\times$   & $\times$   \\
F3 &\checkmark  & N.A.       & N.A.       & \checkmark & \checkmark & $\times$   & $\times$   & \checkmark \\
F4 & \checkmark & N.A.       & N.A.       & \checkmark & \checkmark & $\times$   & \checkmark & $\times$   \\
F5 & \checkmark & $\times$   & \checkmark & $\times$   & $\times$   & \checkmark & \checkmark & \checkmark \\
F6 &\checkmark  & $\times$   & $\times$   & $\times$   & $\times$   & $\times$   & $\times$   & \checkmark \\
F7 & \checkmark & $\times$   & $\times$   & $\times$   & $\times$   & $\times$   & $\times$   & $\times$   \\
F8 & \checkmark & $\times$   & $\times$   & $\times$   & $\times$   & $\times$   & $\times$   & $\times$   \\
F9 & \checkmark & $\times$   & $\times$   & $\times$   & $\times$   & $\times$   & $\times$   & $\times$   \\
F10&\checkmark  & $\times$   & $\times$   & $\times$   & $\times$   & $\times$   & $\times$   & $\times$   \\
\hline
\end{tabular}
\begin{flushleft}

~~~F1: finite automata based model~~F2: IoT-based diagnosis\\
~~~F3: single communication round (user \& server)\\
~~~F4: patient independent query~~~~~F5: privacy preserving\\
~~~F6: strong security level~~~~~~~~~~~~F7: avoid information leakage\\
~~~F8: treatment recommendation~~~~F9: top-$k$ best treatment ranking\\
~~~F10: authorization mechanism\\
\end{flushleft}
\end{table}

The schemes in \cite{GaussKer14,SVM17,PPDP18}
requires a large medical dataset to train the classifier using machine learning algorithm,
and the trained classifier makes decisions using disease prediction algorithm. 
The design conception of P-Med is quite different from the schemes in \cite{GaussKer14,SVM17,PPDP18}: P-Med makes use of medical model (designed by the hospital) and patient's successive illness states to realize diagnosis and treatment.
The schemes in \cite{GaussKer14} and \cite{SVM17} does not realize strong security level since they leak the parameters of the classifier to the server to enable online diagnosis.
In addition, \cite{GaussKer14} and \cite{SVM17} require multiple rounds of communications between patient and cloud in the disease prediction algorithm.
The patients in \cite{GaussKer14,PPDP18} cannot independently query the diagnosis service,
and the hospital is a necessary bridge. In P-Med, the patient could independently issue a medical query and the interaction between patient and CP is a single round.

Furthermore, all the compared schemes \cite{Sasakawa14,Lewis10,Verma17,Kumar18,GaussKer14,SVM17,PPDP18} leak the diagnosis result or intermediate result to the cloud server. On the contrary, P-Med keeps the intermediate calculation result and treatment recommendation secret, and the servers can not distinguish the result comes from which treatment procedure in the selection algorithm.

Although machine learning-based models have great potential for improving healthcare quality, machine learning algorithms usually operate as black boxes, and it is not clear how do they make predictions. The patients and doctors are reluctant to trust a system as a black box. 
On the contrary, DFA/NFA based treatment models are derived from clinical practice guidelines, which is an explainable method to convince professional medical workers. 

On the other hand, machine learning-based privacy-preserving systems train the parameters of the model using forward/backward propagation algorithms. The parameters are repeatedly calculated in the training process, which results in the accuracy decrease. For instance, the scheme in \cite{ZhengY19} designed a privacy-preserving deep neural network for medical image denoising. The peak signal to noise ratio (PSNR) of the security scheme is 0.25-0.27 dB lower than the plaintext method. A privacy-preserving single/multiple layers neural network system was constructed in \cite{LiuX18FGCS} for clinical decision. Since \cite{LiuX18FGCS} utilizes a piecewise polynomial function to fit the nonlinear activation function for neural networks, the performance analysis in \cite{LiuX18FGCS} shows that the error rate increases 0.2\% compared with the plaintext algorithm. Different from these ML-based schemes, the DFA/NFA based treatment models do not involve parameter training process nor nonlinear activation functions. Therefore, the accuracy of the encrypted-domain is the same as for the plaintext-domain in P-Med.
It is a promising research topic to integrate the advantages of DFA/NFA based medical model and machine learning to make the clinical decision process interpretable and more accurate.

\section{Related Work}

A cloud and IoT based disease diagnosis framework is proposed in \cite{Verma17} to analyze the data generated by medical IoT devices and predict the potential disease with its level of severity.
Fuzzy rule based neural classifier is utilized in \cite{Kumar18} to construct a cloud and IoT based mobile healthcare application for monitoring and diagnosing the diseases.
A real-time patient-centric application is constructed in \cite{Alkhaldi18} to assist the treatment of post-discharge patient by a discrete-event dynamic system.
A clinical decision support system is designed in \cite{Ruiz17} to manage the treatment of patients with gestational diabetes, which makes use of finite automata to determine the patient's metabolic condition and generate therapy adjustment recommendations. These schemes realize online diagnosis and treatment based on plaintext medical data, where privacy preserving mechanism is not provided.  

The privacy concerns should essentially be considered to prevent the probable disclosure of the sensitive medical data and diagnosis/treatment result. Yang et al. \cite{YangY_TDSC17} put forth a lightweight traceable scheme for securely sharing electronic health records, which protects the privacy of medical data.
The machine learning methods are introduced in secure medical analytic and diagnosis. 
A support vector machine and Paillier homomorphic  encryption based clinical decision support system was designed in \cite{GaussKer14},
which requires multiple rounds of interaction between the server and clinician in the diagnosis.
A privacy-preserving online medical pre-diagnosis framework was suggested in \cite{SVM17} based on
nonlinear kernel support vector machine, which utilizes multi-party random masking and polynomial aggregation techniques.
Lin et al. \cite{LinJ17} utilized historical medical data of patients to train recurrent neural networks (RNN), and the trained RNN model made predictive diagnosis decisions. The scheme proposed in \cite{LinJ17} leverages Paillier homomorphic encryption to train the healthcare model, and bilinear pairing techniques to authenticate message.
Zhang et al. \cite{PPDP18} presented a privacy-preserving disease prediction system based on single-layer perceptron learning and random matrices algorithm,
which includes disease learning phase and prediction phase. 
A privacy-preserving multiple-layer neural network was designed in \cite{LiuX18FGCS} to support clinical decision, and a secure piecewise polynomial calculation protocol was proposed to fit the non-linear activation function. The deep neural network (DNN) model was introduced to healthcare to construct a secure image denoising system \cite{ZhengY19}, which bridges lightweight additive secret sharing and garbled circuits to execute the multi-party computation. 
Liang et al. \cite{LiangJ19} was suggested a privacy-preserving decision tree classification scheme to provide online diagnosis service. It transforms the outsourced decision tree classification issue to encrypted data retrieval problem, such that searchable encryption can be utilized to search on a set of decision paths.
A secure reinforcement learning system was proposed in \cite{LiuX19TETC} to enable privacy-preserving dynamic treatment decision making, which was constructed based on additive homomorphic encryption primitive.

Personalized medicine \cite{YangY19INS} may analyze the DNA information of the patient to make diagnosis and treatment decisions.
Blanton et al. \cite{Blanton10} constructed a privacy-preserving outsourced error-resilient DNA search scheme via oblivious evaluation of finite automata, where the genetic test pattern is represented as a finite automata and the DNA sequence is deemed as the input. During the test process, both the pattern and DNA sequence are kept secret.
Keshri et al. \cite{Keshri09} presented an automated method of Epileptic Spike detection in Electroencephalogram (EEG), and the system functionality was modeled with DFA.
Lewis et al. \cite{Lewis10} combined DFA and knowledge discovery technology in data mining TV-tree to construct a platform to discover epileptiform activity from Electroencephalograms (EEG), which could predict the interictal spikes within noise to be the predictors of the clinical onset of a seizure.
Mohassel et al. \cite{Mohassel12} designed an oblivious DFA evaluation scheme
with application to secure DNA pattern matching.
Selvakumar et al. \cite{Selvakumar14} utilized DFA to observe the cholesterol metabolism with the accept and reject states and proposed a monitoring procedure based on DFA, which is used to enhance the diagnostic procedures and conventional treatment in cholesterol metabolic disorders. Sasakawa et al. \cite{Sasakawa14} suggested an oblivious evaluation method of NFA based on homomorphic encryption with secure circuit evaluation method, which is applicable to privacy-preserving virus genome detection. However, the solution requires multiple communication rounds between NFA holder and genome data holder.

\section{Conclusion}
In this paper, we proposed a secure medical diagnosis and treatment framework named as P-Med that can be used to recommend therapy methods to the patients according to their illness states. The medical model in P-Med is constructed based on NFA, encrypted and outsourced to cloud. The patient submits successive several days of encrypted mIoT data to issue a query and get the top-$k$ best treatment recommendations using secure selection algorithm.
A secure illness state match protocol is also designed in P-Med to achieve quantitative secure comparison between the state in medical model and patient's illness state that are monitored by mIoT. Moreover, secure NFA evaluation method in P-Med reduces the interaction between cloud and patient to a single round.
Finally, we evaluate the security and performance of P-Med. 

\section*{Acknowledgement}
This work is supported by National Natural Science Foundation of China (61872091, 61932011); Singapore National Research Foundation under the National Satellite of Excellence in Mobile Systems Security and Cloud Security (NRF2018NCR-NSOE004-0001), AXA Research Fund; State Key Laboratory of Integrated Services Networks (Xidian University) (ISN20-17); Guangdong Provincial Key Laboratory of
Data Security and Privacy Protection (2017B030301004-13);
Fujian Provincial Key
Laboratory of Information Processing and Intelligent Control (Minjiang University) (MJUKF-IPIC201908).

%

\begin{IEEEbiography}[{\includegraphics[width=1in,height=1.25in,clip,keepaspectratio]{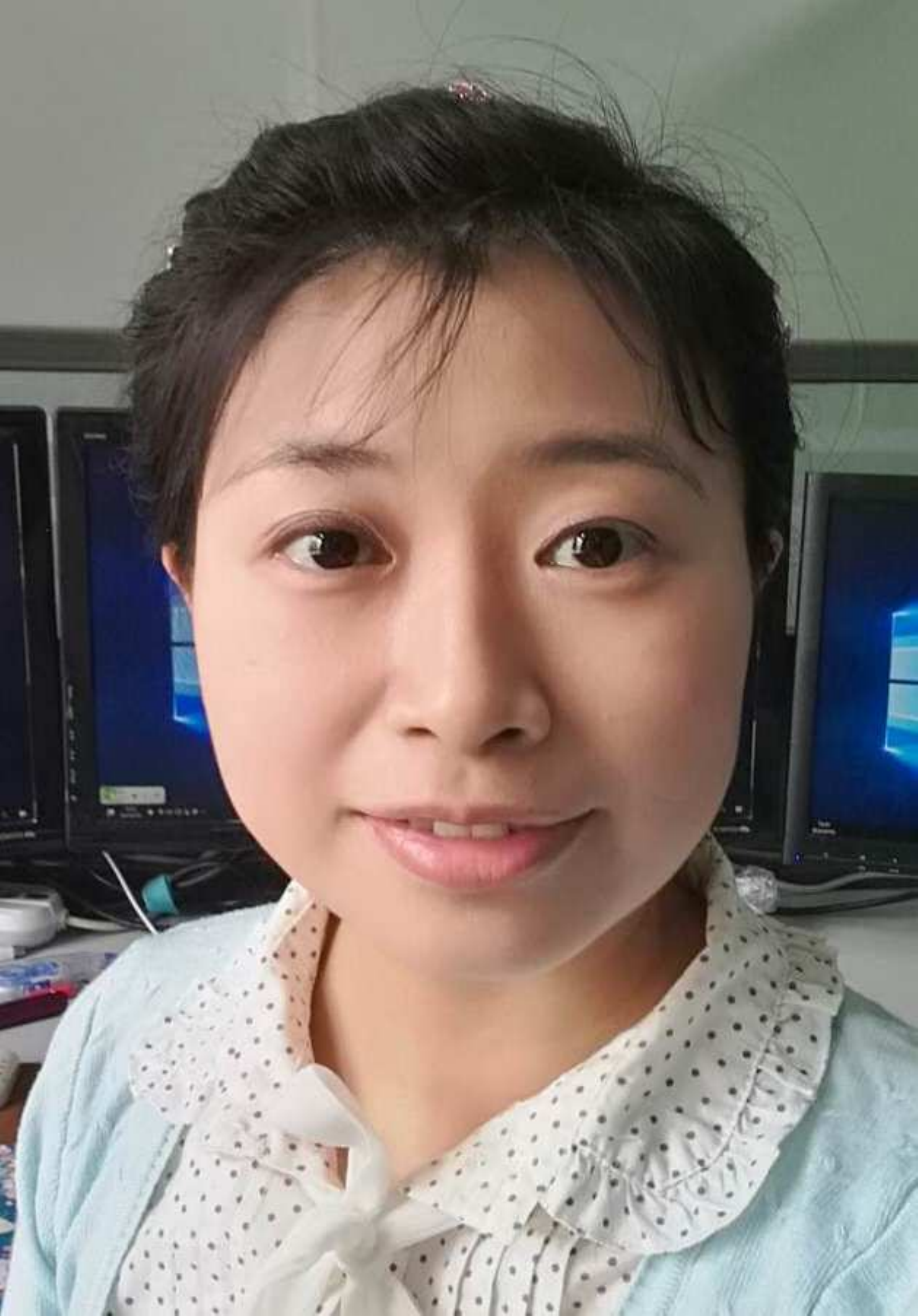}}]
{Yang Yang} received the B.Sc. degree from Xidian University, Xi'an,
China, in 2006 and Ph.D. degrees from Xidian University, China, in
2011. 
She is a research fellow (postdoctor) under supervisor Robert
H. Deng in Singapore Management
University. 
She is also an associate professor in the college of
mathematics and computer science, Fuzhou University. Her research
interests include information security and privacy
protection. She has published more than 60 papers in
IEEE TIFS, IEEE TDSC, IEEE TSC, IEEE TCC, IEEE TII, etc.
\end{IEEEbiography}

\begin{IEEEbiography}[{\includegraphics[width=1in,height=1.25in,clip,keepaspectratio]{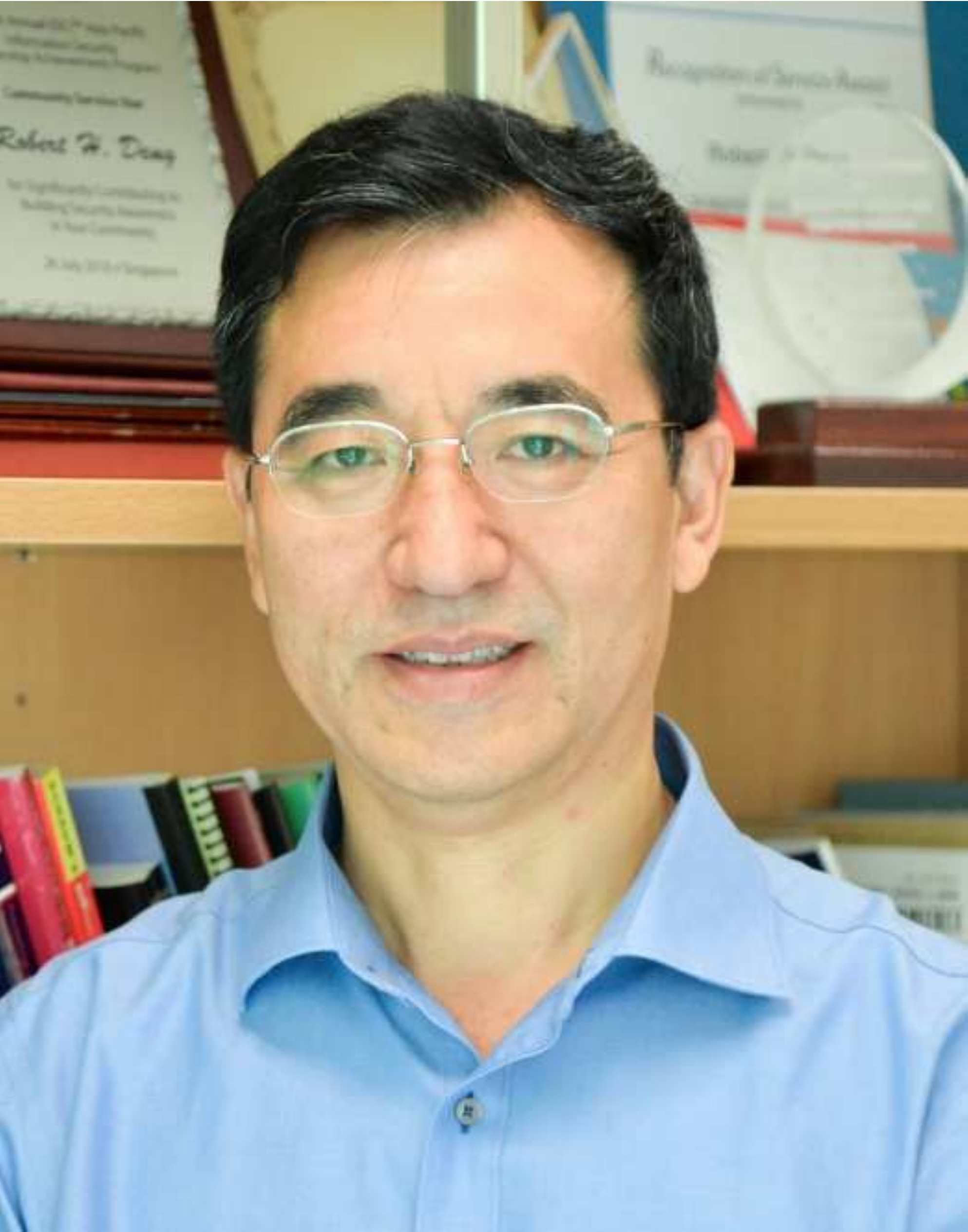}}]
{Robert H. Deng} is an AXA Professor in School of Information Systems,
Singapore Management University. He is Fellow of IEEE. His research
interests include data security and privacy, network and system
security. He has served/is serving on the editorial boards of many
international journals in security, such as IEEE TIFS, IEEE TDSC, the International Journal of Information
Security, and IEEE Security and Privacy Magazine.
\end{IEEEbiography}

\begin{IEEEbiography}[{\includegraphics[width=1in,height=1.25in,clip,keepaspectratio]{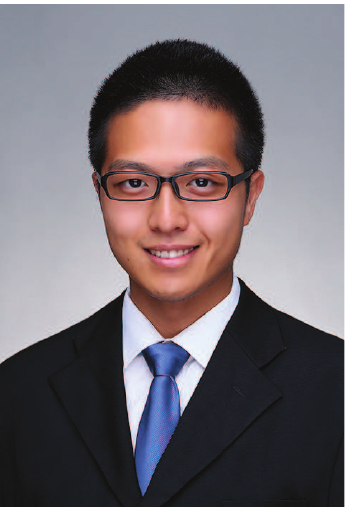}}]
	{Ximeng Liu} received the B.Sc. degree from Xidian University,
	Xi'an, China, in 2010 and Ph.D. degrees from Xidian University,
	China, in 2015. He was the research assistant at School of
	Electrical and Electronic Engineering, Nanyang Technological
	University, Singapore from 2013 to 2014.  Now, he is a professor in the college of
	mathematics and computer science, Fuzhou University. His research interests include cloud security
	and big data security.
\end{IEEEbiography}

\begin{IEEEbiography}[{\includegraphics[width=1in,height=1.25in,clip,keepaspectratio]{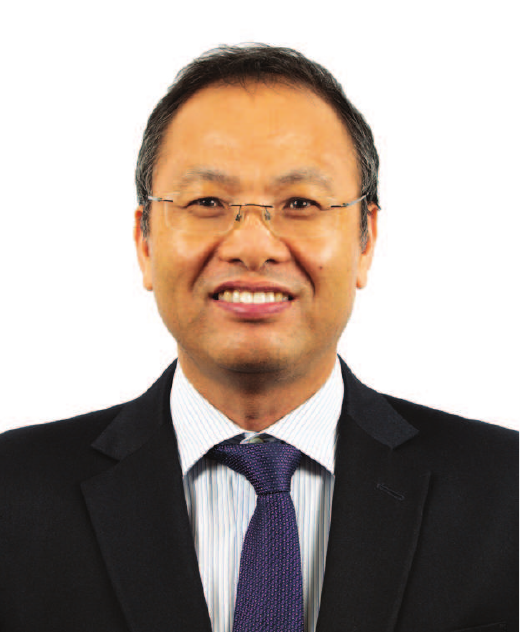}}]
{Yongdong Wu} Received the B.Eng and M.S. from
Beihang University, the Ph.D degree from Institute
of Automation, Chinese Academy of Science, and
Master for Management of Technology from National
University of Singapore. He is a professor with the College of Information Science and Technology, Jinan University.
He has published over 100 papers in IEEE TIFS, IEEE TMM, etc. His research results and proposals was
incorporated in the ISO/IEC JPEG 2000 security standard 15444-8 in 2007.
\end{IEEEbiography}

\begin{IEEEbiography}[{\includegraphics[width=1in,height=1.25in,clip,keepaspectratio]{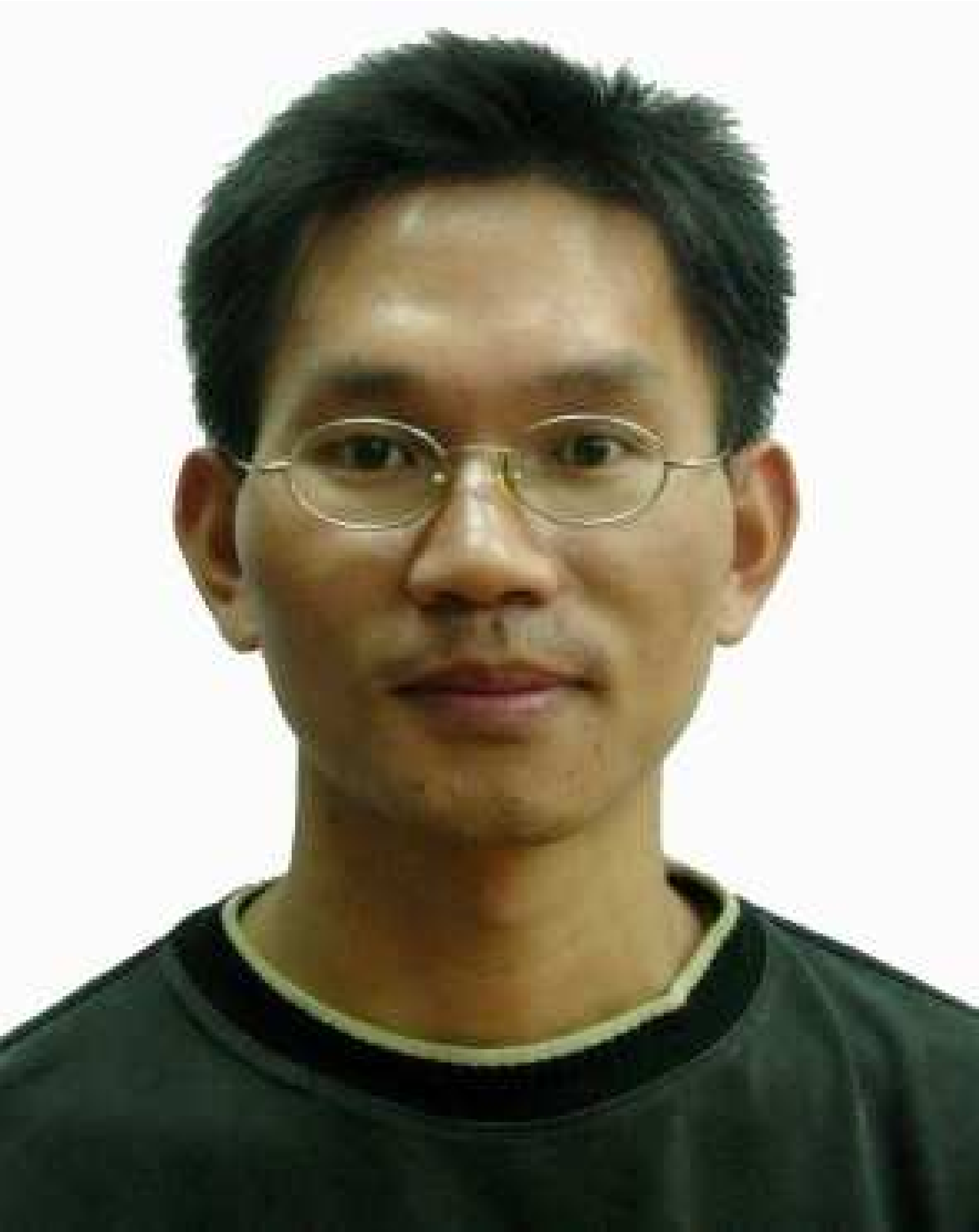}}]
{Jian Weng} received the Ph.D.
degree from Shanghai Jiao Tong University, in 2008.
From April 2008 to March 2010, he was
a post-doctor in the School of Information
Systems, Singapore Management University.
Currently, he is a professor and executive
dean with the College of Information Science and Technology, Jinan University. He has published more than 60 papers in
cryptography conferences and journals such as Eurocryp, Asiacrypt,
PKC, and IEEE TIFS.
\end{IEEEbiography}

\begin{IEEEbiography}[{\includegraphics[width=1in,height=1.25in,clip,keepaspectratio]{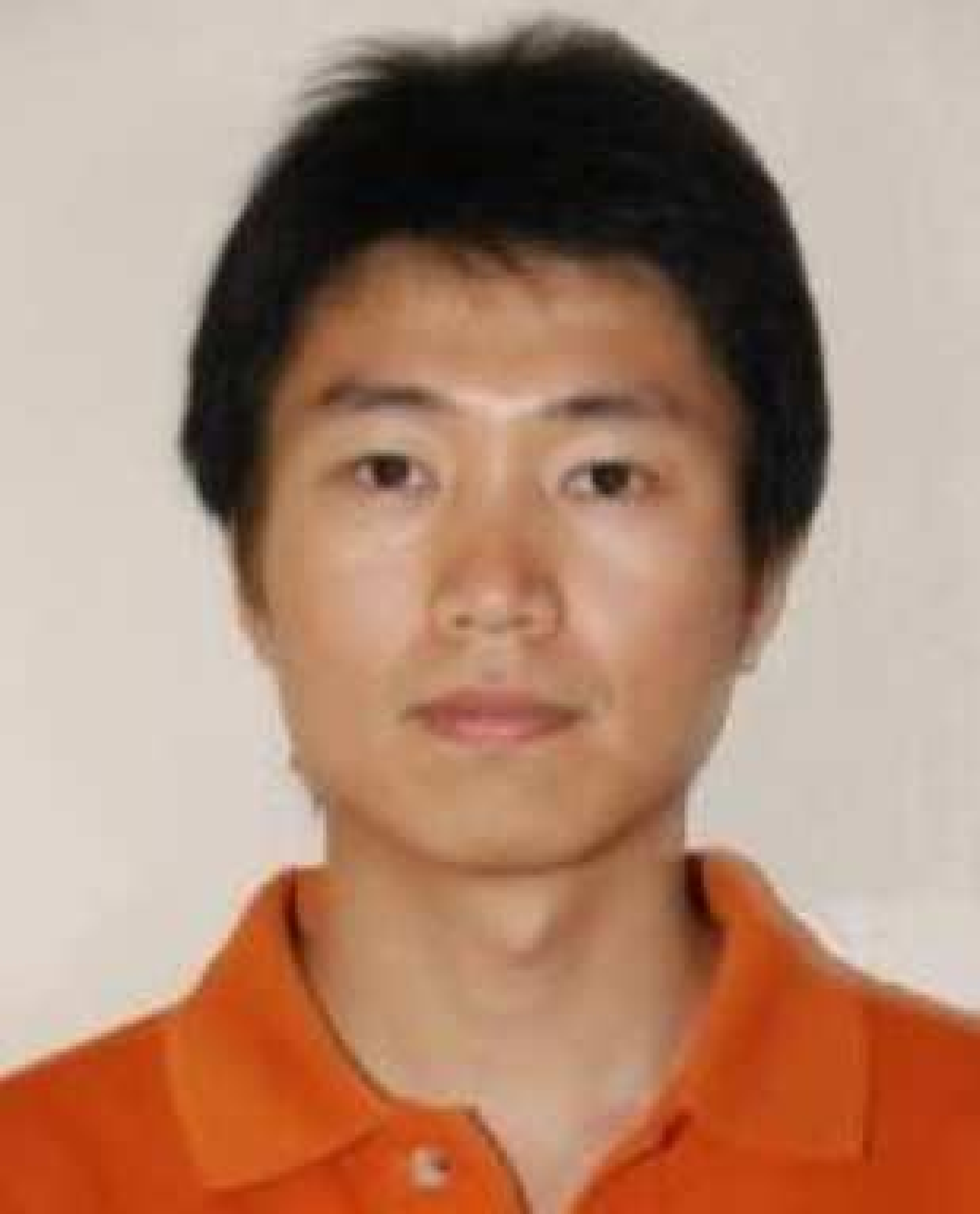}}]
	{Xianghan Zheng} is a professor in the College of Mathematics and Computer Sciences, Fuzhou University, China. He
	received his MSc of Distributed System (2007) and Ph.D of Information Communication Technology (2011) from University of
	Agder, Norway. His current research interests include New Generation Network with special focus on Cloud Computing
	Services and Applications, Big Data Processing and Security.
\end{IEEEbiography}

\begin{IEEEbiography}[{\includegraphics[width=1in,height=1.25in,clip,keepaspectratio]{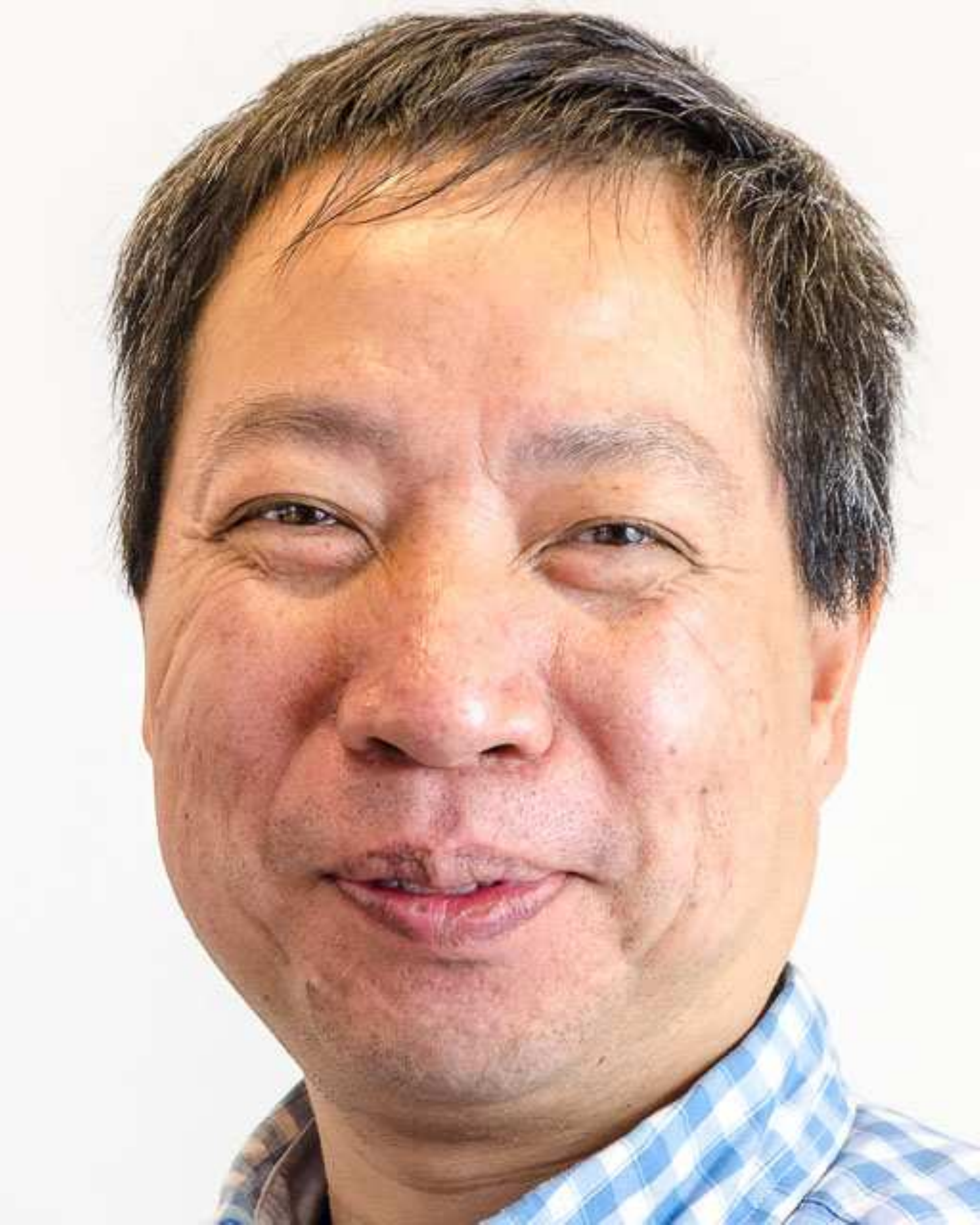}}]
	{Chunming Rong} is a professor and head of the Center for IP-based Service Innovation at University of Stavanger in Norway. His research interests include cloud computing, big data analysis, security and privacy. He is co-founder and chairman of the Cloud Computing Association (CloudCom.org) and its associated conference and workshop series. He is a member of the IEEE Cloud Computing Initiative, and co-Editor-in-Chief of the Springer Journal of Cloud Computing.
\end{IEEEbiography}

\clearpage
\setcounter{page}{1}
\section*{Supplemental Material}
\subsection*{A. Basic Primitives}
\subsubsection*{A-1. Paillier Cryptosystem with Threshold Decryption}
\label{sec:PCTD}
The Paillier cryptosystem with threshold
decryption (PCTD) \cite{Paillier99,Bresson03} is
the basic primitive in P-Med.\vspace{2mm}


\textbf{KeyGen:}  Let $\kappa$ be the security parameter and $p,q$ be two
large prime numbers such that ${\cal{L}}(p)={\cal{L}}(q)=\kappa$. Let
$N=pq$ and $\lambda=lcm(p-1, q-1)$ \footnote{$lcm$ : lowest common
multiple.}. Define a function $L(x)=\frac{x-1}{N}$ and select a
generator $g$ of order $ord(g)=(p-1)(q-1)/2$. The system public
parameter is $PP=(g,N)$. The master secret key of the system is $SK
= \lambda$. A user $i$ in the system is assigned a secret key $sk_i\in Z_N$
and a public key $pk_i =g^{sk_i} \mod N^2$.\vspace{2mm}

\textbf{Encryption} (\texttt{Enc}): On input a plaintext $m \in Z_N$,
a user randomly selects $r \in Z_{N}$ and uses his
public key $pk_i$ to encrypt $m$ to ciphertext
$[\![m]\!]_{pk_i}=(C_1,C_2)$, in which $C_1 = pk_i^r(1+mN) \mod N^2$ and
$C_2=g^r \mod N^2$.\vspace{2mm}

\textbf{Decryption with $sk_i$} (\texttt{WDec}): On input ciphertext
$[\![m]\!]_{pk_i}$ and secret key $sk_i$, the message can
be recovered by computing $m= L({C_1}/{C_2^{sk_i}}\mod N^2).$\vspace{2mm}

\textbf{Decryption with master secret key} (\texttt{SDec}): Using master secret key
$SK=\lambda$ of the system, any ciphertext $[\![m]\!]_{pk_i}$ encrypted by
any public key can be decrypted by computing $C_1^\lambda
=(pk_i^r)^{\lambda}(1+mN \lambda)= (1+mN \lambda)\mod N^2.$ Since
$gcd(\lambda, N)=1$ holds \footnote{$gcd$: greatest common
divider.}, we have $m= L(C_1^\lambda \mod N^2) \lambda^{-1} \mod N.$\vspace{2mm}

\textbf{Master secret key splitting} (\texttt{SKeyS}): The master secret key $SK =
\lambda$ can be randomly split into two parts $SK_1= \lambda_1$
and $SK_2= \lambda_2$ such that $\lambda_1 + \lambda_2 \equiv 0
\mod \lambda$ and $ \lambda_1 + \lambda_2 \equiv 1 \mod N^2$.\vspace{2mm}

\textbf{Partial Decryption with $SK_1$} (\texttt{PD1}):  On input the ciphertext
$[\![m]\!]_{pk_i}=(C_1,C_2)$, we can use
$SK_1=\lambda_1$ to compute
  $C_1^{(1)}  = {(C_1)}^{\lambda_1} = (pk_i^r)^{\lambda_1}(1+ mN \lambda_1) \mod N^2.$\vspace{2mm}

\textbf{Partial Decryption with $SK_2$} (\texttt{PD2}): On input $[\![m]\!]_{pk_i}$ and $C_1^{(1)}$, we
can use $SK_2=\lambda_2$ to compute
  $C_1^{(2)}  = {(C_1)}^{\lambda_2} = (pk_i^r)^{\lambda_2}(1+ mN \lambda_2)  \mod N^2.$
The message can be recovered by computing $m = L(C_1^{(1)}\cdot
C_1^{(2)}).$\vspace{2mm}

\textbf{Ciphertext Refresh} (\texttt{CR}): Refresh a ciphertext
$[\![m]\!]_{pk_i}=(C_1,C_2)$ to a new ciphertext
$[\![m']\!]_{pk_i}=(C_1',C_2')$ such that $m=m'$. It select a random $r'
\in  Z_N$ and calculates $ C_1' =  C_1 \cdot (pk_i)^{r'}\mod
N^2,C_2'=C_2 \cdot g^{r'} \mod N^2.$\vspace{2mm}

It is easy to verify that PCTD is \textbf{additive homomorphic}
$[\![m_1]\!]_{pk_i}\cdot [\![m_2]\!]_{pk_i} = [\![m_1+m_2]\!]_{pk_i}$ and
\textbf{scalar-multiplicative homomorphic} $([\![m]\!]_{pk_i})^r=[\![r\cdot m]\!]_{pk_i}$, $\forall r\in Z_N$.
 Specifically, $([\![m]\!]_{pk_i})^{(N-1)}=[\![-m]\!]_{pk_i}$ when $r=N-1$.

\subsubsection*{A-2. Secure Addition Protocol (\texttt{SAD})}
\label{sec:SMD}
Given two encrypted data $[\![x]\!]_{pk_A}$ and $[\![y]\!]_{pk_B}$ under two different public keys $pk_A$ and $pk_B$, respectively,
{\texttt{SAD}} \cite{LiuX_TIFS16} securely computes $[\![x+y]\!]_{pk_\sigma}$:

\textbf{Step-1(@CP)}: CP selects random $r_x,r_y\in Z_N$, calculates
$X=[\![x]\!]_{pk_A}\cdot[\![r_x]\!]_{pk_A}=[\![x+r_x]\!]_{pk_A}$,
$Y=[\![y]\!]_{pk_B}\cdot[\![r_y]\!]_{pk_B}=[\![y+r_y]\!]_{pk_B}$,
$X_1=\texttt{PD1}_{SK_1}(X)$, $Y_1=\texttt{PD1}_{SK_1}(Y)$,
and sends $X$, $Y$, $X_1$, $Y_1$ to $\text{CSP}$.

\textbf{Step-2($\text{@CSP}$)}: Using partial secret key $SK_2$, CSP calculates
$X_2=\texttt{PD2}_{SK_2} (X,X_1)$, $Y_2=\texttt{PD2}_{SK_2} (Y,Y_1)$,
computes $S=X_2+Y_2$ and sends $[\![S]\!]_{pk_\sigma}$ to CP.

\textbf{Step-3(@CP)}: CP computes  $R=r_x+r_y$
and calculates $[\![S]\!]_{pk_\sigma}\cdot([\![R]\!]_{pk_\sigma})^{N-1}=[\![S-R]\!]_{pk_\sigma}=[\![x+y]\!]_{pk_\sigma}$.

\subsubsection*{A-3. Secure Multiplication Protocol (\texttt{SMD})}
\label{sec:SMD}
Given two encrypted data $[\![x]\!]_{pk_A}$ and $[\![y]\!]_{pk_B}$,
{\texttt{SMD}} \cite{LiuX_TIFS16} securely computes $[\![x\cdot y]\!]_{pk_\sigma}$:

\textbf{Step-1(@CP)}: CP selects random $r_x,r_y, R_x, R_y\in Z_N$, calculates
$X=[\![x]\!]_{pk_A}\cdot[\![r_x]\!]_{pk_A}=[\![x+r_x]\!]_{pk_A}$,
$Y=[\![y]\!]_{pk_B}\cdot[\![r_y]\!]_{pk_B}=[\![y+r_y]\!]_{pk_B}$,
$S=[\![R_x]\!]_{pk_A}\cdot([\![x]\!]_{pk_A})^{N-r_y}=[\![R_x-r_y\cdot x]\!]_{pk_A}$,
$T=[\![R_y]\!]_{pk_B}\cdot([\![y]\!]_{pk_B})^{N-r_x}=[\![R_y-r_x\cdot y]\!]_{pk_B}$.
$X_1=\texttt{PD1}_{SK_1}(X)$, $Y_1=\texttt{PD1}_{SK_1}(Y)$,
$S_1=\texttt{PD1}_{SK_1}(S)$, $T_1=\texttt{PD1}_{SK_1}(T)$,
and sends $X$, $Y$, $S$, $T$, $X_1$, $Y_1$, $S_1$, $T_1$ to $\text{CSP}$.

\textbf{Step-2($\text{@CSP}$)}: Using partial secret key $SK_2$, CSP calculates
$h=\texttt{PD2}_{SK_2} (X,X_1)\cdot\texttt{PD2}_{SK_2} (Y,Y_1)$,
$S_2=\texttt{PD2}_{SK_2} (S,S_1)$, $T_2=\texttt{PD2}_{SK_2} (T,T_1)$,
$H=[\![h]\!]_{pk_\sigma}$, $S_3=[\![S_2]\!]_{pk_\sigma}$, $T_3=[\![T_2]\!]_{pk_\sigma}$
and sends $H,S_3,T_3$ to CP. It is obvious that $h=(x+r_x)(y+r_y)$.

\textbf{Step-3(@CP)}: Once $H,S_3,T_3$ are received, CP computes  $S_4=([\![r_x\cdot r_y]\!]_{pk_\sigma})^{N-1}$,
$S_5=([\![R_x]\!]_{pk_\sigma})^{N-1}$ and $S_6=([\![R_y]\!]_{pk_\sigma})^{N-1}$,
and calculates $H\cdot T_3\cdot S_3\cdot S_4\cdot S_5\cdot S_6=[\![h+(R_x-r_y\cdot x)+(R_y-r_x\cdot y)-r_x\cdot r_y-R_x-R_y]\!]_{pk_\sigma}=[\![x\cdot y]\!]_{pk_\sigma}$.

\subsubsection*{A-4. Secure greater or Equal Protocol (\textbf{\texttt{SGE}})}
\label{SubSec:SLE}

Given $[\![x]\!]_{pk_A}$ and $[\![y]\!]_{pk_B}$,
{\texttt{SGE}} \cite{YangY_INS18} securely computes $[\![u^*]\!]_{pk_\sigma}$ such that $u^*=1$ if  $x\geq y$ and $u^*=0$ if  $x<y$.

 \textbf{Step 1:} CP calculates
$[\![x']\!]_{pk_{A}}= ([\![x]\!]_{pk_A})^2\cdot [\![1]\!]_{pk_{A}} =
[\![2x+1]\!]_{pk_{A}}$
$[\![y']\!]_{pk_{B}}= ([\![y]\!]_{pk_B})^2 = [\![2y]\!]_{pk_{B}}$,
, and chooses random $r_1,r_2$, s.t. $ {\cal{L}}(r_1) <
{\cal{L}}(N)/4-1,{\cal{L}}(r_2) <
{\cal{L}}(N)/8$. Then, CP flips a coin $s\in\{0,1\}$ randomly.
CP and CSP jointly execute the following operations.

If $s=1$,
$[\![\gamma]\!]_{pk_\sigma} \leftarrow  \textbf{\texttt{SAD}}
(([\![x']\!]_{pk_{A}})^{r_1}, ([\![y']\!]_{pk_{B}})^{N-r_1}). $

If $s=0$,
$[\![\gamma]\!]_{pk_\sigma} \leftarrow  \textbf{\texttt{SAD}}
(([\![y']\!]_{pk_{B}})^{r_1}, ([\![x']\!]_{pk_{A}})^{N-r_1}).$

Then, CP calculates $l=[\![\gamma]\!]_{pk_\sigma}\cdot[\![r_2]\!]_{pk_\sigma}$
and $l' = \textbf{\texttt{PD1}}_{SK_1} (l)$ and
sends $(l,l')$ to CSP.

 \textbf{Step 2}:
 CSP decrypts $l'' = \textbf{\texttt{PD2}}_{SK_2} (l,l')$.
If $ {\cal{L}}(l'')  >  {\cal{L}}(N)/2$, CSP denotes $u' =0$; otherwise,
$u'=1$. Then, CSP uses $pk_\sigma$ to encrypt $u'$,
  and sends $ [\![u']\!]_{pk_\sigma} $  to CP.

  \textbf{Step 3}:
Receiving $ [\![u']\!]_{pk_\sigma}$, CP computes as follows: if $s=1$, CP denotes
$  [\![u^*]\!]_{pk_\sigma}= \textbf{\texttt{CR}}([\![u']\!]_{pk_\sigma});$
otherwise, CP computes
$[\![u^*]\!]_{pk_\sigma}= [\![1]\!]_{pk_\sigma}\cdot
([\![u']\!]_{pk_\sigma})^{N-1}= [\![1-u']\!]_{pk_\sigma}$.

\subsubsection*{A-5. Secure Less or Equal Protocol (\textbf{\texttt{SLE}})}
\label{SubSec:SLE}

Given $[\![x]\!]_{pk_A}$ and $[\![y]\!]_{pk_B}$ with ${\cal{L}}(x),{\cal{L}}(y)<{\cal{L}}(N)/8$,
{\texttt{SLE}} \cite{YangY_INS18} securely computes $[\![u^*]\!]_{pk_\sigma}$ such that $u^*=1$ if  $x\leq y$ and $u^*=0$ if  $x>y$.

 \textbf{Step 1:} CP calculates
$[\![x']\!]_{pk_{A}}= ([\![x]\!]_{pk_A})^2 = [\![2x]\!]_{pk_{A}}$,
$[\![y']\!]_{pk_{B}}= ([\![y]\!]_{pk_B})^2\cdot [\![1]\!]_{pk_{B}} =
[\![2y+1]\!]_{pk_{B}}$, and chooses random $r_1,r_2$, s.t. $ {\cal{L}}(r_1) <
{\cal{L}}(N)/4-1,{\cal{L}}(r_2) <
{\cal{L}}(N)/8$. Then, CP flips a coin $s\in\{0,1\}$ randomly.
CP and CSP jointly execute the following operations.

If $s=1$,
$[\![\gamma]\!]_{pk_\sigma} \leftarrow  \textbf{\texttt{SAD}}
(([\![y']\!]_{pk_{B}})^{r_1}, ([\![x']\!]_{pk_{A}})^{N-r_1}). $

If $s=0$,
$[\![\gamma]\!]_{pk_\sigma} \leftarrow  \textbf{\texttt{SAD}}
(([\![x']\!]_{pk_{A}})^{r_1}, ([\![y']\!]_{pk_{B}})^{N-r_1}).$

Then, CP calculates $l=[\![\gamma]\!]_{pk_\sigma}\cdot[\![r_2]\!]_{pk_\sigma}$
and $l' = \textbf{\texttt{PD1}}_{SK_1} (l)$ and
sends $(l,l')$ to CSP.

 \textbf{Step 2}:
 CSP decrypts $l'' = \textbf{\texttt{PD2}}_{SK_2} (l,l')$.
If $ {\cal{L}}(l'')  >  {\cal{L}}(N)/2$, CSP denotes $u' =0$; otherwise, 
$u'=1$. Then, CSP uses $pk_\sigma$ to encrypt $u'$,
  and sends $ [\![u']\!]_{pk_\sigma} $  to CP.

  \textbf{Step 3}:
Receiving $ [\![u']\!]_{pk_\sigma}$, CP computes as follows: if $s=1$, CP denotes
$  [\![u^*]\!]_{pk_\sigma}= \textbf{\texttt{CR}}([\![u']\!]_{pk_\sigma});$
otherwise, CP computes
$[\![u^*]\!]_{pk_\sigma}= [\![1]\!]_{pk_\sigma}\cdot
([\![u']\!]_{pk_\sigma})^{N-1}= [\![1-u']\!]_{pk_\sigma}$.

\subsubsection*{A-6. Secure Less Than Protocol (\textbf{\texttt{SLT}})}
\label{SubSec:SGE}

Given $[\![x]\!]_{pk_A}$ and $[\![y]\!]_{pk_B}$ with ${\cal{L}}(x),{\cal{L}}(y)<{\cal{L}}(N)/8$,
{\texttt{SLT}}  \cite{LiuX_TIFS16,YangY_INS18} securely computes $[\![u^*]\!]_{pk_\sigma}$ such that $u^*=1$ if  $x<y$ and $u^*=0$ if  $x\geq y$.

 \textbf{Step 1} and \textbf{Step 2} are the same as in \textbf{\texttt{SGE}} protocol.

  \textbf{Step 3}:
  Once   $ [\![u']\!]_{pk_\sigma}$ is received, CP  computes as follows: if $s=1$, CP denotes
$[\![u^*]\!]_{pk_\sigma}= [\![1]\!]_{pk_\sigma}\cdot
([\![u']\!]_{pk_\sigma})^{N-1}= [\![1-u']\!]_{pk_\sigma}.$
otherwise, CP computes $  [\![u^*]\!]_{pk_\sigma}= \textbf{\texttt{CR}}([\![u']\!]_{pk_\sigma});$

\subsubsection*{A-7. Secure Greater Than Protocol (\textbf{\texttt{SGT}})}
\label{SubSec:SGE}

Given $[\![x]\!]_{pk_A}$ and $[\![y]\!]_{pk_B}$ with ${\cal{L}}(x),{\cal{L}}(y)<{\cal{L}}(N)/8$,
{\texttt{SGT}} \cite{YangY_INS18} securely computes $[\![u^*]\!]_{pk_\sigma}$ such that $u^*=1$ if  $x>y$ and $u^*=0$ if  $x\leq y$.

 \textbf{Step 1} and \textbf{Step 2} are the same as in \textbf{\texttt{SLE}} protocol.

  \textbf{Step 3}:
  Once   $ [\![u']\!]_{pk_\sigma}$ is received, CP  computes as follows: if $s=1$, CP denotes
$[\![u^*]\!]_{pk_\sigma}= [\![1]\!]_{pk_\sigma}\cdot
([\![u']\!]_{pk_\sigma})^{N-1}= [\![1-u']\!]_{pk_\sigma}.$
otherwise, CP computes $  [\![u^*]\!]_{pk_\sigma}= \textbf{\texttt{CR}}([\![u']\!]_{pk_\sigma});$

\subsubsection*{A-8. Secure Equivalence Testing Protocol (\textbf{\texttt{SET}})}
\label{SubSec:SET}

Given $[\![x]\!]_{pk_A}$ and $[\![y]\!]_{pk_B}$ with ${\cal{L}}(x),{\cal{L}}(y)<{\cal{L}}(N)/8$,
{\texttt{SET}} \cite{YangY_INS18} securely computes $[\![u^*]\!]_{pk_\sigma}$ such that $u^*=1$ if  $x=y$ and $u^*=0$ if  $x\neq y$.
 CP and CSP jointly computes
$[\![u_1]\!]_{pk_\sigma}\leftarrow\textbf{\texttt{SLE}}([\![x]\!]_{pk_A},[\![y]\!]_{pk_B})$,
$[\![u_2]\!]_{pk_\sigma}\leftarrow\textbf{\texttt{SLE}}([\![y]\!]_{pk_B},[\![x]\!]_{pk_A})$,
$[\![u^*]\!]_{pk_\sigma}\leftarrow\textbf{\texttt{SMD}}([\![u_1]\!]_{pk_\sigma},[\![u_2]\!]_{pk_\sigma})$.

\subsubsection*{A-9. Secure Range Comparison Protocol (\textbf{\texttt{SRC}})}

Given $[\![x]\!]_{pk_A}$, $[\![y_1]\!]_{pk_B}$ and $[\![y_2]\!]_{pk_B}$ with ${\cal{L}}(x),{\cal{L}}(y_1),{\cal{L}}(y_2)$$<{\cal{L}}(N)/8$,
{\texttt{SRC}}\footnote{\texttt{SRC}is \texttt{SRT1}
(secure range test protocol type-1) in \cite{YangY_INS18}.}  \cite{YangY_INS18} securely computes $[\![u^*]\!]_{pk_\sigma}\leftarrow\textbf{\texttt{SRC}}([\![x]\!]_{pk_A},[\![y_1]\!]_{pk_B},[\![y_2]\!]_{pk_B})$ such that $u^*=1$ if $y_1\leq x\leq y_2$ and $u^*=0$ otherwise.
 CP and CSP jointly computes
$[\![u_1]\!]_{pk_\sigma}\leftarrow\textbf{\texttt{SGE}}([\![x]\!]_{pk_A},[\![y_1]\!]_{pk_B})$,
$[\![u_2]\!]_{pk_\sigma}\leftarrow\textbf{\texttt{SLE}}([\![x]\!]_{pk_B},[\![y_2]\!]_{pk_A})$,
$[\![u^*]\!]_{pk_\sigma}\leftarrow\textbf{\texttt{SMD}}([\![u_1]\!]_{pk_\sigma},[\![u_2]\!]_{pk_\sigma})$.

\subsection*{B. Security  Model and Proof}
Here we recall the security model for securely realizing an ideal functionality in the presence of non-colluding semi-honest adversaries
\cite{Kamara11,YangY_TSC17}. For simplicity, we do it for the specific scenario of our functionality,
which involve the patient/hospital (a.k.a "$D_1$"),
CP (a.k.a "$S_1$") and CSP (a.k.a "$S_2$").
We refer the readers to \cite{Kamara11} for the general case definitions.\vspace{2mm}

Let $\mathcal{P}=(D_1,S_1,S_2)$ be the set of all protocol parties. We consider three kinds of adversaries
$(\mathcal{A}_{D_1},\mathcal{A}_{S_1},\mathcal{A}_{S_2})$ that
corrupt $D_1$, $S_1$ and $S_2$, respectively. In the real world, $D_1$ runs on input $x$ and $y$
(with additional auxiliary inputs $z_x$ and $z_y$), while $S_1$ and $S_2$ receive auxiliary inputs $z_1$ and $z_2$.
Let $H\subseteq \mathcal{P}$ be the set of honest parties. Then, for every $P\in H$, let $out_P$ be the output
of party $P$, whereas if $P$ is corrupted, i.e. $P\in \mathcal{P}\backslash H$, then $out_P$ denotes the view of $P$ during the protocol $\Pi$.

For every $P^*\in\mathcal{P}$, the partial view of $P^*$ in a real-world execution of protocol $\Pi$ in the presence of
adversaries $\mathcal{A}=(\mathcal{A}_{D_1},\mathcal{A}_{S_1},\mathcal{A}_{S_2})$ is defined as
$$REAL^{P^*}_{\Pi,\mathcal{A},H,z}(\kappa,x,y)=\{out_P:P\in H\}\cup out_{P^*},$$
where $\kappa\in\mathbb{N}$ is the security parameter.\vspace{2mm}

In the ideal world, there is an ideal functionality $\textbf{\texttt{f}}$ for a function $f$ and the parties
interact only with $\textbf{\texttt{f}}$. Here, the challenge user sends $x$ and $y$ to $\textbf{\texttt{f}}$.
If any of $x$ or $y$ is $\bot$, then $\textbf{\texttt{f}}$ returns $\bot$. Finally, $\textbf{\texttt{f}}$ returns $f(x,y)$ to the challenge user.
As before, let $H\subseteq \mathcal{P}$ be the set of honest parties. Then, for every $P\in H$, let $out_P$ be the output
returned by $\textbf{\texttt{f}}$ to party $P$, whereas if $P$ is corrupted, $out_P$ is the same value returned by $P$.

For every $P^*\in\mathcal{P}$, the partial view of $P^*$ in an ideal-world execution in the presence of
independent simulators $Sim=(Sim_{D_1},Sim_{S_1},Sim_{S_2})$ is defined as
$$IDEAL^{P^*}_{\textbf{\texttt{f}},Sim,H,z}(\kappa,x,y)=\{out_P:P\in H\}\cup out_{P^*}.$$

Informally, a protocol $\Pi$ is considered secure against non-colluding semi-honest adversaries if it partially emulates,
in the real world, an execution of $\textbf{\texttt{f}}$ in the ideal world. More formally,

\begin{definition}
\label{Def:Security}
Let $\textbf{\texttt{f}}$ be a deterministic functionality among parties in $\mathcal{P}$. Let $H\subseteq \mathcal{P}$
be the subset of honest parties in $\mathcal{P}$. We say that $\Pi$ securely realizes $\textbf{\texttt{f}}$
if there exists a set $Sim=(Sim_{D_1},Sim_{S_1},Sim_{S_2})$ of PPT transformations ($Sim_{D_1}=Sim_{D_1}(\mathcal{A}_{D_1})$ and so on)
such that for all semi-honest PPT adversaries $\mathcal{A}=(\mathcal{A}_{D_1},\mathcal{A}_{S_1},\mathcal{A}_{S_2})$,
for all inputs $x,y$ and auxiliary inputs $z$, and for all parties $P\in\mathcal{P}$ it holds
\begin{align*}
&~\{REAL^{P^*}_{\Pi,\mathcal{A},H,z}(\kappa,x,y)\}_{\kappa\in\mathbb{N}}\\
\overset{c}{\approx}&~\{IDEAL^{P^*}_{\textbf{\texttt{f}},Sim,H,z}(\kappa,x,y)\}_{\kappa\in\mathbb{N}},
  \end{align*}
where $\overset{c}{\approx}$ denotes computational indistinguishability.
\end{definition}

\subsection*{C. Protocol Elaborations}

\subsubsection*{C-1. Elaboration of \texttt{SSM}}
Secure illness state match protocol (\texttt{SSM}) is elaborated below.
\begin{enumerate}
	\item Line 1. Initializes $[\![u^*]\!]_{pk_\sigma}=[\![1]\!]_{pk_\sigma}$.
	\item Line 2. If the patient $B$'s body temperature $BT_B$ is within the range $[BT_{A,1}, BT_{A,2}]$, i.e., $BT_{A,1}\leq BT_B\leq BT_{A,2}$, we have $[\![u_1]\!]_{pk_\sigma}=[\![1]\!]_{pk_\sigma}$; otherwise, $[\![u_1]\!]_{pk_\sigma}=[\![0]\!]_{pk_\sigma}$.
	\item Line 3-5. Blood pressure is usually expressed in terms of the systolic pressure (maximum during one heart beat) over diastolic pressure (minimum between two heart beats), and is measured in millimeters of mercury (mmHg). If the patient $B$'s blood pressure $BP_{B,1}/BP_{B,2}$ is within the range $[BP_{A,1}/BP_{A,2}$, $BP_{A,3}/BP_{A,4}]$, i.e., $BP_{A,1}\leq BP_{B,1}\leq BP_{A,3}$ and
	$BP_{A,2}\leq BP_{B,2}\leq BP_{A,4}$, we have $[\![u_{2,1}]\!]_{pk_\sigma}=[\![u_{2,2}]\!]_{pk_\sigma}=[\![1]\!]_{pk_\sigma}$ and
	$[\![u_2]\!]_{pk_\sigma}=\texttt{SMD}([\![u_{2,1}]\!]_{pk_\sigma},[\![u_{2,2}]\!]_{pk_\sigma})=[\![u_{2,1}\cdot u_{2,2}]\!]_{pk_\sigma}=[\![1]\!]_{pk_\sigma}$. Otherwise, we have $[\![u_{2,1}]\!]_{pk_\sigma}=[\![0]\!]_{pk_\sigma}$ or $[\![u_{2,2}]\!]_{pk_\sigma}=[\![0]\!]_{pk_\sigma}$, and $[\![u_2]\!]_{pk_\sigma}=[\![0]\!]_{pk_\sigma}$.
	\item Line 6. The blood glucose level is the amount of glucose present in the blood of humans, and is measured in mmol/L (millimoles per litre).
	If $B$'s blood glucose level $BG_B$ is within the range $[BG_{A,1}, BG_{A,2}]$, i.e., $BG_{A,1}\leq BG_B\leq BG_{A,2}$, we have $[\![u_3]\!]_{pk_\sigma}=[\![1]\!]_{pk_\sigma}$; otherwise, $[\![u_3]\!]_{pk_\sigma}=[\![0]\!]_{pk_\sigma}$.
	\item Line 7. The respiratory rate is the rate at which breathing occurs, and it is usually measured in breaths per minute. If the patient $B$'s respiratory rate $RR_B$ is larger than $RR_A$, we have $[\![u_4]\!]_{pk_\sigma}=[\![1]\!]_{pk_\sigma}$; otherwise, $[\![u_4]\!]_{pk_\sigma}=[\![0]\!]_{pk_\sigma}$.
	\item Line 8. The heart rate is the speed of the heartbeat measured by the number of contractions of the heart per minute. If the patient $B$'s heart rate $HR_B$ is less than $HR_A$, we have $[\![u_5]\!]_{pk_\sigma}=[\![1]\!]_{pk_\sigma}$; otherwise, $[\![u_5]\!]_{pk_\sigma}=[\![0]\!]_{pk_\sigma}$.
	\item Line 9-11. A symptom is a departure from normal function or feeling which is noticed by a patient, reflecting the presence of an unusual state, or of a disease. If the patient $B$'s symptoms
	$S_{B,1}=S_{A,1}$ and $S_{B,2}=S_{A,2}$, we have
	$[\![u_{6,1}]\!]_{pk_\sigma}=[\![u_{6,2}]\!]_{pk_\sigma}=[\![1]\!]_{pk_\sigma}$ and
	$[\![u_6]\!]_{pk_\sigma}=[\![u_{6,1}\cdot u_{6,2}]\!]_{pk_\sigma}=[\![1]\!]_{pk_\sigma}$. Otherwise, we have $[\![u_{6,1}]\!]_{pk_\sigma}=[\![0]\!]_{pk_\sigma}$ or $[\![u_{6,2}]\!]_{pk_\sigma}=[\![0]\!]_{pk_\sigma}$, and $[\![u_6]\!]_{pk_\sigma}=[\![0]\!]_{pk_\sigma}$.
	\item Line 12-13. If the patient $B$'s illness state $\phi$ in $[\![\phi]\!]_{pk_B}$ matches
	the state $q$ in $[\![q]\!]_{pk_{A}}$, we have $[\![u_i]\!]_{pk_\sigma}=[\![1]\!]_{pk_\sigma}$ (for $\forall i\in[\![1,6]\!]$)
	and $[\![u^*]\!]_{pk_\sigma}=[\![1]\!]_{pk_\sigma}$. Otherwise, there exists $j\in[\![1,6]\!]$ such that $[\![u_j]\!]_{pk_\sigma}=[\![0]\!]_{pk_\sigma}$, and thus $[\![u^*]\!]_{pk_\sigma}=[\![0]\!]_{pk_\sigma}$.\\
\end{enumerate}

\subsubsection*{C-2. Elaboration of \texttt{TPT}}

\begin{enumerate}
  \item Line 1. The two-dimensional arrays $value(\cdot,\cdot)$ and $weight(\cdot,\cdot)$ are initialized according to the state transition table
  of the encrypted NFA $[\![\mathbb{M}]\!]_{pk_B}$. Take the encrypted NFA shown in Fig. \ref{Fig:Hospital} and the state transition table shown in Table \ref{Tab:STT}
  as an example. $value_{4,1}=[\![y_4]\!]_{pk_A}$ indicates that the encrypted treatment method $[\![y_4]\!]_{pk_A}$ could lead the current state
  $[\![q_4]\!]_{pk_A}$ to transit to the next state $[\![q_1]\!]_{pk_A}$. $weight_{4,1}=[\![w_4]\!]_{pk_A}$ indicates that the
  encrypted transition weight (from $[\![q_4]\!]_{pk_A}$ to $[\![q_1]\!]_{pk_A}$) is $[\![w_4]\!]_{pk_A}$.
  \item Line 2-9. The stacks $Q$, $Y$ and $W$ are initialized to be empty. The total treatment procedure number $n$ is initialized to be zero.
  The arrays $count(\cdot)$ and $visit(\cdot,\cdot,\cdot)$ are initialized to be 0.
  Push $[\![q_0]\!]_{pk_B}$ into $Q$, and add 1 to $count_0$.
  \item Line 10. If the stack $Q$ is not empty, execute the following operations.
  \item Line 11. Let $\alpha$ be top element in $Q$.
  The integer $\beta$ is initialized to be $-1$.
  \item Line 12-14. Execute the operations introduced in step 2 of the basic idea. If such state exists,
  $\beta$ is set to be the index of the found state and $visit_{count[\alpha],\alpha,\beta}$ is set to be 1 indicating the corresponding edge is visited.
  \item Line 15-21. If $\beta=-1$, it indicates that the state (described in step 2 of the basic idea) is not found.
  Then, for the encrypted state $[\![q_{\alpha}]\!]$ that appears the $count_{\alpha}$-th time in the stack $R$,
  set all the transitions starting from $[\![q_{\alpha}]\!]$ to be unvisited.
  Pop $\alpha$ from $Q$, and minus 1 from $count_{\alpha}$. Pop the top element in stacks $Y$ and $W$ if the stacks are not empty.
  \item Line 22-23. If $\beta\neq-1$ and $count_{\beta}<MVisit$, it indicates that the state (described in step 2 of the basic idea) exits and $\beta$ is set to be the number of the found state.
  Push the encrypted state $[\![q_{\beta}]\!]_{pk_A}$ into stack $Q$,
  push the encrypted treatment method $value_{\alpha,\beta}$ into stack $Y$,
  and push the edge weight $w_{\alpha,\beta}$ into stack $W$.
  Then, add 1 to $count_{\beta}$.
  \item Line 24-25. If the stack $Q$ is not empty, denote the top element in $Q$ as $\alpha'$.
  \item Line 26-28. If $[\![q_{\alpha'}]\!]_{pk_A}$ belongs to $[\![\mathcal{F}]\!]_{pk_B}$, it indicates that
  a path from $[\![q_0]\!]_{pk_A}$ to $[\![\mathcal{F}]\!]_{pk_A}$ is found. Add 1 to the total treatment procedure number $n$.
  The encrypted states along the path are stored in $[\![\mathcal{Q}_n]\!]_{pk_A}$, the encrypted treatment methods are stored in $[\![\mathcal{Y}_n]\!]_{pk_A}$,
  and the encrypted transition weights are stored in $[\![\mathcal{W}_n]\!]_{pk_A}$.
  Then, pop the top elements in $Q$, $Y$, $W$ and minus 1 from $count_{\alpha'}$.
  \item Line 29-30. If $[\![q_{\alpha'}]\!]_{pk_A}$ does not belong to $[\![\mathcal{F}]\!]_{pk_A}$
  and $Q$ achieves the maximum state number $MState$,
  pop the top element in stacks $Q$, $Y$, $W$ and minus 1 from $count_{\alpha'}$.\\
\end{enumerate}

\begin{figure*}[htp]
	\begin{center}
		\includegraphics[width=6.5in]{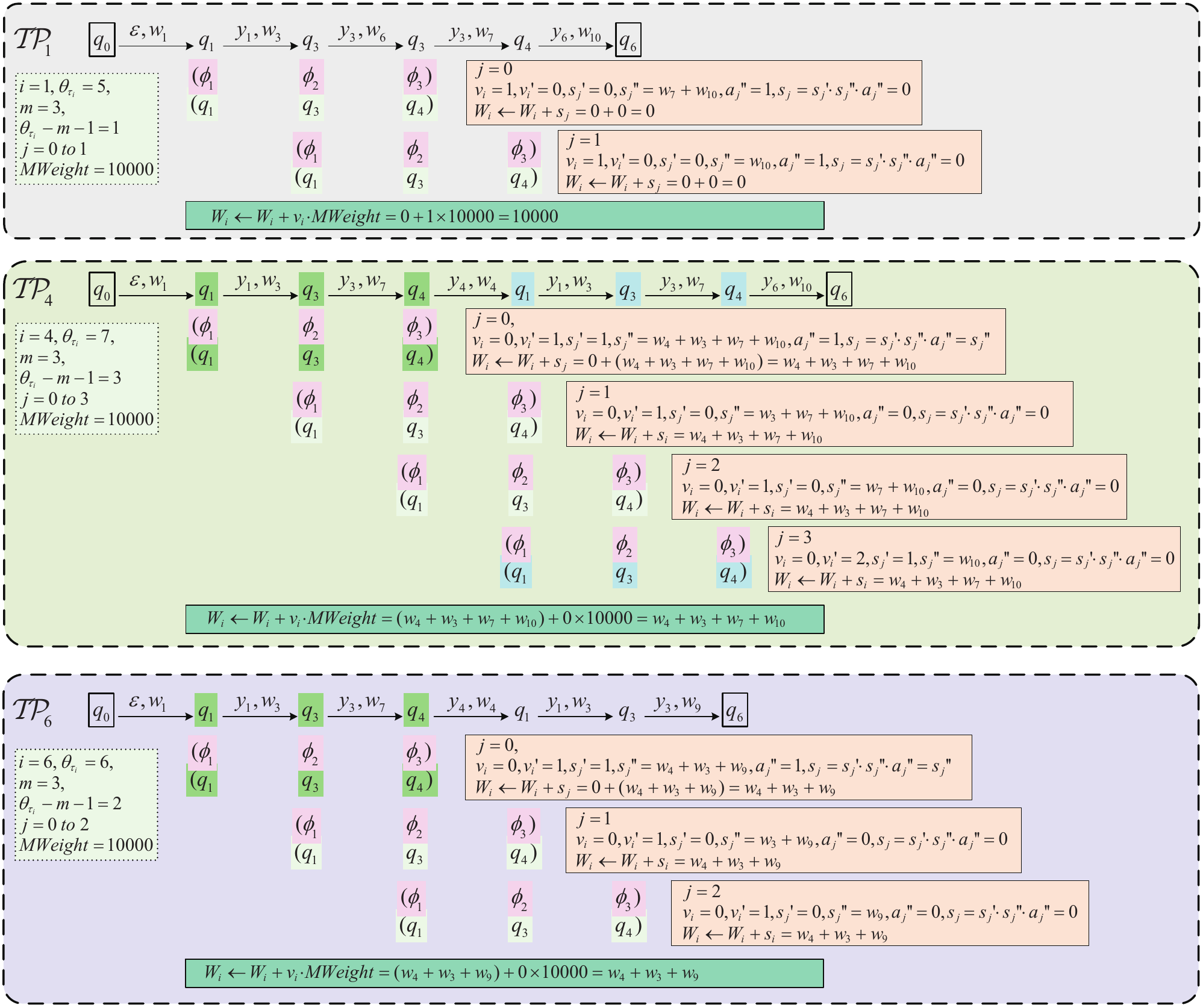}
		\caption{A Toy Example to Illustrate Workflow of \texttt{TPW}}
		\label{Fig:WorkflowTPW}
	\end{center}
\end{figure*}

\subsubsection*{C-3. Elaboration of \texttt{TPW}}

\texttt{TPW} calculates the encrypted weight $[\![W_i]\!]_{pk_{A}}$ for each treatment procedure
$[\![\mathcal{TP}_i]\!]_{pk_A}$ ($1\leq i \leq n$), which is elaborated below.
\begin{enumerate}
  \item Line 1-2. For each treatment procedure $\mathcal{TP}_i$ ($1\leq i\leq n$), initialize the weight $W_i=0$, the temporary variables $v_i=1$ and $v_i'=0$.
  $v_i$ indicates whether $[\![\Phi]\!]_{pk_B}$ is found in $[\![\mathcal{TP}_i]\!]_{pk_A}$. If it is not found, $v_i=1$; otherwise, $v_i=0$.
  $v_i'$ indicates the number of times that $[\![\Phi]\!]_{pk_B}$ is found in $[\![\mathcal{TP}_i]\!]_{pk_A}$.
  \item Line 3 and 21-22. If the intermediate state number in $\mathcal{TP}_i$ is equal or greater than $m$, go to steps 4-20 to search for the match states. Otherwise, the weight $W_i$ is set to be $MWeight$ since $\mathcal{TP}_i$ does not match with $\Phi$.
  \item Line 4. Try to find $m$ successive encrypted illness states $[\![\mathcal{Q}_{i,\theta_{j},m}]\!]_{pk_{A}}=([\![q_{i,\theta_{j}}]\!]_{pk_{A}},\cdots,[\![q_{i,\theta_{j+m-1}}]\!]_{pk_{A}})$ in $\mathcal{TP}_i$ that match $[\![\Phi]\!]_{pk_B}$.
      Since the encrypted illness state set in $\mathcal{TP}_i$ is denoted as
      $[\![\mathcal{Q}_i]\!]_{pk_{A}}=([\![q_0]\!]_{pk_{A}}, [\![q_{i,\theta_1}]\!]_{pk_{A}},\cdots,[\![q_{i,\theta_{\tau_i}}]\!]_{pk_{A}})$,
      \texttt{TPW} searches for the $m$ successive states from $q_0$ to $q_{i,\theta_{\tau_i-m}}$.
  \item Line 5. Initialize the temporary variables $s_j=0$, $s_j'=0$ and $s_j''=0$.
        \begin{itemize}
           \item $s_j'$ indicates whether $[\![\Phi]\!]_{pk_B}$ and $[\![\mathcal{Q}_{i,\theta_{j},m}]\!]_{pk_{A}}$ match. If they
             match, $s_j'=1$; otherwise, $s_j'=0$.
           \item $s_j''$ is utilized to calculate the sum of the transition weights from the illness state $q_{i,\theta_{j+m-1}}$ to the accept illness state $q_{i,\theta_{\tau_i}}$, and the sum equals  $\Sigma_{k=j+m}^{\tau_i}{w_{i,\theta_{k}}}$.
           \item $s_j$ equals to 0 or $\Sigma_{k=j+m}^{\tau_i}{w_{i,\theta_{k}}}$. If $[\![\Phi]\!]_{pk_B}$ and $[\![\mathcal{Q}_{i,\theta_{j},m}]\!]_{pk_{A}}$ match, and $[\![\Phi]\!]_{pk_B}$ appears for the first time in $[\![\mathcal{TP}_i]\!]_{pk_A}$, we have $s_j=\Sigma_{k=j+m}^{\tau_i}{w_{i,\theta_{k}}}$;
               otherwise, $s_j=0$.
        \end{itemize}
  \item Line 6. Initialize the temporary variables $a_j=0$, $a_j'=0$ and $a_j''=1$.
          \begin{itemize}
           \item $a_j$ calculates the number of match states of $[\![\Phi]\!]_{pk_B}$ and $[\![\mathcal{Q}_{i,\theta_{j},m}]\!]_{pk_{A}}$.
           \item $a_j'$ is a temporary variable to compute $[\![s_j]\!]_{pk_\sigma}=[\![s_j'\cdot s_j''\cdot v_i'']\!]_{pk_\sigma}.$
           \item $a_j''$ equals 1 or 0. For $j=0$, we have $a_j''=1$. For $1\leq j\leq \theta_{\tau_i-m}$,
           if there exists $j'\in[0,j)$ satisfying that $[\![\mathcal{Q}_{i,\theta_{j'},m}]\!]_{pk_{A}}$ and $[\![\Phi]\!]_{pk_B}$ match,
           we have $a_j''=0$; otherwise, $a_j''=1$.
        \end{itemize}
  \item Line 7-9. Calculate the number of match states in $[\![\Phi]\!]_{pk_B}$ and $[\![\mathcal{Q}_{i,\theta_{j},m}]\!]_{pk_{A}}$, and the result is stored in $[\![a_j]\!]_{pk_\sigma}$.
  \item Line 10. If $a_j=m$, we have $s_j'=1$ indicating that $[\![\Phi]\!]_{pk_B}$ and $[\![\mathcal{Q}_{i,\theta_{j},m}]\!]_{pk_{A}}$ match. Otherwise, $s_j'=0$.
  \item Line 11. If $v_i'\neq 0$, we have $a_j''=0$, which indicates that $[\![\Phi]\!]_{pk_B}$ has already appeared in
      $[\![\mathcal{TP}_i]\!]_{pk_A}$ before the $j$-th round search. Otherwise, $a_j''=1$.
  \item Line 12. If $[\![\Phi]\!]_{pk_B}$ and $[\![\mathcal{Q}_{i,\theta_{j},m}]\!]_{pk_{A}}$ match,
      add $s_j'=1$ to $v_i'$; otherwise, add $s_j'=0$ to $v_i'$.
  \item Line 13-14. $s_j''$ is the sum of the transition weights from the $\theta_{j+m-1}$-th illness state to the accept illness state, and $s_j''=\Sigma_{k=j+m}^{\tau_i}{w_{i,\theta_{k}}}$ after the for-loop calculation in line 13-14.
  \item Line 15-16. Calculate $[\![s_j]\!]_{pk_\sigma}=[\![s_j'\cdot s_j''\cdot a_j'']\!]_{pk_\sigma}$.
      \begin{itemize}
      \item If $s_j'=0$, we have $[\![s_j]\!]_{pk_\sigma}=[\![0]\!]_{pk_\sigma}$.
      \item If $[\![s_j']\!]_{pk_\sigma}=[\![1]\!]_{pk_\sigma}$ and $[\![a_j'']\!]_{pk_\sigma}=[\![1]\!]_{pk_\sigma}$,
      it indicates that $[\![\Phi]\!]_{pk_B}$ appears for the first time in $[\![\mathcal{TP}_i]\!]_{pk_A}$, and we
      have $[\![s_j]\!]_{pk_\sigma}=[\![s_j'']\!]_{pk_\sigma}=[\![\Sigma_{k=j+m}^{\tau_i}{w_{i,\theta_{k}}}]\!]_{pk_\sigma}.$
      \item If $[\![s_j']\!]_{pk_\sigma}=[\![1]\!]_{pk_\sigma}$ and $[\![a_j'']\!]_{pk_\sigma}=[\![0]\!]_{pk_\sigma}$,
      it indicates that $[\![\Phi]\!]_{pk_B}$ has already appeared in
      $[\![\mathcal{TP}_i]\!]_{pk_A}$ before the $j$-th round search, and we
      have $[\![s_j]\!]_{pk_\sigma}=[\![0]\!]_{pk_\sigma}.$
      \end{itemize}
  \item Line 17. Calculate $[\![W_i+s_j]\!]_{pk_\sigma}$ and the result is stored in $[\![W_i]\!]_{pk_\sigma}$.
   If $[\![\Phi]\!]_{pk_B}$ is found in $\mathcal{TP}_i$, we have
  $[\![W_i]\!]_{pk_\sigma}=[\![\Sigma_{k=\bar{j}+m}^{\tau_i}{w_{i,\theta_{k}}}]\!]_{pk_\sigma}$
   after the for-loop calculation in line 3-16,
   where $[\![\mathcal{Q}_{i,\theta_{\bar{j}},m}]\!]_{pk_{A}}$ is the first-match state set.
   Otherwise, $[\![W_i]\!]_{pk_\sigma}=[\![0]\!]_{pk_\sigma}$.
  \item Line 18. If $v_i'=0$, we have $v_i=1$, which indicates that $[\![\Phi]\!]_{pk_B}$ is not found in $\mathcal{TP}_i$. Otherwise, $v_i=0$.
  \item Line 19. If $[\![\Phi]\!]_{pk_B}$ is not found in $\mathcal{TP}_i$,
  we have
  \begin{align*}
  &[\![W_i]\!]_{pk_\sigma}\cdot([\![v_i]\!]_{pk_\sigma})^{MWeight}\\
  =&[\![0]\!]_{pk_\sigma}\cdot([\![1]\!]_{pk_\sigma})^{MWeight}=[\![MWeight]\!]_{pk_\sigma}.
  \end{align*}
  If it is found, we have
  \begin{align*}
  &[\![W_i]\!]_{pk_\sigma}\cdot([\![v_i]\!]_{pk_\sigma})^{MWeight}\\
  =&[\![W_i]\!]_{pk_\sigma}\cdot([\![0]\!]_{pk_\sigma})^{MWeight}\\
  =&[\![\Sigma_{k=\bar{j}+m}^{\tau_i}{w_{i,\theta_{k}}}]\!]_{pk_\sigma},
  \end{align*}
  where $[\![\mathcal{Q}_{i,\theta_{\bar{j}},m}]\!]_{pk_{A}}$ is the first-match state set.
  \item Line 20. Set the $i$-th weighted treatment procedure as $[\![\mathcal{WTP}_i]\!]_{pk_A}=([\![\mathcal{Q}_i]\!]_{pk_{A}}, [\![\mathcal{Y}_i]\!]_{pk_{A}}, [\![W_i]\!]_{pk_\sigma})$.\\
\end{enumerate}

Next, we use the toy example (in Fig. \ref{Fig:M-IoT}) to illustrate the workflow of \texttt{TPW} in Fig. \ref{Fig:WorkflowTPW}. Three treatment procedures ($\mathcal{TP}_1$, $\mathcal{TP}_4$ and $\mathcal{TP}_6$) in Section IV.D are selected to analyze the change of internal variables in security protocol \texttt{TPW}. Since patient's illness state set $\Phi=(\phi_1,\phi_2,\phi_3)$ matches with $(q_1, q_3, q_4)$ in Fig. \ref{Fig:M-IoT}, $\mathcal{TP}_1$ is a treatment procedure does not match with $\Phi$. The match state set $(q_1, q_3, q_4)$ appears two times in $\mathcal{TP}_4$ and one time in $\mathcal{TP}_6$. These three cases are analyzed in detail below. \\

The treatment procedure $\mathcal{TP}_1$ is depicted as
$q_0
\xrightarrow{\epsilon,\textcolor[rgb]{0,0,1}{w_1}}q_1
\xrightarrow{y_1,\textcolor[rgb]{0,0,1}{w_3}}q_3
\xrightarrow{y_3,\textcolor[rgb]{0,0,1}{w_6}}q_3
\xrightarrow{y_3,\textcolor[rgb]{0,0,1}{w_7}}q_4
\xrightarrow{y_6,\textcolor[rgb]{0,0,1}{w_{10}}}q_6$. According to the symbolic representation in Algorithm \ref{Algo:TPW}, we have $i=1$, $\theta_{\tau_i}=5$, $m=3$, $\theta_{\tau_i}-m-1=1$. Then, the variable $j$ should traverse from 0 to 1.

\begin{enumerate}
	\item For $j=0$, line 5-17 checks whether $q_1
	\xrightarrow{y_1,\textcolor[rgb]{0,0,1}{w_3}}q_3
	\xrightarrow{y_3,\textcolor[rgb]{0,0,1}{w_6}}q_3$ in $\mathcal{TP}_1$ matches with $\Phi=(\phi_1,\phi_2,\phi_3)$. Since they mismatch with each other, the internal variable changes: 
	\begin{itemize}
		\item $v_i=1$: indicates $\Phi$ is not yet found in $\mathcal{TP}_1$;
		\item $v_i'=0$: indicates the number of times that $\Phi$ is found in $\mathcal{TP}_1$ is still 0;
		\item $s_j'=0$: indicates $\Phi$ and $\mathcal{Q}_{1,\theta_0,3}=(q_1,q_3,q_3)$ (in $\mathcal{TP}_1$) do not match;
		\item $s_j''=w_7+w_{10}$: calculates the sum of weights in $q_3\xrightarrow{y_3,\textcolor[rgb]{0,0,1}{w_7}}q_4
		\xrightarrow{y_6,\textcolor[rgb]{0,0,1}{w_{10}}}q_6$;
		\item $a_j''=1$: indicates $\Phi$ is not yet found in $\mathcal{TP}_1$ for even one time;
		\item $s_j=s_j'\cdot s_j''\cdot a_j''=0$: is calculated according to Line 15-16;
		\item $W_i\leftarrow W_i+s_j=0$: is calculated according to Line 17.
	\end{itemize}

	\item For $j=1$, line 5-17 checks whether $q_3
	\xrightarrow{y_3,\textcolor[rgb]{0,0,1}{w_6}}q_3
	\xrightarrow{y_3,\textcolor[rgb]{0,0,1}{w_7}}q_4$ in $\mathcal{TP}_1$ matches with $\Phi=(\phi_1,\phi_2,\phi_3)$. Since they mismatch with each other, the internal variable changes: 
	\begin{itemize}
		\item $v_i=1$: indicates $\Phi$ is not yet found in $\mathcal{TP}_1$;
		\item $v_i'=0$: indicates the number of times that $\Phi$ is found in $\mathcal{TP}_1$ is still 0;
		\item $s_j'=0$: indicates $\Phi$ and $\mathcal{Q}_{1,\theta_1,3}=(q_3,q_3,q_4)$ (in $\mathcal{TP}_1$) do not match;
		\item $s_j''=w_{10}$: calculates the sum of weights in $q_4
		\xrightarrow{y_6,\textcolor[rgb]{0,0,1}{w_{10}}}q_6$;
		\item $a_j''=1$: indicates $\Phi$ is not yet found in $\mathcal{TP}_1$ for even one time;
		\item $s_j=s_j'\cdot s_j''\cdot a_j''=0$: is calculated according to Line 15-16;
		\item $W_i\leftarrow W_i+s_j=0$: is calculated according to Line 17. 
	\end{itemize}

	\item After the loops, the weight of $\mathcal{TP}_1$ is calculated as $W_1\leftarrow W_1+v_i\cdot MWeight = 0+1\times 10000=10000$ (according to Line 19).\\
\end{enumerate}

The treatment procedure $\mathcal{TP}_4$ is depicted as
$q_0
\xrightarrow{\epsilon,\textcolor[rgb]{0,0,1}{w_1}}
\textcolor[rgb]{1,0,0}{q_1}
\xrightarrow{y_1,\textcolor[rgb]{0,0,1}{w_3}}
\textcolor[rgb]{1,0,0}{q_3}
\xrightarrow{y_3,\textcolor[rgb]{0,0,1}{w_7}}
\textcolor[rgb]{1,0,0}{q_4}
\xrightarrow{y_4,\textcolor[rgb]{0,0,1}{w_4}}
\textcolor[RGB]{148,0,211}{q_1}
\xrightarrow{y_1,\textcolor[rgb]{0,0,1}{w_3}}
\textcolor[RGB]{148,0,211}{q_3}
\xrightarrow{y_3,\textcolor[rgb]{0,0,1}{w_7}}
\textcolor[RGB]{148,0,211}{q_4}
\xrightarrow{y_6,\textcolor[rgb]{0,0,1}{w_{10}}}q_6$. According to the symbolic representation in Algorithm \ref{Algo:TPW}, we have $i=4$, $\theta_{\tau_i}=7$, $m=3$, $\theta_{\tau_i}-m-1=3$. Then, the variable $j$ should traverse from 0 to 3.

\begin{enumerate}
	\item For $j=0$, line 5-17 checks whether $\textcolor[rgb]{1,0,0}{q_1}
	\xrightarrow{y_1,\textcolor[rgb]{0,0,1}{w_3}}
	\textcolor[rgb]{1,0,0}{q_3}
	\xrightarrow{y_3,\textcolor[rgb]{0,0,1}{w_7}}
	\textcolor[rgb]{1,0,0}{q_4}$ in $\mathcal{TP}_4$ matches with $\Phi=(\phi_1,\phi_2,\phi_3)$. Since they match, the internal variable changes: 
	\begin{itemize}
		\item $v_i=0$: indicates $\Phi$ is found in $\mathcal{TP}_4$;
		\item $v_i'=1$: indicates the number of times that $\Phi$ is found in $\mathcal{TP}_4$ is 1;
		\item $s_j'=1$: indicates $\Phi$ and $\mathcal{Q}_{4,\theta_0,3}=(q_1,q_3,q_4)$ (in $\mathcal{TP}_4$) match;
		\item $s_j''=w_4+w_3+w_7+w_{10}$: calculates the sum of weights in $\textcolor[rgb]{1,0,0}{q_4}
		\xrightarrow{y_4,\textcolor[rgb]{0,0,1}{w_4}}
		\textcolor[RGB]{148,0,211}{q_1}
		\xrightarrow{y_1,\textcolor[rgb]{0,0,1}{w_3}}
		\textcolor[RGB]{148,0,211}{q_3}
		\xrightarrow{y_3,\textcolor[rgb]{0,0,1}{w_7}}
		\textcolor[RGB]{148,0,211}{q_4}
		\xrightarrow{y_6,\textcolor[rgb]{0,0,1}{w_{10}}}q_6$;
		\item $a_j''=1$: indicates $\Phi$ is found in $\mathcal{TP}_4$ for the first time ($a_j''$ will change to 0 in the following loops);
		\item $s_j=s_j'\cdot s_j''\cdot a_j''=s_j''$: is calculated according to Line 15-16;
		\item $W_i\leftarrow W_i+s_j=w_4+w_3+w_7+w_{10}$: is calculated according to Line 17.
	\end{itemize}
	
	\item For $j=1$, line 5-17 checks whether $\textcolor[rgb]{1,0,0}{q_3}
	\xrightarrow{y_3,\textcolor[rgb]{0,0,1}{w_7}}
	\textcolor[rgb]{1,0,0}{q_4}
	\xrightarrow{y_4,\textcolor[rgb]{0,0,1}{w_4}}
	\textcolor[RGB]{148,0,211}{q_1}$ in $\mathcal{TP}_4$ matches with $\Phi=(\phi_1,\phi_2,\phi_3)$. Since they mismatch with each other, the internal variable changes: 
	\begin{itemize}
		\item $v_i=0$: indicates $\Phi$ is found in $\mathcal{TP}_4$;
		\item $v_i'=1$: indicates the number of times that $\Phi$ is found in $\mathcal{TP}_4$ is 1;
		\item $s_j'=0$: indicates $\Phi$ and $\mathcal{Q}_{4,\theta_1,3}=(q_3,q_4,q_1)$ (in $\mathcal{TP}_4$) do not match;
		\item $s_j''=w_3+w_7+w_{10}$: calculates the sum of weights in 
		$\textcolor[RGB]{148,0,211}{q_1}
		\xrightarrow{y_1,\textcolor[rgb]{0,0,1}{w_3}}
		\textcolor[RGB]{148,0,211}{q_3}
		\xrightarrow{y_3,\textcolor[rgb]{0,0,1}{w_7}}
		\textcolor[RGB]{148,0,211}{q_4}
		\xrightarrow{y_6,\textcolor[rgb]{0,0,1}{w_{10}}}q_6$;
		\item $a_j''=0$: indicates $\Phi$ is already found in $\mathcal{TP}_4$ for at least one time;
		\item $s_j=s_j'\cdot s_j''\cdot a_j''=0$: is calculated according to Line 15-16;
		\item $W_i\leftarrow W_i+s_j=w_4+w_3+w_7+w_{10}$: is calculated according to Line 17. 
	\end{itemize}

	\item For $j=2$, line 5-17 checks whether $\textcolor[rgb]{1,0,0}{q_4}
	\xrightarrow{y_4,\textcolor[rgb]{0,0,1}{w_4}}
	\textcolor[RGB]{148,0,211}{q_1}
	\xrightarrow{y_1,\textcolor[rgb]{0,0,1}{w_3}}
	\textcolor[RGB]{148,0,211}{q_3}$ matches with $\Phi=(\phi_1,\phi_2,\phi_3)$. Since they mismatch with each other, the internal variable changes: 
	\begin{itemize}
		\item $v_i=0$: indicates $\Phi$ is found in $\mathcal{TP}_4$;
		\item $v_i'=1$: indicates the number of times that $\Phi$ is found in $\mathcal{TP}_4$ is 1;
		\item $s_j'=0$: indicates $\Phi$ and $\mathcal{Q}_{4,\theta_2,3}=(q_4,q_1,q_3)$ (in $\mathcal{TP}_4$) do not match;
		\item $s_j''=w_7+w_{10}$: calculates the sum of weights in 
		$\textcolor[RGB]{148,0,211}{q_3}
		\xrightarrow{y_3,\textcolor[rgb]{0,0,1}{w_7}}
		\textcolor[RGB]{148,0,211}{q_4}
		\xrightarrow{y_6,\textcolor[rgb]{0,0,1}{w_{10}}}q_6$;
		\item $a_j''=0$: indicates $\Phi$ is already found in $\mathcal{TP}_4$ for at least one time;
		\item $s_j=s_j'\cdot s_j''\cdot a_j''=0$: is calculated according to Line 15-16;
		\item $W_i\leftarrow W_i+s_j=w_4+w_3+w_7+w_{10}$: is calculated according to Line 17. 
	\end{itemize}
	
	\item For $j=3$, line 5-17 checks whether $\textcolor[RGB]{148,0,211}{q_1}
	\xrightarrow{y_1,\textcolor[rgb]{0,0,1}{w_3}}
	\textcolor[RGB]{148,0,211}{q_3}
	\xrightarrow{y_3,\textcolor[rgb]{0,0,1}{w_7}}
	\textcolor[RGB]{148,0,211}{q_4}$ in $\mathcal{TP}_4$ matches with $\Phi=(\phi_1,\phi_2,\phi_3)$. Since they match, the internal variable changes: 
	\begin{itemize}
		\item $v_i=0$: indicates $\Phi$ is found in $\mathcal{TP}_4$;
		\item $v_i'=1$: indicates the number of times that $\Phi$ is found in $\mathcal{TP}_4$ is 2;
		\item $s_j'=1$: indicates $\Phi$ and $\mathcal{Q}_{4,\theta_3,3}=(q_1,q_3,q_4)$ (in $\mathcal{TP}_4$) match;
		\item $s_j''=w_{10}$: calculates the sum of weights in 
		$\textcolor[RGB]{148,0,211}{q_4}
		\xrightarrow{y_6,\textcolor[rgb]{0,0,1}{w_{10}}}q_6$;
		\item $a_j''=0$: indicates $\Phi$ is already found in $\mathcal{TP}_4$ for at least one time;
		\item $s_j=s_j'\cdot s_j''\cdot a_j''=0$: is calculated according to Line 15-16;
		\item $W_i\leftarrow W_i+s_j=w_4+w_3+w_7+w_{10}$: is calculated according to Line 17.
	\end{itemize}
	
	\item After the loops, the weight of $\mathcal{TP}_4$ is calculated as $W_4\leftarrow W_4+v_i\cdot MWeight = (w_4+w_3+w_7+w_{10})+0\times 10000=w_4+w_3+w_7+w_{10}$.\\
\end{enumerate}

The treatment procedure $\mathcal{TP}_6$ is depicted as
$q_0
\xrightarrow{\epsilon,\textcolor[rgb]{0,0,1}{w_1}}
\textcolor[rgb]{1,0,0}{q_1}
\xrightarrow{y_1,\textcolor[rgb]{0,0,1}{w_3}}
\textcolor[rgb]{1,0,0}{q_3}
\xrightarrow{y_3,\textcolor[rgb]{0,0,1}{w_7}}
\textcolor[rgb]{1,0,0}{q_4}
\xrightarrow{y_4,\textcolor[rgb]{0,0,1}{w_4}}q_1
\xrightarrow{y_1,\textcolor[rgb]{0,0,1}{w_3}}q_3
\xrightarrow{y_3,\textcolor[rgb]{0,0,1}{w_9}}q_6$. According to the symbolic representation in Algorithm \ref{Algo:TPW}, we have $i=6$, $\theta_{\tau_i}=6$, $m=3$, $\theta_{\tau_i}-m-1=2$. Then, the variable $j$ should traverse from 0 to 2.

\begin{enumerate}
	\item For $j=0$, line 5-17 checks whether $\textcolor[rgb]{1,0,0}{q_1}
	\xrightarrow{y_1,\textcolor[rgb]{0,0,1}{w_3}}
	\textcolor[rgb]{1,0,0}{q_3}
	\xrightarrow{y_3,\textcolor[rgb]{0,0,1}{w_7}}
	\textcolor[rgb]{1,0,0}{q_4}$ in $\mathcal{TP}_6$ matches with $\Phi=(\phi_1,\phi_2,\phi_3)$. Since they match, the internal variable changes: 
	\begin{itemize}
		\item $v_i=0$: indicates $\Phi$ is found in $\mathcal{TP}_6$;
		\item $v_i'=1$: indicates the number of times that $\Phi$ is found in $\mathcal{TP}_6$ is 1;
		\item $s_j'=1$: indicates $\Phi$ and $\mathcal{Q}_{6,\theta_0,3}=(q_1,q_3,q_4)$ (in $\mathcal{TP}_6$) match;
		\item $s_j''=w_4+w_3+w_9$: calculates the sum of weights in $\textcolor[rgb]{1,0,0}{q_4}
		\xrightarrow{y_4,\textcolor[rgb]{0,0,1}{w_4}}q_1
		\xrightarrow{y_1,\textcolor[rgb]{0,0,1}{w_3}}q_3
		\xrightarrow{y_3,\textcolor[rgb]{0,0,1}{w_9}}q_6$;
		\item $a_j''=1$: indicates $\Phi$ is found in $\mathcal{TP}_6$ for the first time ($a_j''$ will change to 0 in the following loops);
		\item $s_j=s_j'\cdot s_j''\cdot a_j''=s_j''$: is calculated according to Line 15-16;
		\item $W_i\leftarrow W_i+s_j=w_4+w_3+w_9$: is calculated according to Line 17.
	\end{itemize}
	
	\item For $j=1$, line 5-17 checks whether $\textcolor[rgb]{1,0,0}{q_3}
	\xrightarrow{y_3,\textcolor[rgb]{0,0,1}{w_7}}
	\textcolor[rgb]{1,0,0}{q_4}
	\xrightarrow{y_4,\textcolor[rgb]{0,0,1}{w_4}}
	\textcolor[RGB]{0,0,1}{q_1}$ in $\mathcal{TP}_6$ matches with $\Phi=(\phi_1,\phi_2,\phi_3)$. Since they mismatch with each other, the internal variable changes: 
	\begin{itemize}
		\item $v_i=0$: indicates $\Phi$ is found in $\mathcal{TP}_6$;
		\item $v_i'=1$: indicates the number of times that $\Phi$ is found in $\mathcal{TP}_6$ is 1;
		\item $s_j'=0$: indicates $\Phi$ and $\mathcal{Q}_{6,\theta_1,3}=(q_3,q_4,q_1)$ (in $\mathcal{TP}_6$) do not match;
		\item $s_j''=w_3+w_9$: calculates the sum of weights in $q_1
		\xrightarrow{y_1,\textcolor[rgb]{0,0,1}{w_3}}q_3
		\xrightarrow{y_3,\textcolor[rgb]{0,0,1}{w_9}}q_6$;
		\item $a_j''=0$: indicates $\Phi$ is already found in $\mathcal{TP}_6$ for at least one time;
		\item $s_j=s_j'\cdot s_j''\cdot a_j''=0$: is calculated according to Line 15-16;
		\item $W_i\leftarrow W_i+s_j=w_4+w_3+w_9$: is calculated according to Line 17. 
	\end{itemize}
	
	\item For $j=2$, line 5-17 checks whether $\textcolor[rgb]{1,0,0}{q_4}
	\xrightarrow{y_4,\textcolor[rgb]{0,0,1}{w_4}}q_1
	\xrightarrow{y_1,\textcolor[rgb]{0,0,1}{w_3}}q_3$ matches with $\Phi=(\phi_1,\phi_2,\phi_3)$. Since they mismatch with each other, the internal variable changes: 
	\begin{itemize}
		\item $v_i=0$: indicates $\Phi$ is found in $\mathcal{TP}_6$;
		\item $v_i'=1$: indicates the number of times that $\Phi$ is found in $\mathcal{TP}_6$ is 1;
		\item $s_j'=0$: indicates $\Phi$ and $\mathcal{Q}_{6,\theta_2,3}=(q_4,q_1,q_3)$ (in $\mathcal{TP}_6$) do not match;
		\item $s_j''=w_9$: calculates the sum of weights in $q_3
		\xrightarrow{y_3,\textcolor[rgb]{0,0,1}{w_9}}q_6$;
		\item $a_j''=0$: indicates $\Phi$ is already found in $\mathcal{TP}_6$ for at least one time;
		\item $s_j=s_j'\cdot s_j''\cdot a_j''=0$: is calculated according to Line 15-16;
		\item $W_i\leftarrow W_i+s_j=w_4+w_3+w_9$: is calculated according to Line 17. 
	\end{itemize}
		
	\item After the loops, the weight of $\mathcal{TP}_6$ is calculated as $W_6\leftarrow W_6+v_i\cdot MWeight = (w_4+w_3+w_9)+0\times 10000=w_4+w_3+w_9$.	\\
\end{enumerate}

\begin{figure*}[htp]
	\begin{center}
		\includegraphics[width=6.5in]{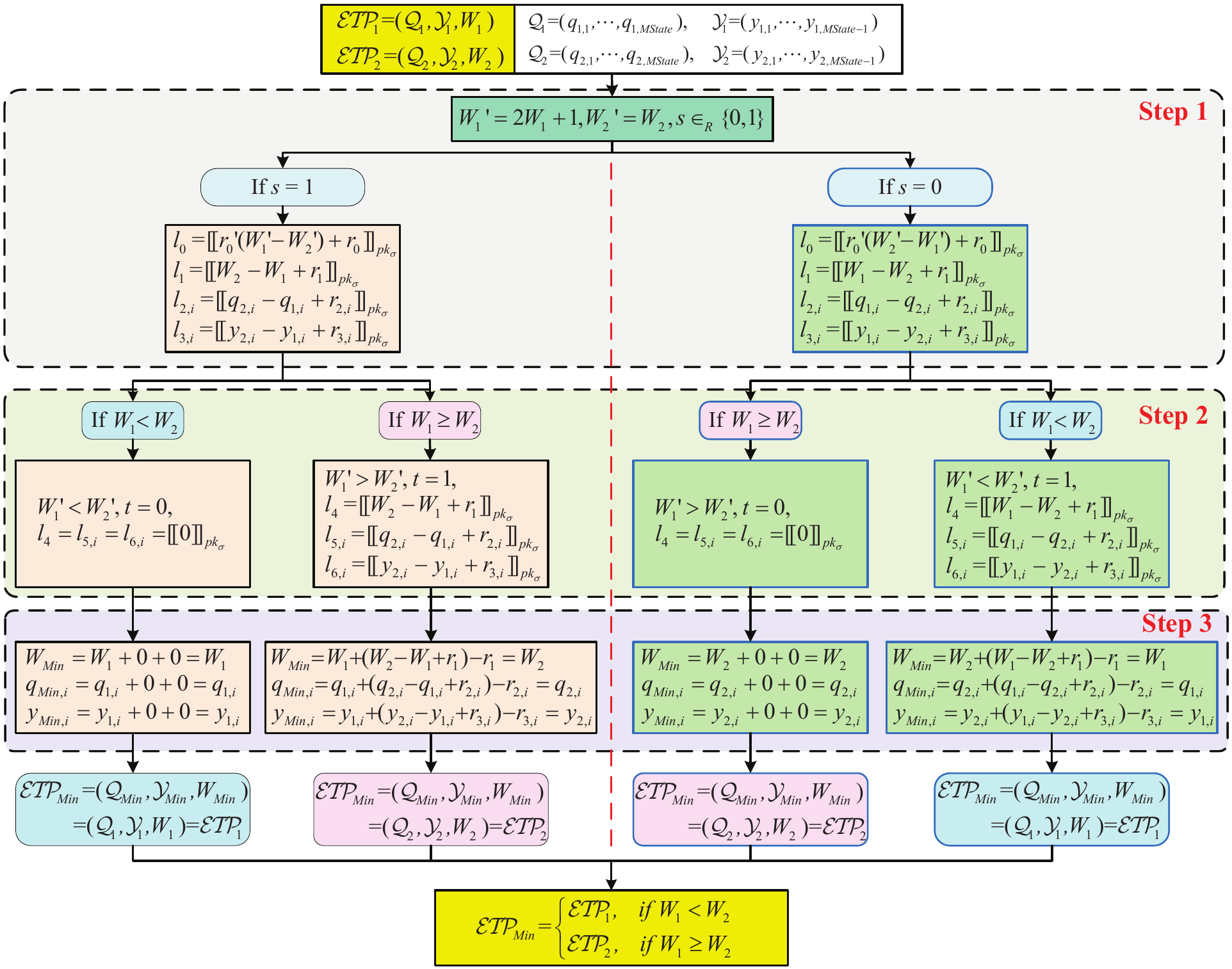}
		\caption{Workflow of \texttt{SMin}}
		\label{Fig:WorkflowSMin}
	\end{center}
\end{figure*}

\subsubsection*{C-4. Elaboration of \texttt{SMin}}
\texttt{SMin} selects the best treatment procedure (with the lowest weight) from two procedures,
and its correctness is elaborated below.

At the beginning of step 1, CP calculates
\begin{align*}
[\![W_1']\!]_{pk_\sigma}&=[\![W_1]\!]_{pk_\sigma}^2\cdot[\![1]\!]_{pk_\sigma}=[\![2W_1+1]\!]_{pk_\sigma},\\
[\![W_2']\!]_{pk_\sigma}&=[\![W_2]\!]_{pk_\sigma}^2=[\![2W_2]\!]_{pk_\sigma}.
\end{align*}
It is obvious that if $W_1<W_2$, we have $W_1'<W_2'$; if $W_1\geq W_2$, we have $W_1'>W_2'$.

(1) When the flipped random coin $s=1$, we have
\begin{align*}
l_0&=([\![W_1']\!]_{pk_\sigma})^{r_0'}\cdot([\![W_2']\!]_{pk_\sigma})^{N-r_0'}\cdot[\![r_0]\!]_{pk_\sigma}\\
   &=[\![r_0'(W_1'-W_2')+r_0]\!]_{pk_\sigma},\\\\
l_1&=[\![W_2]\!]_{pk_\sigma}\cdot([\![W_1]\!]_{pk_\sigma})^{N-1}\cdot[\![r_1]\!]_{pk_\sigma}\\
   &=[\![W_2-W_1+r_1]\!]_{pk_\sigma},\\\\
l_{2,i}&\leftarrow\texttt{SAD}([\![q_{2,i}]\!]_{pk_{A}}\cdot([\![q_{1,i}]\!]_{pk_{A}})^{N-1},[\![r_{2,i}]\!]_{pk_\sigma})\\
       &=[\![q_{2,i}-q_{1,i}+r_{2,i}]\!]_{pk_\sigma},\\\\
l_{3,i}&\leftarrow\texttt{SAD}([\![y_{2,i}]\!]_{pk_{A}}\cdot([\![y_{1,i}]\!]_{pk_{A}})^{N-1},[\![r_{3,i}]\!]_{pk_\sigma})\\
       &=[\![y_{2,i}-y_{1,i}+r_{3,i}]\!]_{pk_\sigma}.
\end{align*}

{\scriptsize$\bullet$} If $W_1<W_2$, then $W_1'<W_2'$,
$l_0''>\mathcal{L}(N)/2$, $t=0$ and
$l_4=[\![0]\!]_{pk_\sigma},l_{5,i}=[\![0]\!]_{pk_\sigma},l_{6,i}=[\![0]\!]_{pk_\sigma}.$
The tuple $([\![W_{Min}]\!]_{pk_\sigma}$, $[\![q_{Min,i}]\!]_{pk_\sigma}$,
$[\![y_{Min,i}]\!]_{pk_\sigma})$ are calculated as
\begin{align*}
[\![W_{Min}]\!]_{pk_\sigma}&=[\![W_1]\!]_{pk_\sigma}\cdot l_4\cdot([\![t]\!]_{pk_\sigma})^{N-r_1}\\
&=[\![W_1]\!]_{pk_\sigma}\cdot [\![0]\!]_{pk_\sigma}\cdot([\![0]\!]_{pk_\sigma})^{N-r_1}\\
&=[\![W_1]\!]_{pk_\sigma},\\\\
[\![q_{Min,i}]\!]_{pk_\sigma}&\leftarrow\texttt{SAD}([\![q_{1,i}]\!]_{pk_{A}},l_{5,i})\cdot([\![t]\!]_{pk_\sigma})^{N-r_{2,i}}\\
&=\texttt{SAD}([\![q_{1,i}]\!]_{pk_{A}},[\![0]\!]_{pk_\sigma})\cdot([\![0]\!]_{pk_\sigma})^{N-r_{2,i}}\\
&=[\![q_{1,i}]\!]_{pk_\sigma},\\\\
[\![y_{Min,i}]\!]_{pk_\sigma}&\leftarrow\texttt{SAD}([\![y_{1,i}]\!]_{pk_{A}},l_{6,i})\cdot([\![t]\!]_{pk_\sigma})^{N-r_{3,i}}\\
&=\texttt{SAD}([\![y_{1,i}]\!]_{pk_{A}},[\![0]\!]_{pk_\sigma})\cdot([\![0]\!]_{pk_\sigma})^{N-r_{3,i}}\\
&=[\![y_{1,i}]\!]_{pk_\sigma}.
\end{align*}

{\scriptsize$\bullet$} If $W_1\geq W_2$, then $W'_1>W'_2$, $l_0''<\mathcal{L}(N)/2$,  $t=1$
and
$l_4=\texttt{CR}(l_1),l_{5,i}=\texttt{CR}(l_{2,i}),l_{6,i}=\texttt{CR}(l_{3,i}).$
The tuple $([\![W_{Min}]\!]_{pk_\sigma}$, $[\![q_{Min,i}]\!]_{pk_\sigma}$,
$[\![y_{Min,i}]\!]_{pk_\sigma})$ is calculated as
\begin{align*}
&[\![W_{Min}]\!]_{pk_\sigma}\\
=&[\![W_1]\!]_{pk_\sigma}\cdot l_4\cdot([\![t]\!]_{pk_\sigma})^{N-r_1}\\
=&[\![W_1]\!]_{pk_\sigma}\cdot [\![W_2-W_1+r_1]\!]_{pk_\sigma}\cdot([\![1]\!]_{pk_\sigma})^{N-r_1}\\
=&[\![W_2]\!]_{pk_\sigma},\\\\
&[\![q_{Min,i}]\!]_{pk_\sigma}\\
\leftarrow&\texttt{SAD}([\![q_{1,i}]\!]_{pk_{A}},l_{5,i})\cdot([\![t]\!]_{pk_\sigma})^{N-r_{2,i}}\\
=&\texttt{SAD}([\![q_{1,i}]\!]_{pk_{A}},[\![q_{2,i}-q_{1,i}+r_{2,i}]\!]_{pk_\sigma})\cdot([\![1]\!]_{pk_\sigma})^{N-r_{2,i}}\\
=&[\![q_{2,i}]\!]_{pk_\sigma},\\\\
&[\![y_{Min,i}]\!]_{pk_\sigma}\\
\leftarrow&\texttt{SAD}([\![y_{1,i}]\!]_{pk_{A}},l_{6,i})\cdot([\![t]\!]_{pk_\sigma})^{N-r_{3,i}}\\
=&\texttt{SAD}([\![y_{1,i}]\!]_{pk_{A}},[\![y_{2,i}-y_{1,i}+r_{3,i}]\!]_{pk_\sigma})\cdot([\![1]\!]_{pk_\sigma})^{N-r_{3,i}}\\
=&[\![y_{2,i}]\!]_{pk_\sigma},
\end{align*}

(2) When the flipped random coin $s=0$, we have
\begin{align*}
l_0&=([\![W_2']\!]_{pk_\sigma})^{r_0'}\cdot([\![W_1']\!]_{pk_\sigma})^{N-r_0'}\cdot[\![r_0]\!]_{pk_\sigma}\\
   &=[\![r_0'(W_2'-W_1')+r_0]\!]_{pk_\sigma},\\\\
l_1&=[\![W_1]\!]_{pk_\sigma}\cdot([\![W_2]\!]_{pk_\sigma})^{N-1}\cdot[\![r_1]\!]_{pk_\sigma}\\
   &=[\![W_1-W_2+r_1]\!]_{pk_\sigma},\\\\
l_{2,i}&\leftarrow\texttt{SAD}([\![q_{1,i}]\!]_{pk_{A}}\cdot([\![q_{2,i}]\!]_{pk_{A}})^{N-1},[\![r_{2,i}]\!]_{pk_\sigma})\\
       &=[\![q_{1,i}-q_{2,i}+r_{2,i}]\!]_{pk_\sigma},\\\\
l_{3,i}&\leftarrow\texttt{SAD}([\![y_{1,i}]\!]_{pk_{A}}\cdot([\![y_{2,i}]\!]_{pk_{A}})^{N-1},[\![r_{3,i}]\!]_{pk_\sigma})\\
       &=[\![y_{1,i}-y_{2,i}+r_{3,i}]\!]_{pk_\sigma}.
\end{align*}

{\scriptsize$\bullet$} If $W_1< W_2$, then $W'_1<W'_2$, $l_0''<\mathcal{L}(N)/2$,  $t=1$
and
$l_4=\texttt{CR}(l_1),l_{5,i}=\texttt{CR}(l_{2,i}),l_{6,i}=\texttt{CR}(l_{3,i}).$
The tuple $([\![W_{Min}]\!]_{pk_\sigma}$, $[\![q_{Min,i}]\!]_{pk_\sigma}$,
$[\![y_{Min,i}]\!]_{pk_\sigma})$ is calculated as
\begin{align*}
&[\![W_{Min}]\!]_{pk_\sigma}\\
=&[\![W_2]\!]_{pk_\sigma}\cdot l_4\cdot([\![t]\!]_{pk_\sigma})^{N-r_1}\\
=&[\![W_2]\!]_{pk_\sigma}\cdot [\![W_1-W_2+r_1]\!]_{pk_\sigma}\cdot([\![1]\!]_{pk_\sigma})^{N-r_1}\\
=&[\![W_1]\!]_{pk_\sigma},\\\\
&[\![q_{Min,i}]\!]_{pk_\sigma}\\
\leftarrow&\texttt{SAD}([\![q_{2,i}]\!]_{pk_{A}},l_{5,i})\cdot([\![t]\!]_{pk_\sigma})^{N-r_{2,i}}\\
=&\texttt{SAD}([\![q_{2,i}]\!]_{pk_{A}},[\![q_{1,i}-q_{2,i}+r_{2,i}]\!]_{pk_\sigma})\cdot([\![1]\!]_{pk_\sigma})^{N-r_{2,i}}\\
=&[\![q_{1,i}]\!]_{pk_\sigma},\\\\
&[\![y_{Min,i}]\!]_{pk_\sigma}\\
\leftarrow&\texttt{SAD}([\![y_{2,i}]\!]_{pk_{A}},l_{6,i})\cdot([\![t]\!]_{pk_\sigma})^{N-r_{3,i}}\\
=&\texttt{SAD}([\![y_{2,i}]\!]_{pk_{A}},[\![y_{1,i}-y_{2,i}+r_{3,i}]\!]_{pk_\sigma})\cdot([\![1]\!]_{pk_\sigma})^{N-r_{3,i}}\\
=&[\![y_{1,i}]\!]_{pk_\sigma},
\end{align*}

{\scriptsize$\bullet$} If $W_1\geq W_2$, then $W_1'>W_2'$,
$l_0''>\mathcal{L}(N)/2$, $t=0$ and
$l_4=[\![0]\!]_{pk_\sigma},l_{5,i}=[\![0]\!]_{pk_\sigma},l_{6,i}=[\![0]\!]_{pk_\sigma}.$
The tuple $([\![W_{Min}]\!]_{pk_\sigma}$, $[\![q_{Min,i}]\!]_{pk_\sigma}$,
$[\![y_{Min,i}]\!]_{pk_\sigma})$ are calculated as
\begin{align*}
[\![W_{Min}]\!]_{pk_\sigma}&=[\![W_2]\!]_{pk_\sigma}\cdot l_4\cdot([\![t]\!]_{pk_\sigma})^{N-r_1}\\
&=[\![W_2]\!]_{pk_\sigma}\cdot [\![0]\!]_{pk_\sigma}\cdot([\![0]\!]_{pk_\sigma})^{N-r_1}\\
&=[\![W_2]\!]_{pk_\sigma},\\\\
[\![q_{Min,i}]\!]_{pk_\sigma}&\leftarrow\texttt{SAD}([\![q_{2,i}]\!]_{pk_{A}},l_{5,i})\cdot([\![t]\!]_{pk_\sigma})^{N-r_{2,i}}\\
&=\texttt{SAD}([\![q_{2,i}]\!]_{pk_{A}},[\![0]\!]_{pk_\sigma})\cdot([\![0]\!]_{pk_\sigma})^{N-r_{2,i}}\\
&=[\![q_{2,i}]\!]_{pk_\sigma},\\\\
[\![y_{Min,i}]\!]_{pk_\sigma}
&\leftarrow\texttt{SAD}([\![y_{2,i}]\!]_{pk_{A}},l_{6,i})\cdot([\![t]\!]_{pk_\sigma})^{N-r_{3,i}}\\
&=\texttt{SAD}([\![y_{2,i}]\!]_{pk_{A}},[\![0]\!]_{pk_\sigma})\cdot([\![0]\!]_{pk_\sigma})^{N-r_{3,i}}\\
&=[\![y_{2,i}]\!]_{pk_\sigma}.
\end{align*}

Since the above analysis includes several case by case discussions (``$s=1$ vs. $s=0$", and ``$W_1<W_2$ vs. $W_1\geq W_2$"), we depict different situations in Figure \ref{Fig:WorkflowSMin} to make them easy to understand. 
The above analysis demonstrates that \texttt{SMin}
could correctly output $\mathcal{ETP}_{Min}$ (with the lowest weight)
from the two encrypted treatment procedures.\\

\begin{figure*}[htp]
	\begin{center}
		\includegraphics[width=6.5in]{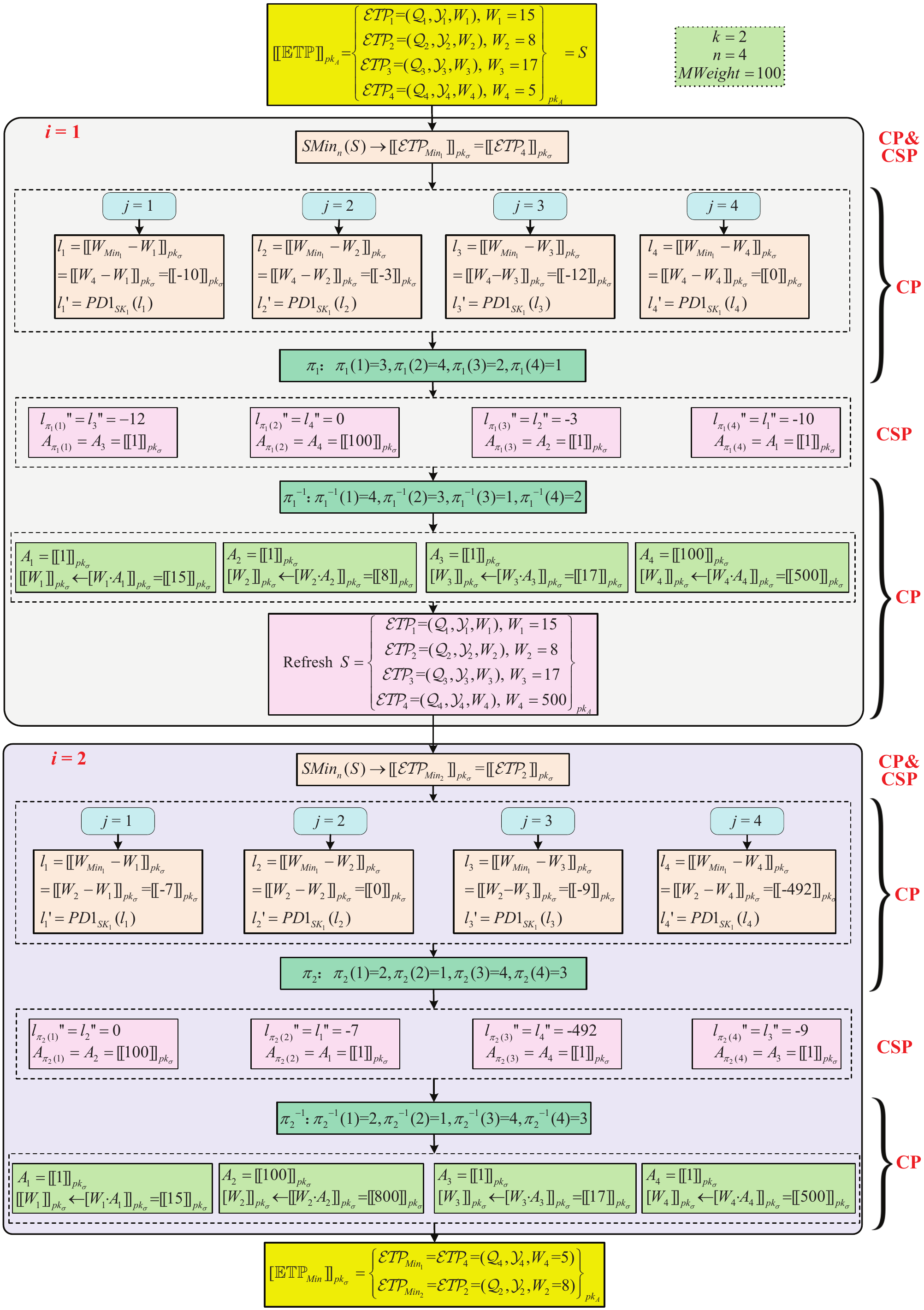}
		\caption{A Toy Example to Illustrate Workflow of \texttt{BPS}-$k$}
		\label{Fig:WorkflowBPSK}
	\end{center}
\end{figure*}

\subsubsection*{C-5. Elaboration of \texttt{BPS-$k$}}
\texttt{BPS-$k$} selects the top-$k$ most recommended treatment procedures (with the top-$k$ lowest weights) in a privacy-preserving way, which is elaborated below.
\begin{enumerate}
  \item Line 1. The set $S$ is assigned with $[\![\mathbb{ETP}]\!]_{pk_A}$.
  \item Line 2. \texttt{BPS-$k$} needs $k$ rounds to get the
result. In each round, the protocol picks up the expanded treatment procedure with the lowest weight.
  \item Line 3. Run $\texttt{SMin}_n$ to get
  the best encrypted treatment procedure $[\![\mathcal{ETP}_{Min_i}]\!]_{pk_\sigma}$ (with the lowest weight) in the $i$-th round.

  \item Line 4-5. For each element in $S$, CP calculates
  \begin{align*}
  l_j&=([\![W_{Min_i}]\!]_{pk_\sigma})^{r_j}\cdot([\![W_{j}]\!]_{pk_\sigma})^{N-r_j}\\
  &=[\![r_j(W_{Min_i}-W_j)]\!]_{pk_\sigma}.
\end{align*}
If $W_j=W_{Min_i}$, we have $l_j=[\![0]\!]_{pk_\sigma}$. Otherwise,
$l_j\neq[\![0]\!]_{pk_\sigma}$. Then, CP partially decrypts $l_j$ and
stores the result in $l_j'$.

  \item Line 6. CP uses
  a permutation $\pi_i$ to disorder $(l_1,\cdots,l_n)$ and $(l_1',\cdots,l_n')$.
  Then, $\{(l_{\pi_i(j)},l_{\pi_i(j)}')\}$ for $1\leq j \leq n$ are sends to CSP.

  \item Line 7-8. CSP decrypts $(l_{\pi_i(j)},l_{\pi_i(j)}')$
  to get $l_{\pi_i(j)}''$ for $1\leq j \leq n$. If $l_{\pi_i(j)}''=0$, CSP sets
  $A_{\pi_i(j)}= [\![MWeight]\!]_{pk_\sigma}$.
  Otherwise, $A_{\pi_i(j)}= [\![1]\!]_{pk_\sigma}$.

  \item Line 9. Receiving $(A_{\pi_i(1)},\cdots,A_{\pi_i(n)})$, CP utilizes
  the inverse permutation function $\pi_i^{-1}$ to recover
  the order and obtains $(A_1,\cdots,A_n)$.
  \begin{itemize}
    \item It easy to find that
  the origin $[\![\mathcal{ETP}_{\zeta}]\!]_{pk_A}$ of $[\![\mathcal{ETP}_{Min_i}]\!]_{pk_\sigma}$
  has $A_\zeta = [\![MWeight]\!]_{pk_\sigma}$.
    \item For $1\leq j \leq n$ and $j\neq \zeta$, $[\![\mathcal{ETP}_{j}]\!]_{pk_A}$ has $A_j=[\![1]\!]_{pk_\sigma}$.
  \end{itemize}

  \item Line 10. The set $([\![W_1]\!]_{pk_\sigma},\cdots,[\![W_n]\!]_{pk_\sigma})$ in $S$ is
  refreshed.
  \begin{itemize}
    \item The element $[\![W_\zeta]\!]_{pk_\sigma}$
  in the origin tuple $[\![\mathcal{ETP}_{\zeta}]\!]_{pk_A}$ of $[\![\mathcal{ETP}_{Min_i}]\!]_{pk_\sigma}$
  is set to $[\![W_\zeta\cdot MWeight]\!]_{pk_\sigma}$.
    \item For $1\leq j \leq  n$ and $j\neq \zeta$, $[\![W_j]\!]_{pk_\sigma}$ is unchanged since
  $[\![W_j]\!]_{pk_\sigma}\leftarrow\texttt{SMD}([\![W_j]\!]_{pk_\sigma}, A_j)
  =\texttt{SMD}([\![W_j]\!]_{pk_\sigma},[\![1]\!]_{pk_\sigma})
  =[\![W_j\times1]\!]_{pk_\sigma}=[\![W_j]\!]_{pk_\sigma}.$\\
  \end{itemize}
\end{enumerate}

We give an example in Figure \ref{Fig:WorkflowBPSK} to describe the working process of \texttt{BPS}-$k$.
Suppose the expanded treatment procedure set $[\![\mathbb{ETP}]\!]_{pk_A}$ contains four ($n=4$) procedures $[\![\mathcal{ETP}_i]\!]_{pk_A}=([\![\mathcal{Q}_i]\!]_{pk_A},[\![\mathcal{Y}_i]\!]_{pk_A},[\![W_i]\!]_{pk_A})$, where $i\in\{1,2,3,4\}$, $W_1=15$, $W_2=8$, $W_3=17$ and $W_4=5$. Let $k=2$ to select the top-2 best treatment procedures. Set the maximum weight $MWeight=100$.

\begin{enumerate}
	\item To select the first best treatment procedure ($i=1$), CP and CSP executes line 3-10 in \texttt{BPS}-$k$.
	
	\begin{itemize}
		\item CSP and CP invoke \texttt{SMin}$_n$ protocol to get $[\![\mathcal{ETP}_{Min_1}]\!]_{pk_{\sigma}}$ in a privacy preserving way, which is $[\![\mathcal{ETP}_4]\!]_{pk_{\sigma}}$ in our example.
		
		\item Since $n=4$, CP calculates four tuples of $(l_j,l_j')$ for $j=\{1,2,3,4\}$:
		\begin{align*}
			l_1&=[\![W_{Min_1}-W_1]\!]_{pk_{\sigma}}=[\![-10]\!]_{pk_{\sigma}},\\
			l_2&=[\![W_{Min_1}-W_2]\!]_{pk_{\sigma}}=[\![-3]\!]_{pk_{\sigma}},\\
			l_3&=[\![W_{Min_1}-W_3]\!]_{pk_{\sigma}}=[\![-12]\!]_{pk_{\sigma}},\\
			l_4&=[\![W_{Min_1}-W_4]\!]_{pk_{\sigma}}=[\![0]\!]_{pk_{\sigma}}.
		\end{align*}
		and $l_j'=\texttt{PK1}_{SK_1}(l_j)$.
		
		\item CP chooses a permutation function $\pi_1:$ $\pi_1(1)=3,~\pi_1(2)=4,~\pi_1(3)=2,~\pi_1(4)=1$ to scramble the above four tuples, which are sent to CSP.
		
		\item Receiving the disordered tuples,CSP decrypts $l_{\pi_1(1)}''=-12$, $l_{\pi_1(2)}''=0$, $l_{\pi_1(3)}''=-3$, $l_{\pi_1(4)}''=-10$. According to the rule in line 8, CSP computes $A_{\pi_1(1)}=[\![1]\!]_{pk_{\sigma}}$,
		$A_{\pi_1(2)}=[\![100]\!]_{pk_{\sigma}}$,
		$A_{\pi_1(3)}=[\![1]\!]_{pk_{\sigma}}$,
		$A_{\pi_1(4)}=[\![1]\!]_{pk_{\sigma}}$ and sends them to CP.
		
		\item The restore the order, CP using $\pi_1^{-1}:$ $\pi_1^{-1}(1)=4,~\pi_1^{-1}(2)=3,~\pi_1^{-1}(3)=1,~\pi_1^{-1}(4)=2$ to recover $A_1=[\![1]\!]_{pk_{\sigma}}$,
		$A_2=[\![1]\!]_{pk_{\sigma}}$,
		$A_3=[\![1]\!]_{pk_{\sigma}}$,
		$A_4=[\![100]\!]_{pk_{\sigma}}$.
		
		\item Then, CP calculates 
		\begin{align*}
			W_1&\leftarrow W_1\cdot A_1=15,\\
			W_2&\leftarrow W_2\cdot A_2=8,\\
			W_3&\leftarrow W_3\cdot A_3=17,\\
			W_4&\leftarrow W_4\cdot A_4=500.
		\end{align*}
		It is obvious that the weights $(W_1,W_2,W_3)$ remain the same; and the weight of $[\![\mathcal{ETP}_{Min_1}]\!]_{pk_{\sigma}}=[\![\mathcal{ETP}_4]\!]_{pk_{\sigma}}$ is set to be the largest one. The weights in set $S$ (containing $[\![\mathcal{ETP}]\!]_{pk_{\sigma}}$ ) are refreshed to the new ones.				
	\end{itemize}

	\item To select the second best treatment procedure ($i=2$), CP and CSP executes line 3-10 in \texttt{BPS}-$k$.

	\begin{itemize}
		\item CSP and CP invoke \texttt{SMin}$_n$ protocol to get $[\![\mathcal{ETP}_{Min_2}]\!]_{pk_{\sigma}}$ in a privacy preserving way, which is $[\![\mathcal{ETP}_2]\!]_{pk_{\sigma}}$ in our example.
		
		\item Since $n=4$, CP calculates four tuples of $(l_j,l_j')$ for $j=\{1,2,3,4\}$:
		\begin{align*}
		l_1&=[\![W_{Min_1}-W_1]\!]_{pk_{\sigma}}=[\![-7]\!]_{pk_{\sigma}},\\
		l_2&=[\![W_{Min_1}-W_2]\!]_{pk_{\sigma}}=[\![0]\!]_{pk_{\sigma}},\\
		l_3&=[\![W_{Min_1}-W_3]\!]_{pk_{\sigma}}=[\![-9]\!]_{pk_{\sigma}},\\
		l_4&=[\![W_{Min_1}-W_4]\!]_{pk_{\sigma}}=[\![-492]\!]_{pk_{\sigma}}.
		\end{align*}
		and $l_j'=\texttt{PK1}_{SK_1}(l_j)$.
		
		\item CP chooses a permutation function $\pi_2:$ $\pi_2(1)=2,~\pi_2(2)=1,~\pi_2(3)=4,~\pi_2(4)=3$ to scramble the above four tuples, which are sent to CSP.
		
		\item Receiving the disordered tuples,CSP decrypts $l_{\pi_2(1)}''=-7$, $l_{\pi_2(2)}''=0$, $l_{\pi_2(3)}''=-9$, $l_{\pi_2(4)}''=-492$. According to the rule in line 8, CSP computes $A_{\pi_2(1)}=[\![100]\!]_{pk_{\sigma}}$,
		$A_{\pi_2(2)}=[\![1]\!]_{pk_{\sigma}}$,
		$A_{\pi_2(3)}=[\![1]\!]_{pk_{\sigma}}$,
		$A_{\pi_2(4)}=[\![1]\!]_{pk_{\sigma}}$ and sends them to CP.
		
		\item The restore the order, CP using $\pi_2^{-1}:$ $\pi_2^{-1}(1)=2,~\pi_2^{-1}(2)=1,~\pi_2^{-1}(3)=4,~\pi_2^{-1}(4)=3$ to recover $A_1=[\![1]\!]_{pk_{\sigma}}$,
		$A_2=[\![100]\!]_{pk_{\sigma}}$,
		$A_3=[\![1]\!]_{pk_{\sigma}}$,
		$A_4=[\![1]\!]_{pk_{\sigma}}$.
		
		\item Then, CP calculates 
		\begin{align*}
		W_1&\leftarrow W_1\cdot A_1=15,\\
		W_2&\leftarrow W_2\cdot A_2=800,\\
		W_3&\leftarrow W_3\cdot A_3=17,\\
		W_4&\leftarrow W_4\cdot A_4=500.
		\end{align*}
		It is obvious that the weights $(W_1,W_2,W_3)$ remain the same; and the weight of $[\![\mathcal{ETP}_{Min_2}]\!]_{pk_{\sigma}}=[\![\mathcal{ETP}_2]\!]_{pk_{\sigma}}$ is set to be the largest one. The weights in set $S$ (containing $[\![\mathcal{ETP}]\!]_{pk_{\sigma}}$ ) are refreshed to the new ones.				
	\end{itemize}
		
	\item Finally, the \texttt{BPS}-$k$ protocol outputs $[\![\mathcal{ETP}_{Min}]\!]_{pk_{\sigma}}=\{[\![\mathcal{ETP}_4]\!]_{pk_{\sigma}},[\![\mathcal{ETP}_2]\!]_{pk_{\sigma}}\}$ ($W_4=5$ and $W_2=8$).	
\end{enumerate}

\subsection*{D. Protocol Proof}

\subsubsection*{D-1. Proof of \texttt{SMin}}

\textbf{Theorem 4-1.} \emph{\texttt{SMin} is secure in the presence of semi-honest (non-colluding) attackers
$\mathcal{A}=(\mathcal{A}_{D_1},\mathcal{A}_{S_1},\mathcal{A}_{S_2})$.}

\emph{\textbf{Proof}}. We now construct the following independent
simulators $(Sim_{D_1},Sim_{S_1},Sim_{S_2})$.

$Sim_{D_1}$ receives $\mathcal{ETP}_1$ and $\mathcal{ETP}_2$ as
inputs and simulates $\mathcal{A}_{D_1}$ as following, where
\begin{align*}
\mathcal{ETP}_i&=(\mathcal{Q}_i, \mathcal{Y}_i, W_i),\\
\mathcal{Q}_i&=(q_0, q_{i,\theta_1},\cdots,q_{i,MState-1}),\\
\mathcal{Y}_i&=(y_{i,\theta_1},\cdots,y_{i,MState-1}),
\end{align*}
for $i\in\{1,2\}$.

It generates the ciphertext $[\![\mathcal{ETP}_1]\!]_{pk_A}$ and
$[\![\mathcal{ETP}_2]\!]_{pk_A}$, where
\begin{align*}
[\![\mathcal{ETP}_i]\!]_{pk_A}&=([\![\mathcal{Q}_i]\!]_{pk_{A}}, [\![\mathcal{Y}_i]\!]_{pk_{A}}, [\![W_i]\!]_{pk_\sigma}),\\
[\![\mathcal{Q}_i]\!]_{pk_{A}}&=([\![q_0]\!]_{pk_{A}}, [\![q_{i,\theta_1}]\!]_{pk_{A}},\cdots,[\![q_{i,MState-1}]\!]_{pk_{A}}),\\
[\![\mathcal{Y}_i]\!]_{pk_{A}}&=([\![y_{i,\theta_1}]\!]_{pk_{A}},\cdots,[\![y_{i,MState-1}]\!]_{pk_{A}}),
\end{align*}
for $i\in\{1,2\}$.

The entire view
of $\mathcal{A}_{D_1}$ is the received tuples and the encrypted
data. The IND-CPA security of PCTD ensures that
$\mathcal{A}_{D_1}$'s view is indistinguishable from its view in the
real world execution.

$Sim_{S_1}$ simulates $\mathcal{A}_{S_1}$ as following. It randomly
selects $\hat{\mathcal{ETP}_1}$ and
$\hat{\mathcal{ETP}_2}$ and encrypts them to
$[\![\hat{\mathcal{ETP}_1}]\!]_{pk_A}$ and
$[\![\hat{\mathcal{ETP}_2}]\!]_{pk_A}$. Then, it computes
$[\![\hat{W}_1]\!]_{pk_\sigma}=([\![\hat{W}_1]\!]_{pk_\sigma})^2\cdot[\![1]\!]_{pk_\sigma}$,
$[\![\hat{W}_2]\!]_{pk_\sigma}=([\![\hat{W}_2]\!]_{pk_\sigma})^2$.
$Sim_{S_1}$ flips a random coin $s\in\{0,1\}$
and computes
$\hat{l}_0,\hat{l}_1,\hat{l}_{2,i}, \hat{l}_{3,i}$.
Then,
utilizing \texttt{PD1} algorithm, it computes $\hat{l}_0'$.
It randomly selects
$\hat{t}\in\{0,1\}$ and computes
$[\![\hat{t}]\!]_{pk_\sigma}$. It generates
random ciphertext
$\hat{l}_4,\hat{l}_{5,i},\hat{l}_{6,i}$.
According to the randomly flipped coins $s\in\{0,1\}$,
it inputs
$([\![\hat{q}_{1,i}]\!]_{pk_A},[\![\hat{q}_{2,i}]\!]_{pk_A})$,
$([\![\hat{y}_{1,i}]\!]_{pk_A},[\![\hat{y}_{2,i}]\!]_{pk_A})$,
$l_{5,i}$ and $l_{6,i}$
into $Sim_{S_1}^\texttt{SAD}$ in
Ref. \cite{LiuX_TIFS16} and gets $[\![\hat{q}_{Min,i}]\!]_{pk_\sigma},[\![\hat{y}_{Min,i}]\!]_{pk_\sigma}$.
Then, $Sim_{S_1}$ sends
$(\hat{l}_0',\hat{l}_0,\hat{l}_1,\hat{l}_{2,i}, \hat{l}_{3,i},\hat{l}_4,\hat{l}_{5,i},\hat{l}_{6,i},[\![\hat{t}]\!]_{pk_\sigma})$ and the intermediate
encrypted data of $Sim_{S_1}^\texttt{SAD}$ to
$\mathcal{A}_{S_1}$. If $\mathcal{A}_{S_1}$ replies with $\bot$,
then $Sim_{S_1}$ outputs $\bot$. The IND-CPA security of PCTD
ensures that $\mathcal{A}_{D_1}$'s view is indistinguishable from
its view in the real world execution.

$Sim_{S_2}$ simulates $\mathcal{A}_{S_2}$ as following.
It selects random $\hat{t}\in\{0,1\}$. If
$\hat{t}=0$, it calculates $\hat{l}_4=[\![0]\!]_{pk_\sigma}$,
$\hat{C}_{5,i}=[\![0]\!]_{pk_\sigma}$, $\hat{l}_{6,i}=[\![0]\!]_{pk_\sigma}$. If
$\hat{t}=1$, it generates random encryptions to be
$(\hat{l}_4,\hat{l}_{5,i},\hat{l}_{6,i})$. For a certain $\hat{t}$, the
generated ciphertexts are computationally indistinguishable from the
real world due to the IND-CPA security of PCTD. In real and
ideal world, the views of $\mathcal{A}_{S_2}$ are indistinguishable.
\hfill$\square$\vspace{2mm}

\textbf{Theorem 4-2.} \emph{\texttt{SMin} is secure against the
 adversary $\mathcal{A}^*$ defined in the attack model.}

\emph{\textbf{Proof}}. The adversary $\mathcal{A}^*$ is assumed to have the following abilities.

\begin{enumerate}
  \item [(1)] $\mathcal{A}^*$ is assumed to be an outside adversary and eavesdrop
  all the communications to get the transmitted data. As $\mathcal{A}^*$ is assumed to be an outside adversary, $\mathcal{A}^*$ cannot get hospital $A$'s secret key
  $sk_A$, patient $B$'s secret key $sk_B$ and $B$'s authorization secret key $sk_\sigma$.
  $\mathcal{A}^*$ also cannot get CP's partial strong key $SK_1$ and CSP's partial strong key $SK_2$.

  If $\mathcal{A}^*$ eavesdrops the communication channel, $\mathcal{A}^*$ could get the encrypted tuples $[\![\mathcal{ETP}_1]\!]_{pk_{A}}$ and $[\![\mathcal{ETP}_2]\!]_{pk_{A}}$ that
  are transmitted at the beginning of the \texttt{SMin} protocol,
  and $[\![\mathcal{ETP}_{Min}]\!]_{pk_\sigma}$ that is transmitted at the end of the protocol.
  Since $[\![\mathcal{ETP}_1]\!]_{pk_{A}}$, $[\![\mathcal{ETP}_2]\!]_{pk_{A}}$ and $[\![\mathcal{ETP}_{Min}]\!]_{pk_\sigma}$
   are encrypted using PCTD,
  the adversary $\mathcal{A}^*$ cannot recover $(\mathcal{Q}_1, \mathcal{Y}_1, W_1)$,
   $(\mathcal{Q}_2, \mathcal{Y}_2, W_2)$ and $(\mathcal{Q}_{Min}, \mathcal{Y}_{Min}, W_{Min})$ due to the IND-CPA security of PCTD.

  If $\mathcal{A}^*$ eavesdrops the communication channel between CP and
  CSP, $\mathcal{A}^*$ could get $(l_0',l_0,l_1,l_{2,i},l_{3,i})$ in the end of step 1,
  and $([\![t]\!]_{pk_\sigma},l_4,l_{5,i},l_{6,i})$ at the end of step 2.
  In \texttt{SMin}, these ciphertexts
  $(l_0',l_0,l_1,l_{2,i},l_{3,i},[\![t]\!]_{pk_\sigma},l_4,l_{5,i},l_{6,i})$ are all
  encrypted using the public key $pk_\sigma$.
Since the adversary $\mathcal{A}^*$ does know the data owner $B$'s authorization secret key $sk_\sigma$ and
the CSP's partial strong key $SK_2$, $\mathcal{A}^*$ cannot recover the underlying plaintexts.

  \item [(2)] $\mathcal{A}^*$ is assumed to compromise CP and get CP's partial strong key $SK_1$.
  But $\mathcal{A}^*$ cannot get CSP's partial strong key $SK_2$.
  $\mathcal{A}^*$ also cannot get hospital $A$'s secret key
  $sk_A$, patient $B$'s secret key $sk_B$ and $B$'s authorization secret key $sk_\sigma$.

  In step 1 of \texttt{SMin}, $\mathcal{A}^*$ obtains $[\![\mathcal{ETP}_1]\!]_{pk_{A}}$ and $[\![\mathcal{ETP}_2]\!]_{pk_{A}}$.
  $\mathcal{A}^*$ cannot recover $(\mathcal{Q}_1, \mathcal{Y}_1, W_1)$ and
   $(\mathcal{Q}_2, \mathcal{Y}_2, W_2)$  without the secret keys
  $sk_\sigma,sk_{A}$. In step 3, $\mathcal{A}^*$ obtains $([\![t]\!]_{pk_\sigma},l_4,l_{5,i},l_{6,i})$ from CSP.

  If $s=1$,
  \begin{eqnarray*}
  &&[\![W_{Min}]\!]_{pk_\sigma}\\
  &=&[\![W_1]\!]_{pk_\sigma}\cdot l_4\cdot([\![t]\!]_{pk_\sigma})^{N-r_1}\\
  &=&\left\{
  \begin{aligned}
  &[\![W_1+(W_2-W_1+r_1)-r_1]\!]_{pk_\sigma},  & if~t =1,\\
  &[\![W_1+0-0]\!]_{pk_\sigma},  & if~t =0,
  \end{aligned}
  \right.\\
  &=&\left\{
  \begin{aligned}
  &[\![W_2]\!]_{pk_\sigma},  & ~~~~~~~~~~~~~~~~~~~~~~~~~~~~~if~t =1,\\
  &[\![W_1]\!]_{pk_\sigma},  & if~t =0,\\
  \end{aligned}
  \right.\\\\\\
  &&[\![q_{Min,i}]\!]_{pk_\sigma} \\
  &\leftarrow&\texttt{SAD}([\![q_{1,i}]\!]_{pk_{A}},l_{5,i})\cdot([\![t]\!]_{pk_\sigma})^{N-r_{2,i}}\\
  &=&\left\{
  \begin{aligned}
  &[\![q_{1,i}+(q_{2,i}-q_{1,i}+r_{2,i})-r_{2,i}]\!]_{pk_\sigma},\\
    & ~~~~~~~~~~~~~~~~~~~~~~~~~~~~~~~~~~~~~~~~~~if~t =1,\\
  &[\![q_{1,i}+0-0]\!]_{pk_\sigma},~~~~~~~~~~~~~~~~~~~~~if~t =0,
  \end{aligned}
  \right.\\\\
  &=&\left\{
  \begin{aligned}
  &[\![q_{2,i}]\!]_{pk_\sigma},  & ~~~~~~~~~~~~~~~~~~~~~~~~~~~~~if~t =1,\\
  &[\![q_{1,i}]\!]_{pk_\sigma},  & if~t =0,
  \end{aligned}
  \right.\\\\\\
  &&[\![y_{Min,i}]\!]_{pk_\sigma} \\
  &\leftarrow&\texttt{SAD}([\![y_{1,i}]\!]_{pk_{A}},l_{6,i})\cdot([\![t]\!]_{pk_\sigma})^{N-r_{3,i}}\\
  &=&\left\{
  \begin{aligned}
  &[\![y_{1,i}+(y_{2,i}-y_{1,i}+r_{3,i})-r_{3,i}]\!]_{pk_\sigma},\\
    & ~~~~~~~~~~~~~~~~~~~~~~~~~~~~~~~~~~~~~~~~~~if~t =1,\\
  &[\![y_{1,i}+0-0]\!]_{pk_\sigma},~~~~~~~~~~~~~~~~~~~~if~t =0,
  \end{aligned}
  \right.\\\\
  &=&\left\{
  \begin{aligned}
  &[\![y_{2,i}]\!]_{pk_\sigma},  & ~~~~~~~~~~~~~~~~~~~~~~~~~~~~if~t =1,\\
  &[\![y_{1,i}]\!]_{pk_\sigma},  & if~t =0,
  \end{aligned}
  \right.\\
  \end{eqnarray*}

If $s=0$,
  \begin{eqnarray*}
  &&[\![W_{Min}]\!]_{pk_\sigma}\\
  &=&[\![W_2]\!]_{pk_\sigma}\cdot l_4\cdot([\![t]\!]_{pk_\sigma})^{N-r_1}\\
  &=&\left\{
  \begin{aligned}
  &[\![W_2+(W_1-W_2+r_1)-r_1]\!]_{pk_\sigma},  & if~t =1,\\
  &[\![W_2+0-0]\!]_{pk_\sigma},  & if~t =0,
  \end{aligned}
  \right.\\
  &=&\left\{
  \begin{aligned}
  &[\![W_1]\!]_{pk_\sigma},  & ~~~~~~~~~~~~~~~~~~~~~~~~~~~~~if~t =1,\\
  &[\![W_2]\!]_{pk_\sigma},  & if~t =0,\\
  \end{aligned}
  \right.\\\\\\
  &&[\![q_{Min,i}]\!]_{pk_\sigma} \\
  &\leftarrow&\texttt{SAD}([\![q_{2,i}]\!]_{pk_{A}},l_{5,i})\cdot([\![t]\!]_{pk_\sigma})^{N-r_{2,i}}\\
  &=&\left\{
  \begin{aligned}
  &[\![q_{2,i}+(q_{1,i}-q_{2,i}+r_{2,i})-r_{2,i}]\!]_{pk_\sigma},\\
    & ~~~~~~~~~~~~~~~~~~~~~~~~~~~~~~~~~~~~~~~~~~if~t =1,\\
  &[\![q_{2,i}+0-0]\!]_{pk_\sigma},~~~~~~~~~~~~~~~~~~~~~if~t =0,
  \end{aligned}
  \right.\\\\
  &=&\left\{
  \begin{aligned}
  &[\![q_{1,i}]\!]_{pk_\sigma},  & ~~~~~~~~~~~~~~~~~~~~~~~~~~~~~if~t =1,\\
  &[\![q_{2,i}]\!]_{pk_\sigma},  & if~t =0,
  \end{aligned}
  \right.\\\\\\
  &&[\![y_{Min,i}]\!]_{pk_\sigma} \\
  &\leftarrow&\texttt{SAD}([\![y_{2,i}]\!]_{pk_{A}},l_{6,i})\cdot([\![t]\!]_{pk_\sigma})^{N-r_{3,i}}\\
  &=&\left\{
  \begin{aligned}
  &[\![y_{2,i}+(y_{1,i}-y_{2,i}+r_{3,i})-r_{3,i}]\!]_{pk_\sigma},\\
    & ~~~~~~~~~~~~~~~~~~~~~~~~~~~~~~~~~~~~~~~~~~if~t =1,\\
  &[\![y_{2,i}+0-0]\!]_{pk_\sigma},~~~~~~~~~~~~~~~~~~~~if~t =0,
  \end{aligned}
  \right.\\\\
  &=&\left\{
  \begin{aligned}
  &[\![y_{1,i}]\!]_{pk_\sigma},  & ~~~~~~~~~~~~~~~~~~~~~~~~~~~~if~t =1,\\
  &[\![y_{2,i}]\!]_{pk_\sigma},  & if~t =0,
  \end{aligned}
  \right.\\
  \end{eqnarray*}

  Since $t$ is unknown to CP, $\mathcal{A}^*$ cannot decide the result $[\![\mathcal{ETP}_{Min}]\!]_{pk_\sigma}$
  comes from $[\![\mathcal{ETP}_1]\!]_{pk_{A}}$ or $[\![\mathcal{ETP}_2]\!]_{pk_{A}}$.\\

  \item [(3)] $\mathcal{A}^*$ is assumed to compromise CSP and get CSP's partial strong key $SK_2$.
  But $\mathcal{A}^*$ cannot get CP's partial strong key $SK_1$.
  $\mathcal{A}^*$ also cannot get hospital $A$'s secret key
  $sk_A$, patient $B$'s secret key $sk_B$ and $B$'s authorization secret key $sk_\sigma$.

  In step 2 of the \texttt{SMin} protocol, $\mathcal{A}^*$ obtains
  $(l_0',l_0,l_1,l_{2,i}, l_{3,i})$ transmitted by CP.
  Since $\mathcal{A}^*$ knows CSP's partial strong key $SK_2$,
  $\mathcal{A}^*$ decrypts $l_0''=\texttt{PD2}_{SK_2}(l_0,l_0')$.
  If $\mathcal{L}(l_0'')<\mathcal{L}(N)/2$, CSP sets $t=0$ and computes
$$l_4=[\![0]\!]_{pk_\sigma},l_{5,i}=[\![0]\!]_{pk_\sigma},l_{6,i}=[\![0]\!]_{pk_\sigma}.$$

If $\mathcal{L}(l_0'')>\mathcal{L}(N)/2$, CSP sets $t=1$ and computes
$$l_4=\texttt{CR}(l_1),l_{5,i}=\texttt{CR}(l_{2,i}),l_{6,i}=\texttt{CR}(l_{3,i}).$$

  Since $\mathcal{A}^*$ can get the plaintext $l_0''$,
  $\mathcal{A}^*$ cannot deduce the relationship of size between $W_1$ and $W_2$. The reason is explained below.

  In step 1, CP flips a random coins $s\in\{0,1\}$ and calculates
  $(l_0,l_1,l_{2,i}, l_{3,i})$ according to $s$.
 The adversary $\mathcal{A}^*$ gets
  \begin{eqnarray*}
  l_0 &=&
  \left\{
\begin{aligned}
&[\![r_0'(W_1'-W_2')+r_0]\!]_{pk_\sigma},  & if~s =1,\\
&[\![r_0'(W_2'-W_1')+r_0]\!]_{pk_\sigma}, & if~s =0,
\end{aligned}
\right.\\\\\\
    l_1 &=&
  \left\{
\begin{aligned}
&[\![W_2-W_1+r_1]\!]_{pk_\sigma}, & if~s =1,\\
&[\![W_1-W_2+r_1]\!]_{pk_\sigma}, & if~s =0,
\end{aligned}
\right.\\\\\\
l_{2,i} &=&
  \left\{
\begin{aligned}
&[\![q_{2,i}-q_{1,i}+r_{2,i}]\!]_{pk_\sigma}, & if~s =1,\\
&[\![q_{1,i}-q_{2,i}+r_{2,i}]\!]_{pk_\sigma}, & if~s =0,
\end{aligned}
\right.\\\\\\
l_{3,i} &=&
  \left\{
\begin{aligned}
&[\![y_{2,i}-y_{1,i}+r_{3,i}]\!]_{pk_\sigma}, & if~s =1,\\
&[\![y_{1,i}-y_{2,i}+r_{3,i}]\!]_{pk_\sigma}, & if~s =0,
\end{aligned}
\right.
\end{eqnarray*}

  Due to the randomness of $s$, $\mathcal{A}^*$ cannot deduce the
  relationship of size between $W_1$ and $W_2$.\\

   \item [(4)] $\cal A^*$ is assumed to be a set of collude malicious patients $(B_1,\cdots,B_n)$ (except the
  challenge patient $B^*$), and $\cal A^*$ gets their secret keys $(sk_{B_1},\cdots,sk_{B_n})$.
  $\cal A^*$ wants to get the data that belongs to
  the challenge patient $B^*$. Suppose the returned result is $[\![\mathcal{ETP}_{Min}]\!]_{pk_{\Sigma^*}}=([\![\mathcal{Q}_{Min}]\!]_{pk_{\Sigma^*}}, [\![\mathcal{Y}_{Min}]\!]_{pk_{\Sigma^*}}, [\![W_{Min}]\!]_{pk_{\Sigma^*}})$,
  where $pk_{\Sigma^*}$ is the authorize public key from hospital $A$ to challenge patient $B^*$.
  Since the patient's secret keys are independently generated, the adversary $\cal A^*$
  cannot utilize $(sk_{B_1},\cdots,sk_{B_n})$ to deduce the challenge patient $B^*$'s secret key $sk_{B^*}$.
  $\cal A^*$ also cannot get the authorization secret key $sk_{\Sigma^*}$. Thus, $\cal A^*$ cannot
  recover $(\mathcal{Q}_{Min},\mathcal{Y}_{Min},W_{Min})$.\\
  \end{enumerate}

According to the analysis, \texttt{SMin} is secure against the
 adversary $\mathcal{A}^*$ defined in the attack model.
 \hfill$\square$\vspace{2mm}

\subsubsection*{D-2. Proof of \texttt{BPS-$k$}}

\textbf{Theorem 6-1.} \emph{\texttt{BPS-$k$}
is secure to select the top-$k$ best encrypted treatment procedures
(with lowest weights) in the presence of semi-honest (non-colluding) attackers
$\mathcal{A}=(\mathcal{A}_{D_1},\mathcal{A}_{S_1},\mathcal{A}_{S_2})$.}

\emph{\textbf{Proof}}. We now construct the following independent
simulators $(Sim_{D_1},Sim_{S_1},Sim_{S_2})$.

$Sim_{D_1}$ receives $\mathcal{ETP}_i$ (for $1\leq i \leq n$) as
inputs and simulates $\mathcal{A}_{D_1}$ as following, where
\begin{align*}
\mathcal{ETP}_i&=(\mathcal{Q}_i, \mathcal{Y}_i, W_i),\\
\mathcal{Q}_i&=(q_0, q_{i,\theta_1},\cdots,q_{i,MState-1}),\\
\mathcal{Y}_i&=(y_{i,\theta_1},\cdots,y_{i,MState-1}),
\end{align*}
for $1\leq i \leq n$.

It generates $[\![\mathcal{ETP}_i]\!]_{pk_A}$ for $1\leq i \leq n$, where
\begin{align*}
[\![\mathcal{ETP}_i]\!]_{pk_A}&=([\![\mathcal{Q}_i]\!]_{pk_{A}}, [\![\mathcal{Y}_i]\!]_{pk_{A}}, [\![W_i]\!]_{pk_\sigma}),\\
[\![\mathcal{Q}_i]\!]_{pk_{A}}&=([\![q_0]\!]_{pk_{A}}, [\![q_{i,\theta_1}]\!]_{pk_{A}},\cdots,[\![q_{i,MState-1}]\!]_{pk_{A}}),\\
[\![\mathcal{Y}_i]\!]_{pk_{A}}&=([\![y_{i,\theta_1}]\!]_{pk_{A}},\cdots,[\![y_{i,MState-1}]\!]_{pk_{A}}).
\end{align*}

The entire view
of $\mathcal{A}_{D_1}$ is the received tuples and the encrypted
data. The IND-CPA security of PCTD ensures that
$\mathcal{A}_{D_1}$'s view is indistinguishable from its view in the
real world execution.

$Sim_{S_1}$ simulates $\mathcal{A}_{S_1}$ as following.
It randomly generates $[\![\mathbb{ETP}]\!]_{pk_A}=([\![\mathcal{ETP}_1]\!]_{pk_{A}},\cdots, [\![\mathcal{ETP}_n]\!]_{pk_{A}})$ and inputs them into $Sim_{S_1}^{\texttt{SMin}_n}$ in
Theorem \ref{Theorem:SMinN} and gets
$[\![\mathcal{ETP}_{Min_i}]\!]_{pk_\sigma}=([\![\mathcal{Q}_{Min_i}]\!]_{pk_\sigma}, [\![\mathcal{Y}_{Min_i}]\!]_{pk_\sigma}, [\![W_{Min_i}]\!]_{pk_\sigma}),$
where $W_{Min_i}$ is the lowest weight in the $i$-th round.
$Sim_{S_1}$ randomly
selects $r_j',r_j \in Z_N$, and computes
$l_j=([\![W_{Min_i}]\!]_{pk_\sigma})^{r_j}\cdot([\![W_{j}]\!]_{pk_\sigma})^{N-r_j}$. Then, $Sim_{S_1}$ partially decrypts $l_j$ to
$l_j'=\texttt{PD1}_{SK_1}(l_j)$, and
permutes $(l_j,l_j')$ using the permutation function $\pi_i$.
The result is denoted as $(l_{\pi_i(j)},l_{\pi_i(j)}')$.
$Sim_{S_1}$ sends $(l_{\pi_i(j)},l_{\pi_i(j)}')$ and the intermediate
encrypted data of $Sim_{S_1}^{\texttt{SMin}_n}$ to
$\mathcal{A}_{S_1}$. If $\mathcal{A}_{S_1}$ replies with $\bot$,
then $Sim_{S_1}$ outputs $\bot$. The IND-CPA security of PCTD
ensures that $\mathcal{A}_{S_1}$'s view is indistinguishable from
its view in the real world execution.

$Sim_{S_2}$ simulates $\mathcal{A}_{S_2}$ as following.
It randomly generates $\rho\in\{0,1\}$.
If $\rho=0$, it sets  $A_{\pi_i(j)}= [\![MWeight]\!]_{pk_\sigma}$;
otherwise, it sets  $A_{\pi_i(j)}= [\![1]\!]_{pk_\sigma}$.
The
generated ciphertexts $A_{\pi_i(j)}$ is computationally indistinguishable from the
real world due to the IND-CPA security of PCTD. In both real and
ideal world, the views of $\mathcal{A}_{S_2}$ are indistinguishable.
\hfill$\square$\vspace{2mm}

\textbf{Theorem 6-2.} \emph{\texttt{BPS-$k$} is secure against the
 adversary $\mathcal{A}^*$ defined in the attack model.}

\emph{\textbf{Proof}}. The adversary $\mathcal{A}^*$ is assumed to have the following abilities.
\begin{enumerate}
  \item [(1)] $\mathcal{A}^*$ is assumed to be an outside adversary and eavesdrop
  all the communications to get the transmitted data. As $\mathcal{A}^*$ is assumed to be an outside adversary, $\mathcal{A}^*$ cannot get hospital $A$'s secret key
  $sk_A$, patient $B$'s secret key $sk_B$ and $B$'s authorization secret key $sk_\sigma$.
  $\mathcal{A}^*$ also cannot get CP's partial strong key $SK_1$ and CSP's partial strong key $SK_2$.
  If $\mathcal{A}^*$ eavesdrops the communication channel between system user and CP, $\mathcal{A}^*$ could get $[\![\mathbb{ETP}]\!]_{pk_A}=([\![\mathcal{ETP}_1]\!]_{pk_\sigma},\cdots,[\![\mathcal{ETP}_n]\!]_{pk_\sigma})$ that
  is transmitted at the beginning of \texttt{BPS-$k$},
  and $[\![\mathbb{ETP}_{Min}]\!]_{pk_\sigma}=([\![\mathcal{ETP}_{Min_1}]\!]_{pk_\sigma},\cdots,[\![\mathcal{ETP}_{Min_k}]\!]_{pk_\sigma})$ that is sent at the end of \texttt{BPS-$k$},
  where $[\![\mathcal{ETP}_{Min_i}]\!]_{pk_A}=([\![\mathcal{Q}_{Min_i}]\!]_{pk_{A}}, [\![\mathcal{Y}_{Min_i}]\!]_{pk_{A}}, [\![W_{Min_i}]\!]_{pk_\sigma})$ for $i\in[\![1,k]\!]$.
  Since the elements in $[\![\mathbb{ETP}]\!]_{pk_A}$ and $[\![\mathbb{ETP}_{Min}]\!]_{pk_\sigma}$ are encrypted using the PCTD algorithm,
  the adversary $\mathcal{A}^*$ cannot recover $\mathcal{ETP}_i=(\mathcal{Q}_i, \mathcal{Y}_i, \mathcal{W}_i)$
  for $i\in[\![1,n]\!]$ and the plaintext underlying $[\![\mathbb{ETP}_{Min}]\!]_{pk_\sigma}$ due to the IND-CPA security of PCTD.

  If $\mathcal{A}^*$ eavesdrops the communication channel between CP and
  CSP, $\mathcal{A}^*$ could get $(l_{\pi_i(j)},l_{\pi_i(j)}')$ in line 6,
  and $A_{\pi_i(j)}$ in line 8 of \texttt{BPS-$k$},
  where $l_j=([\![W_{Min_i}]\!]_{pk_\sigma})^{r_j}\cdot([\![W_{j}]\!]_{pk_\sigma})^{N-r_j}$,
$l_j'=\texttt{PD1}_{SK_1}(l_j)$ and $A_{\pi_i(j)}= [\![MWeight]\!]_{pk_\sigma}$ or $A_{\pi_i(j)}= [\![1]\!]_{pk_\sigma}$.
  In \texttt{BPS-$k$}, the ciphertexts $(l_{\pi_i(j)},l_{\pi_i(j)}',A_{\pi_i(j)})$ are all
  encrypted using the public key $pk_\sigma$.
Since the adversary $\mathcal{A}^*$ does know the patient $B$'s authorization secret key $sk_\sigma$ and
the CSP's partial strong key $SK_2$, $\mathcal{A}^*$ cannot recover the underlying plaintexts.\\

  \item [(2)] $\mathcal{A}^*$ is assumed to compromise CP and get CP's partial strong key $SK_1$.
  But $\mathcal{A}^*$ cannot get CSP's partial strong key $SK_2$.
  $\mathcal{A}^*$ also cannot get hospital $A$'s secret key
  $sk_A$, patient $B$'s secret key $sk_B$ and $B$'s authorization secret key $sk_\sigma$.

  In line 6 of \texttt{BPS-$k$}, $\mathcal{A}^*$ obtains
  \begin{eqnarray*}
  l_j&=&([\![W_{Min_i}]\!]_{pk_\sigma})^{r_j}\cdot([\![W_j]\!]_{pk_\sigma})^{N-r_j}\\
  &=&[\![r_j(W_{Min_i}-W_j)]\!]_{pk_\sigma}
  \end{eqnarray*}

  $\mathcal{A}^*$ cannot recover $r_j(W_{Min_i}-W_j)$ without the secret key
  $sk_\sigma$. In line 8, $\mathcal{A}^*$ obtains $A_{\pi_i(j)}$ from CSP, where
  \begin{eqnarray*}
  A_{\pi_i(j)}
  &=&\left\{
  \begin{aligned}
  &[\![MWeight]\!]_{pk_\sigma},  & ~~~if~l_{\pi_i(j)}''=0,\\
  &[\![1]\!]_{pk_\sigma},          & otherwise.
  \end{aligned}
  \right.
  \end{eqnarray*}

  Since $l_{\pi_i(j)}''$ and $sk_\sigma$ are unknown to CP, $\mathcal{A}^*$ cannot decrypt $A_{\pi_i(j)}$ nor distinguish $[\![\mathcal{ETP}_{Min_i}]\!]_{pk_\sigma}$
  comes from which element in $[\![\mathbb{ETP}]\!]_{pk_A}$.\\

  \item [(3)] $\mathcal{A}^*$ is assumed to compromise CSP and get CSP's partial strong key $SK_2$.
  But $\mathcal{A}^*$ cannot get CP's partial strong key $SK_1$.
  $\mathcal{A}^*$ also cannot get hospital $A$'s private key
  $sk_A$, patient $B$'s private key $sk_B$ and $B$'s authorization secret key $sk_\sigma$.

  In line 6 of \texttt{BPS-$k$}, $\mathcal{A}^*$ obtains $(l_{\pi_i(j)},l_{\pi_i(j)}')$ transmitted by CP.
  Since $\mathcal{A}^*$ knows CSP's partial strong key $SK_2$,
  $\mathcal{A}^*$ decrypts
  \begin{eqnarray*}
  l_{\pi_i(j)}''&=&\texttt{PD2}_{SK_2}(l_{\pi_i(j)},l_{\pi_i(j)}')\\
  &=&r_{\pi_i(j)}(W_{Min_i}-W_{\pi_i(j)})
  \end{eqnarray*}
  Although $\mathcal{A}^*$ knows whether $r_{\pi_i(j)}(W_{Min_i}-W_{\pi_i(j)})$ equals 0,
  the adversary $\mathcal{A}^*$ cannot distinguish $W_{\pi_i(j)}$ comes from which element in $(W_1,\cdots,W_n)$.
  The reason is that CP utilizes a permutation function $\pi_i$ to disrupt the order of $(l_1,\cdots,l_n)$ and $(l_1',\cdots,l_n')$
  in line 6 of \texttt{BPS-$k$}. Thus, $\mathcal{A}^*$ cannot distinguish $[\![\mathcal{ETP}_{Min_i}]\!]_{pk_\sigma}$
  comes from which element in $[\![\mathbb{ETP}]\!]_{pk_A}=([\![\mathcal{ETP}_1]\!]_{pk_\sigma},\cdots,[\![\mathcal{ETP}_n]\!]_{pk_\sigma})$.\\

   \item [(4)] $\cal A^*$ is assumed to be the collude malicious patients $(B_1,\cdots,B_n)$ (except the
  challenge patient $B^*$), and $\cal A^*$ gets their secret keys $(sk_{B_1},\cdots,sk_{B_n})$.
  $\cal A^*$ wants to get the information that belongs to
  the challenge patient $B^*$. Suppose the returned result is
  $[\![\mathbb{ETP}_{Min}]\!]_{pk_{\Sigma^*}}=
  ([\![\mathcal{ETP}_{Min_1}]\!]_{pk_{\Sigma^*}},\cdots,[\![\mathcal{ETP}_{Min_k}]\!]_{pk_{\Sigma^*}})$
  that is sent at the end of the protocol,
  where $[\![\mathcal{ETP}_{Min_i}]\!]_{pk_{\Sigma^*}}=([\![\mathcal{Q}_{Min_i}]\!]_{pk_{\Sigma^*}}, [\![\mathcal{Y}_{Min_i}]\!]_{pk_{\Sigma^*}}, [\![W_{Min_i}]\!]_{pk_{\Sigma^*}})$
  $(i\in[\![1,k]\!])$,
  and $pk_{\Sigma^*}$ is the authorize public key from hospital $A$ to challenge patient $B^*$.
  Since the patient's secret keys are independently generated, the adversary $\cal A^*$
  cannot utilize $(sk_{B_1},\cdots,sk_{B_n})$ to deduce the challenge patient $B^*$'s secret key $sk_{B^*}$.
  $\cal A^*$ also cannot get $sk_{\Sigma^*}$. Thus, $\cal A^*$ cannot
  recover $(\mathcal{Q}_{Min_i},\mathcal{Y}_{Min_i},W_{Min_i})$ for $i\in[\![1,k]\!]$.  \vspace{2mm}
  \end{enumerate}

According to the above analysis, \texttt{BPS-$k$} is secure against the
 adversary $\mathcal{A}^*$ defined in the attack model.
 \hfill$\square$\\

\subsection*{E. Detailed Experiment Data}

This section presents the detailed experiment data of the performance analysis shown in Section \ref{Sec:Performance}.
Table \ref{Table:SSM-TPW-SMin} shows the experiment data of Fig. \ref{SubFig:SSM-TPW-SMin-cal}-\ref{SubFig:SSM-TPW-SMin-com}.
Table \ref{Table:TPW} presents the detailed data of Fig. \ref{SubFig:TPW-cal}-\ref{SubFig:TPW-com}.
Table \ref{Table:SMinN-BPSK} shows the experiment data of Fig. \ref{SubFig:SMinN-cal}-\ref{SubFig:BPSK-com}.
Table \ref{Table:P-Med} gives out the detailed data of Fig. \ref{SubFig:100-cal}-\ref{SubFig:100-com}.
Table \ref{Table:P-Gene} gives out the detailed data of Fig. \ref{SubFig:PGene-cal}-\ref{SubFig:PGene-com}.


\renewcommand\arraystretch{1.3}
\begin{table*}[htbp]
\caption{Performance of \texttt{SSM}, \texttt{TPW} and \texttt{SMin} ($MState=10$, $m=3$, $n=1$)}
\label{Table:SSM-TPW-SMin} \centering
\begin{tabular}{Ic||c|c|c|c|c|c|c||c|c|c|c|c|c|cI}
  \thickhline
   & \multicolumn{7}{c||}{\textbf{Computation Cost (s)}} & \multicolumn{7}{cI}{\textbf{Communication Cost (MB)}}\\
  \hline
  $\boldsymbol{\mathcal{L}(N)}$ & \textbf{512} & \textbf{768} & \textbf{\textcolor[rgb]{0,0,1}{1024}} & \textbf{1280} & \textbf{1536} & \textbf{1792} & \textbf{2048} & \textbf{512} & \textbf{768} & \textbf{\textcolor[rgb]{0,0,1}{1024}} & \textbf{1280} & \textbf{1536} & \textbf{1792} & \textbf{2048}\\
  \thickhline
  \texttt{SSM} & 0.361 &  1.079 & \textcolor[rgb]{0,0,1}{2.343}  & 4.162 &  6.658  &  12.005  & 18.012
  & 0.052 & 0.078 & \textcolor[rgb]{0,0,1}{0.104} & 0.130 & 0.156 & 0.182 & 0.208 \\
  \hline
  \texttt{TPW} & 6.423 & 20.155 & 	\textcolor[rgb]{0,0,1}{42.427}& 	81.127	& 127.588	& 220.552     & 291.764
  & 1.248 & 1.876 & \textcolor[rgb]{0,0,1}{2.503} & 3.130 & 3.757 & 4.382 & 5.010 \\
  \hline
  \texttt{SMin} & 0.777	& 2.194	& \textcolor[rgb]{0,0,1}{4.418}	& 10.154	& 16.211	& 25.062	& 38.442
  & 0.181	&0.272	& \textcolor[rgb]{0,0,1}{0.363}	&0.455	&0.546	&0.636	&0.728\\
  \thickhline
\end{tabular}
\end{table*}

\renewcommand\arraystretch{1.3}
\begin{table}[htbp]
\caption{Performance of \texttt{TPW} ($\mathcal{L}(N)=1024$)}
\label{Table:TPW} \centering
\newcommand{\tabincell}[2]{\begin{tabular}{@{}#1@{}}#2\end{tabular}}
\begin{tabular}{Ic|c|c|cI}
  \thickhline
  \multirow{2}*{$\mathcal{AVG}(\mathbb{TP})$} & \multirow{2}*{$\boldsymbol{m}$} & \textbf{Computation} & \textbf{Communication} \\
   & & \textbf{(min)} & \textbf{(MB)}\\
  \thickhline						
  \multirow{5}{*}{\tabincell{c}{10}} &1 & 0.653	&1.356 \\
  \cline{2-4}						
   &2 &  0.674	&2.033  \\
  \cline{2-4}						
   &3 &  0.690	&2.503   \\
  \cline{2-4}						
   &4 &  0.712	&2.765   \\
   \cline{2-4}						
   &5 &  0.719	&2.823   \\
  \hline
  \multirow{5}{*}{\tabincell{c}{20}} &1 &  1.859 	&3.050 \\
  \cline{2-4}						
   &2 &  1.934 	&4.749  \\
  \cline{2-4}						
   &3 &  1.975 	&6.243 \\
  \cline{2-4}						
   &4 &  1.997 	&7.529  \\
  \cline{2-4}						
   &5 &  2.052 	&8.598   \\
  \hline
  \multirow{5}{*}{\tabincell{c}{30}} &1 &  3.296	&4.946 \\
  \cline{2-4}						
   &2 &  3.368  &  7.670 \\
  \cline{2-4}						
   &3 &  3.411	&10.189 \\
  \cline{2-4}						
   &4 &  3.438 &12.495   \\
  \cline{2-4}						
   &5 &  3.478	&14.592   \\
  \hline
  \multirow{5}{*}{\tabincell{c}{40}} &1 &  4.735	&17.66 \\
  \cline{2-4}						
   &2 &  4.926	&10.786  \\
  \cline{2-4}						
   &3 &  5.164	&14.333 \\
  \cline{2-4}						
   &4 &  5.331	&17.66   \\
  \cline{2-4}						
   &5 & 5.436	&20.789   \\
  \hline
  \multirow{5}{*}{\tabincell{c}{50}} &1 &  6.670	&9.357 \\
  \cline{2-4}						
   &2 &  6.903	&14.118 \\
  \cline{2-4}						
   &3 &  7.031	&18.669 \\
  \cline{2-4}						
   &4 &  7.263	&23.041   \\
  \cline{2-4}						
   &5 &  7.461	&27.201   \\
  \thickhline
\end{tabular}
\end{table}

\renewcommand\arraystretch{1.2}
\begin{table}[htbp]
\caption{Performance of \texttt{SMin$_n$} and \texttt{BPS$_k$} ($k=1$)}
\label{Table:SMinN-BPSK} \centering
\newcommand{\tabincell}[2]{\begin{tabular}{@{}#1@{}}#2\end{tabular}}
\begin{tabular}{Ic|cIc|cIc|cI}
  \thickhline						
  \multicolumn{2}{IcI}{\multirow{2}*{\textbf{Parameter}}}& \multicolumn{2}{cI}{\textbf{Computation}} & \multicolumn{2}{cI}{\textbf{Communication}}\\
  \multicolumn{2}{IcI}{} & \multicolumn{2}{cI}{\textbf{(min)}} & \multicolumn{2}{cI}{\textbf{(MB)}}\\
  \thickhline
  $\boldsymbol{n}$ & $\boldsymbol{MState}$ & \textbf{\texttt{SMin$\boldsymbol{_n}$}} & \textbf{\texttt{BPS$\boldsymbol{_k}$}}& \textbf{\texttt{SMin$\boldsymbol{_n}$}} & \textbf{\texttt{BPS$\boldsymbol{_k}$}} \\
  \thickhline
  \multirow{5}{*}{\tabincell{c}{20}} &10 & 0.957	&1.018        &6.910    &7.042\\
  \cline{2-6}						
   &20 &  1.765	&1.829        &13.704    &13.838\\
  \cline{2-6}						
   &30 &  2.521	&2.965       &20.506    &20.654\\
  \cline{2-6}						
   &40 &  3.520	&3.695        &27.312    &27.443\\
   \cline{2-6}						
   &50 &  4.331	&4.587        &34.123    &34.250\\
  \hline\hline
  \multirow{5}{*}{\tabincell{c}{40}} &10 &  1.719 	&1.805        &14.196    &14.443\\
  \cline{2-6}						
   &20 &  3.199 	&3.512        &28.140     &28.396\\
  \cline{2-6}						
   &30 &  5.078	&5.388        &41.112    &42.396\\
  \cline{2-6}						
   &40 &  5.976 	&6.237        &56.036    &56.284 \\
  \cline{2-6}						
   &50 &  8.644 	&9.028        &70.046    &70.274\\
  \hline\hline
  \multirow{5}{*}{\tabincell{c}{60}} &10 &  2.077	&2.498        &21.457    &21.871\\
  \cline{2-6}						
   &20 &  4.096  &4.773        &42.595       &42.979\\
  \cline{2-6}						
   &30 &  6.909	&7.279        &63.633      &64.091\\
  \cline{2-6}						
   &40 &  9.667 &10.112        &84.810        &85.270\\
  \cline{2-6}						
   &50 &  11.741	&12.244        &105.864     &106.342\\
  \hline\hline
  \multirow{5}{*}{\tabincell{c}{80}} &10 &  2.907	&3.383        &28.753     &29.279\\
  \cline{2-6}						
   &20 &  6.730	&7.039        &57.030      &57.508\\
  \cline{2-6}						
   &30 &  9.342	&9.670        &85.213      &85.864\\
  \cline{2-6}						
   &40 &  11.905	&12.896        &113.574     &114.056\\
  \cline{2-6}						
   &50 & 15.025	&16.634        &141.774     &142.311 \\
  \hline\hline
  \multirow{5}{*}{\tabincell{c}{100}} &10 &  3.821 	&4.025        &35.991     &36.666\\
  \cline{2-6}						
   &20 &  7.981	&8.271        &71.471        &72.115\\
  \cline{2-6}						
   &30 &  10.527	 &11.605        &106.810      &107.449\\
  \cline{2-6}						
   &40 &  15.261	 &16.386        &142.310      &142.955\\
  \cline{2-6}						
   &50 &  19.018	 &21.531        &177.625      &178.266\\
  \thickhline
\end{tabular}
\end{table}

\renewcommand\arraystretch{1.2}
\begin{table*}[htbp]
\caption{Performance of P-Med ($\mathcal{L}(N)=1024,k=1$)}
\label{Table:P-Med} \centering
\newcommand{\tabincell}[2]{\begin{tabular}{@{}#1@{}}#2\end{tabular}}
\begin{tabular}{Ic|c|c||c|cIc|c|c||c|cI}
  \thickhline
  \multirow{2}*{$\boldsymbol{n}$} & \multirow{2}*{$\mathcal{AVG}(\mathbb{TP})$} & \multirow{2}*{$\boldsymbol{m}$} & \textbf{Computation} & \textbf{Communication}
  & \multirow{2}*{$\boldsymbol{n}$} & \multirow{2}*{$\mathcal{AVG}(\mathbb{TP})$} & \multirow{2}*{$\boldsymbol{m}$} & \textbf{Computation} & \textbf{Communication} \\
   & & & \textbf{(min)} & \textbf{(MB)}& & & & \textbf{(min)} & \textbf{(MB)}\\
  \thickhline						
    \multirow{25}{*}{\tabincell{c}{\textbf{40}}} &\multirow{5}{*}{\tabincell{c}{10}} &1 & 2.778	& 68.683
    &\multirow{25}{*}{\tabincell{c}{\textbf{60}}} &\multirow{5}{*}{\tabincell{c}{10}} &1 & 3.711	& 103.231\\
  \cline{3-5}\cline{8-10}						
   &&2 &  2.799 	& 95.763  &&&2 &  3.732	& 143.851 \\
  \cline{3-5}\cline{8-10}						
   &&3 &  2.815 	& 114.563  &&&3 &  3.748 	& 172.051   \\
  \cline{3-5}\cline{8-10}						
   &&4 &  2.837 	& 125.043  &&&4 &  3.770 	& 187.771    \\
  \cline{3-5}\cline{8-10}						
   &&5 &  2.844 	& 127.363  &&&5 &  3.777 	& 119.251  \\
  \cline{2-5}\cline{7-10}
  &\multirow{5}{*}{\tabincell{c}{20}} &1 &  6.231 	& 150.396 &&\multirow{5}{*}{\tabincell{c}{20}}&1 &  7.596 	& 225.979 \\
  \cline{3-5}\cline{8-10}						
   &&2 &  6.306 	& 218.356  &&&2 & 7.671 	& 327.919 \\
  \cline{3-5}\cline{8-10}						
   &&3 &  6.347	&278.116  &&&3 &  7.712	&417.559 \\
  \cline{3-5}\cline{8-10}						
   &&4 &  6.369	&329.556  &&&4 &  7.734	&494.719 \\
  \cline{3-5}\cline{8-10}						
   &&5 &  6.424	&372.316  &&&5 &  7.789	&558.859    \\
  \cline{2-5}\cline{7-10}
  &\multirow{5}{*}{\tabincell{c}{30}} &1 &  9.995	&240.236 &&\multirow{5}{*}{\tabincell{c}{30}}&1 &  11.809	&360.851 \\
  \cline{3-5}\cline{8-10}						
   &&2 &  10.067	&349.196  &&&2 &  11.881	&524.291 \\
  \cline{3-5}\cline{8-10}						
   &&3 &  10.11	&449.956  &&&3 &   11.924	&675.431 \\
  \cline{3-5}\cline{8-10}						
   &&4 &  10.137	&542.196 &&&4 &  11.951	&813.791   \\
  \cline{3-5}\cline{8-10}						
   &&5 &  10.177	&626.076 &&&5 &  11.991	&939.611   \\
  \cline{2-5}\cline{7-10}
  &\multirow{5}{*}{\tabincell{c}{40}} &1 &  12.985	&338.244 &&\multirow{5}{*}{\tabincell{c}{40}}&1 &  17.081	&508.210 \\
  \cline{3-5}\cline{8-10}						
   &&2 &  13.176	&487.724   &&&2 &  17.272	&732.43  \\
  \cline{3-5}\cline{8-10}						
   &&3 &  13.414	&629.604 &&&3 &  17.51	&945.25 \\
  \cline{3-5}\cline{8-10}						
   &&4 &  13.581	&762.684   &&&4 &  17.677	&1144.87  \\
  \cline{3-5}\cline{8-10}						
  &&5 & 13.686	&887.844   &&&5 & 17.782	&1332.61   \\
  \cline{2-5}\cline{7-10}
  &\multirow{5}{*}{\tabincell{c}{50}} &1 &  18.78	&444.554
   &&\multirow{5}{*}{\tabincell{c}{50}}&1 &  22.37	&667.762 \\
  \cline{3-5}\cline{8-10}						
  &&2 &  19.013	&634.994  &&&2 &  22.603	&953.422 \\
  \cline{3-5}\cline{8-10}						
    &&3 &  19.141	&817.034 &&&3 &  22.731	&1226.482 \\
  \cline{3-5}\cline{8-10} 						
   &&4 &  19.373	&991.914   &&&4 &  22.963	&1488.802   \\
  \cline{3-5}\cline{8-10}					
   &&5 &  19.571	&1158.314   &&&5 &  23.161	&1738.402  \\
  \thickhline
    \multirow{25}{*}{\tabincell{c}{\textbf{80}}} &\multirow{5}{*}{\tabincell{c}{10}} &1 & 5.049	&137.759

    &\multirow{25}{*}{\tabincell{c}{\textbf{100}}} &\multirow{5}{*}{\tabincell{c}{10}} &1 & 5.71	&172.266
\\
  \cline{3-5}\cline{8-10}						
   &&2 &  5.07	&191.919  &&&2 &  5.731	&239.966  \\
  \cline{3-5}\cline{8-10}						
   &&3 &  5.086	&229.519  &&&3 &  5.747	&286.966   \\
  \cline{3-5}\cline{8-10}						
   &&4 &  5.108	&250.479  &&&4 &  5.769	&313.166    \\
  \cline{3-5}\cline{8-10}						
   &&5 &  5.115	&255.119  &&&5 &  5.776	&318.966  \\
  \cline{2-5}\cline{7-10}
  &\multirow{5}{*}{\tabincell{c}{20}} &1 &  10.552	&301.508 &&\multirow{5}{*}{\tabincell{c}{20}}&1 &  12.452	&377.115 \\
  \cline{3-5}\cline{8-10}						
   &&2 &  10.627	&437.428  &&&2 &  12.527	&547.015 \\
  \cline{3-5}\cline{8-10}						
   &&3 &  10.668	&556.948  &&&3 &  12.568	&696.415 \\
  \cline{3-5}\cline{8-10}						
   &&4 &  10.690	&659.828  &&&4 &  12.593	&825.015 \\
  \cline{3-5}\cline{8-10}						
   &&5 &  10.745	&745.348  &&&5 &  12.645	&931.915    \\
  \cline{2-5}\cline{7-10}
  &\multirow{5}{*}{\tabincell{c}{30}} &1 &  15.624	&481.544 &&\multirow{5}{*}{\tabincell{c}{30}}&1 &  17.912	&602.049 \\
  \cline{3-5}\cline{8-10}						
   &&2 &  15.696	&699.464  &&&2 &  17.984	&874.449 \\
  \cline{3-5}\cline{8-10}						
   &&3 &  15.739	&900.984  &&&3 &   18.027	&1126.349 \\
  \cline{3-5}\cline{8-10}						
   &&4 &  15.766	&1085.464 &&&4 &  18.054	&1356.949   \\
  \cline{3-5}\cline{8-10}						
   &&5 &  15.806	&1253.224 &&&5 &  18.094	&1566.649   \\
  \cline{2-5}\cline{7-10}
  &\multirow{5}{*}{\tabincell{c}{40}} &1 &  20.982	&677.976 &&\multirow{5}{*}{\tabincell{c}{40}}&1 &  25.357	&847.855 \\
  \cline{3-5}\cline{8-10}						
   &&2 &  21.173	&976.936   &&&2 &  25.548	&1221.555  \\
  \cline{3-5}\cline{8-10}						
   &&3 &  21.411	&1260.696 &&&3 &  25.786	&1576.255 \\
  \cline{3-5}\cline{8-10}						
   &&4 &  21.578	&1526.856   &&&4 &  25.953	&1908.955  \\
  \cline{3-5}\cline{8-10}						
  &&5 & 21.683	&1777.176   &&&5 & 26.058	&2221.855   \\
  \cline{2-5}\cline{7-10}
  &\multirow{5}{*}{\tabincell{c}{50}} &1 &  27.986	&890.871 &&\multirow{5}{*}{\tabincell{c}{50}}&1 &  32.898	&1113.966 \\
  \cline{3-5}\cline{8-10}						
  &&2 &  28.219	&1271.751  &&&2 &  33.131	&1590.066 \\
  \cline{3-5}\cline{8-10}						
    &&3 &  28.347	&1635.831 &&&3 &  33.259	&2045.166 \\
  \cline{3-5}\cline{8-10} 						
   &&4 &  28.579	&1985.591   &&&4 &  33.491	&2482.366   \\
  \cline{3-5}\cline{8-10}				
   &&5 &  28.777	&2318.391   &&&5 &  33.689	&2898.366   \\
  \thickhline
\end{tabular}
\end{table*}

\renewcommand\arraystretch{1.3}
\begin{table}[htbp]
	\caption{Performance of P-Gene ($\mathcal{L}(N)=1024$)}
	\label{Table:P-Gene} \centering
	\newcommand{\tabincell}[2]{\begin{tabular}{@{}#1@{}}#2\end{tabular}}
	\begin{tabular}{Ic|c|c||c|cIc|c|c||c|cI}
		\thickhline
		\multirow{2}*{$\boldsymbol{m}$} & \multirow{2}*{$\boldsymbol{\mu}$} & \multirow{2}*{$\boldsymbol{n}$} & \textbf{Computation} & \textbf{Communication}\\
		& & & \textbf{(min)} & \textbf{(MB)}\\
		\thickhline						
		\multirow{3}{*}{\tabincell{c}{10}} &\multirow{3}{*}{\tabincell{c}{2}} &$m-\mu=8$ & 1.097	& 4.294\\
		\cline{3-5}					
		&&$m=10$ &  1.371 	& 5.368  \\
		\cline{3-5}					
		&&$m+\mu=12$ &  1.645 	& 6.442  \\	
		\hline\hline
		\multirow{3}{*}{\tabincell{c}{20}} &\multirow{3}{*}{\tabincell{c}{4}} &$m-\mu=16$ & 3.544	& 30.269\\
		\cline{3-5}					
		&&$m=20$ &  4.430 	& 37.837 \\
		\cline{3-5}					
		&&$m+\mu=24$ &  5.316 	&  45.404 \\	
		\hline\hline
		\multirow{3}{*}{\tabincell{c}{30}} &\multirow{3}{*}{\tabincell{c}{6}} &$m-\mu=24$ & 9.696	& 97.559\\
		\cline{3-5}					
		&&$m=30$ &  12.121 	& 121.948  \\
		\cline{3-5}					
		&&$m+\mu=36$ &  14.545 	& 146.338  \\	
		\hline\hline
		\multirow{3}{*}{\tabincell{c}{40}} &\multirow{3}{*}{\tabincell{c}{8}} &$m-\mu=32$ & 19.484	& 225.797\\
		\cline{3-5}					
		&&$m=40$ &  24.355 	& 282.246  \\
		\cline{3-5}					
		&&$m+\mu=48$ &  29.226 	& 338.697  \\
			\hline\hline
		\multirow{3}{*}{\tabincell{c}{50}} &\multirow{3}{*}{\tabincell{c}{10}} &$m-\mu=40$ & 35.197	& 434.619\\
		\cline{3-5}					
		&&$m=50$ &  43.996 	& 543.274  \\
		\cline{3-5}					
		&&$m+\mu=60$ &  52.795 	& 651.929  \\
		\thickhline
	\end{tabular}
\end{table}

\end{document}